\documentclass{aa}  

\usepackage{graphicx}
\usepackage{amssymb} 
\usepackage{multirow} 
\usepackage{longtable}
\usepackage[dvipsnames]{xcolor}
\usepackage{soul}
\usepackage{txfonts}
\usepackage{xcolor}
\usepackage{pict2e}
\newsavebox{\ORCIDlogo}
\savebox{\ORCIDlogo}{
\setlength{\unitlength}{\dimexpr 1em/256\relax}
\begin{picture}(256,256)
  \color[HTML]{A6CE39}\put(128,128){\circle*{256}}
  \color{white}
  \put(78.6,199.2){\circle*{20}}
  \moveto(70.9,176,9)\lineto(86.3,176,9)\lineto(86.3,69.8)\lineto(70.9,69.8)
  \closepath\fillpath
  \moveto(108.9,176.9)\lineto(150.5,176.9)
  \curveto(190.1,176.9)(207.5,148.6)(207.5 ,123.3)
  \curveto(207.5,95,8)(186,69.7)(150.7,69.7)
  \lineto(108.9,69.7)
  \closepath\fillpath
  \color[HTML]{A6CE39}
  \moveto(124.3,83.6)\lineto(148.8,83.6)
  \curveto(183.7,83.6)(191.7,110.1)(191.7,123.3)
  \curveto(191.7,144.8)(178,163)(148,163)
  \lineto(124.3,163)
  \closepath\fillpath
\end{picture}
}
\newcommand\orcid[1]{\href{https://orcid.org/#1}{\usebox{\ORCIDlogo}}}
\usepackage{hyperref} 

\begin{document} 

\titlerunning{GRB taxonomy in the rest frame}
\authorrunning{Tsvetkova et al.}

\title{Gamma-ray burst taxonomy: looking for the third class on the spectral peak energy-duration plane in the rest frame}

\author{
    Anastasia~Tsvetkova\thanks{tsvetkova.lea@gmail.com}\inst{\ref{unica},\ref{ioffe}}\orcid{0000-0003-0292-6221}
    \and Lorenzo Amati\inst{\ref{inaf-oas}}\orcid{0000-0001-5355-7388}
    \and Mattia~Bulla\inst{\ref{unife},\ref{infn-fe},\ref{inaf-oaab}}\orcid{0000-0002-8255-5127}
    \and Luciano~Burderi\inst{\ref{inaf-iasf}, \ref{unica}, \ref{infn-ca}}\orcid{0000-0001-5458-891X}
    \and Dmitry~Frederiks\inst{\ref{ioffe}}\orcid{0000-0002-1153-6340}
    \and Filippo~Frontera\inst{\ref{unife}, \ref{inaf-oas}}\orcid{0000-0003-2284-571X}
    \and Cristiano~Guidorzi\inst{\ref{unife},\ref{inaf-oas},\ref{infn-fe}}\orcid{0000-0001-6869-0835}
    \and Alessandro~Riggio\inst{\ref{unica}, \ref{infn-ca}, \ref{inaf-iasf}}\orcid{0000-0002-6145-9224}
    \and Tiziana~di~Salvo\inst{\ref{unipa}}\orcid{0000-0002-3220-6375}
    \and Andrea~Sanna\inst{\ref{unica}, \ref{infn-ca}, \ref{inaf-oac}}\orcid{0000-0002-0118-2649}
    \and Fyodor~Sviridov\inst{\ref{au}}\orcid{0009-0001-4543-6460}
}

\institute{
Dipartimento di Fisica, Universit\`{a} degli Studi di Cagliari, SP Monserrato-Sestu, km 0.7, I-09042 Monserrato, Italy \label{unica}
\and
Ioffe Institute, Politekhnicheskaya 26, 194021 St. Petersburg, Russia\label{ioffe}
\and 
INAF -- Osservatorio di Astrofisica e Scienza dello Spazio di Bologna, Via Piero Gobetti 93/3, 40129 Bologna, Italy \label{inaf-oas}
\and 
Department of Physics and Earth Science, University of Ferrara, Via Saragat 1, I-44122 Ferrara, Italy \label{unife}
\and  
INFN -- Sezione di Ferrara, Via Saragat 1, 44122 Ferrara, Italy \label{infn-fe}
\and
INAF, Osservatorio Astronomico d’Abruzzo, via Mentore Maggini snc, 64100 Teramo, Italy \label{inaf-oaab}
\and 
INFN, Sezione di Cagliari, Cittadella Universitaria, 09042 Monserrato, CA, Italy \label{infn-ca}
\and
INAF - Istituto di Astrofisica Spaziale e Fisica Cosmica di Palermo, Via U. La Malfa 153, 90146 Palermo, Italy \label{inaf-iasf}
\and 
INAF - Osservatorio Astronomico di Cagliari, via della Scienza 5, 09047 Selargius (CA), Italy \label{inaf-oac}
\and
Dipartimento di Fisica e Chimica, Universit\`{a} degli Studi di Palermo, via Archirafi 36, 90123 Palermo, Italy \label{unipa}
\and 
Alferov Academic University, Khlopina St. 8/3, 194021, St. Petersburg, Russia
\label{au}
}

\date{\today}
 
\abstract
{Two classes of Gamma-ray bursts (GRBs), corresponding to the short/hard and the long/soft events, with a putative intermediate class, are typically considered in the observer frame. 
However, when considering GRB characteristics in the cosmological rest frame, the boundary between the classes becomes blurred.}
{The goal of this research is to check for the evidences of a third ``intermediate'' class of GRBs and investigate how the transformation from the observer to the rest frame affects the hardness-duration-based   classification.}
{We applied  fits with skewed and non-skewed (symmetric) Gaussian and Student distributions to the sample of 409 GRBs with reliably measured redshifts to cluster the bursts on the hardness ($E_p$) - duration ($T_{90}$) plane.}
{We found that, based on AIC/BIC criteria, the statistically preferred number of clusters on the GRB rest-frame hardnesses--duration plane does not exceed two. We also assess the robustness of the clustering technique.}
{We did not find any solid evidence of an intermediate GRB class on the rest-frame hardness-duration plane.}


\maketitle

\section{Introduction}
Numerous attempts have recently been made  to classify Gamma-ray bursts (GRBs); however, no definite classification has been established yet.
GRBs are usually divided into two classes based on their prompt-emission parameter distributions in the observer frame.
The bimodality of the GRB duration was discovered for the sample of 85 bursts observed by Konus detectors onboard Venera-11 and Venera-12 missions \citep{Mazets1981}.
\citet{Norris1984} confirmed this finding based on 123 events detected by Konus and 24 bursts detected in the Goddard ISEE-3 experiment.
Later, \citet{Dezalay1991} found that the GRBs lasting less than 2~seconds among the 66 GRBs detected by the Phoebus instrument onboard the Granat observatory exhibit harder spectra.
Presently, the division of GRBs into two classes based on the $T_{90} \simeq 2$~s threshold suggested by \citet{Kouveliotou1993} is considered as a classic GRB taxonomy (however, the value of the $T_{90}$ threshold can be instrument-dependent , see, e.g. \citealt{Svinkin2019}). 

Nowadays, the ``physical'' classification based on the complex of GRB properties including hardness and duration is becoming more popular: Type~I are the merger-origin \citep{Blinnikov1984, Paczynski1986, Eichler1989, Paczynski1991, Narayan1992}, typically short/hard bursts, and Type~II are the collapsar-origin \citep{Woosley1993, Paczynski1998, MacFadyen1999, Woosley2006}, typically long/soft GRBs (see, e.g., \citet{Zhang2009} for more information on this classification scheme).
The noticeable exceptions are the shortest collapsar GRB~200826A \citep{Ahumada2021, Rossi2022}, long mergers GRB~211211A \citep{Troja2022, Rastinejad2022, Yang2022}, GRB~230307A \citep{Levan2024, Gillanders2023, Yang2024}, and the short GRBs with extended emission \citep{Norris2006, Svinkin2016}. 

\citet{Horvath1998} discovered the third peak lying between the two peaks corresponding to the short and the long bursts in the $\log T_{90}$ distribution of BATSE bursts, attributing it to a presumed ``intermediate'' class of GRBs. 
Some subsequent studies confirmed this claim, while others denied it, i.e. the result varied depending on the GRB sample size, the set of the GRB parameters under investigation, their reference frame, and the instrument that collected the data.
In \citet{deUgartePostigo2011}, the properties of the presumed intermediate GRBs were carefully analysed, and it was shown that intermediate bursts differ from short bursts, but do not exhibit any significant differences from long bursts apart from their lower brightness, so that they might be simply a low-luminosity tail of the long GRB class.
The authors suggested that the physical difference between intermediate and long bursts could be explained by being produced by similar progenitors being the ejecta thin shells (intermediate GRBs) and the thick shells (long GRBs).
Intermediate GRBs may also relate to short GRBs with extended emission \citep{Norris2006, Dichiara2021}.

As the GRB classification on the hardness--duration plane is considered to be more robust than the one based only on $T_{90}$, the bivariate classification using multicomponent Gaussian mixture models was extensively applied to the BATSE, RHESSI, and Swift samples.
However, yielded from the same data, two or three clusters were favoured in different studies.
See, e.g., \citet{Tarnopolski2019} for a comprehensive list and brief description of the main studies examining the reliability of the third GRB class in both the one- and two-dimensional cases.

Originally, to obtain GRB classification, the GRB parameter distributions were fitted with a set of several (typically, from one to three) 1D or 2D Gaussian components.
Nevertheless, the logarithmic duration distribution should not necessarily be normal or symmetric \citep{Koen2012, Tarnopolski2015}. 
The asymmetry, or skewness, can originate from, e.g., an asymmetric distribution of the progenitor envelope masses \citep{Zitouni2015}. 
Furthermore, modelling an inherently skewed distribution with a mixture of non-skewed (symmetric) ones may lead to the overfitting of the data, i.e. the addition of excessive components \citep{Koen2012}.
Hence, the BATSE, Swift, and Fermi data sets were reanalysed using different distributions, including the skewed ones \citep{Tarnopolski2016b, Tarnopolski2016c, Kwong2018}.
\citet{Tarnopolski2016b} and \cite{Tarnopolski2019} found that fits with a two-component mixture of skewed distributions outperform the approximations with three-component symmetric models.

Recently, different works presented an ML approach to GRB classification. 
\citet{Jespersen2020} applied a ML dimensionality reduction algorithm, t-distributed stochastic neighbourhood embedding (t-SNE), to the \textit{Swift}/BAT (BAT) data. 
\citet{Salmon2022a} employed principal component analysis (PCA), t-SNE, and wavelet-based decomposition technique to do the \textit{CGRO}/BATSE, \textit{Swift}/BAT (BAT), and \textit{Fermi}/GBM (GBM) burst classification based on their light curves, while \citet{Salmon2022b} utilised the Gaussian-mixture model (GMM) with an entropy criterion on the hardness-duration plane of the BAT and GBM GRB data. 
In all three cases the authors yield two GRB clusters. 
At the same time, \citet{Dimple2023, Dimple2024} applied t-SNE to the BAT and GBM data and obtained five GRB groups.

While in some works (e.g., \citealt{Huja2009, Zitouni2015, Tarnopolski2016a, Yang2016, Zhang2016, Kulkarni2017, Acuner2018}), the rest-frame parameter distributions were explored, the majority of GRB classification studies are based on the observer-frame parameter samples, which are subject to the Malmquist bias that favours the brightest objects against faint objects at large distances leading to the sample incompleteness, as well as the biases, introduced by cosmological time-dilation and spectral-softening effects, see Sections~\ref{sec:durations}, \ref{sec:T90_vs_T50}, and \ref{sec:obs_vs_rest} for more detail. The importance of the correction for the burst redshift for GRB parameter distributions was demonstrated in \citet{Tsvetkova2017}, where it was shown that the boundary between the long and the short GRB clusters becomes blurred after the transformation of GRB hardnesses and durations from the observer to the rest frame. 

In many studies, including the ones that employ the data on the hardness-duration plane in the GRB source frame, the ratio of counts collected in different energy ranges of a given detector is used as a proxy to the GRB spectral hardness. This approach may be justified, in particular, when  
the \textit{Swift}/BAT data are used, which cover the low-energy part of the GRB spectrum that is typically well described by a power-law function.
For broad-band data, however, the spectral peak energy $E_p$ (the maximum of a $\nu F_{\nu}$ spectrum) becomes a more robust hardness estimator of ``curved'' (i.e., ``broken'' or ``cutoff'') GRB spectra. 
The advantages of employing $E_p$ to describe the general hardness of the GRB spectrum include its independence from the limits of the spectral range of the instrument and the fact that this parameter is included in the majority of phenomenological spectral models used to describe GRB prompt emission.

This research is aimed at the study of GRB classification in the hardness-duration plane, specifically, at the search for the third, ``intermediate'' class of bursts. 
For this study, we use the most recent and complete sample of GRBs with well-measured redshifts, which have their spectral data collected in a wide energy range allowing reliable estimates of their rest-frame peak energies and, consequently, hardness. The sample comprises 409 bursts observed from the beginning of the cosmological GRB era in 1997 up to August 12, 2023, by Konus-\textit{Wind} (KW; \citealt{Aptekar1995}) alone and jointly with \textit{Swift}/BAT (BAT; \citealt{Barthelmy2005}), and \textit{Fermi}/GBM (GBM; \citealt{Meegan2009}).

The paper is organised as follows.
We start with a description of the burst sample in Section~\ref{sec:sample}.
We present the results in Section~\ref{sec:results}.
In Section~\ref{sec:discussion}, we discuss the derived GRB classification and the possible effect of the instrumental biases on it as well as test the robustness of the applied technique.
Section~\ref{sec:conclusions} concludes this paper with a brief summary.
In Appendices, we provide a brief description of the instruments that collected the data on the GRBs under consideration, some remarks on the differences in their duration computation procedures, and the formalism of the statistical methods used in this research.
The multivariate clustering of GRBs will be considered in the forthcoming paper.

\section{The burst sample} \label{sec:sample}
The full sample of GRBs with known redshifts (hereafter, referred to as ``Total'') exploited in this work comprises 409 events which can be split into three subsamples: GRBs detected by KW in the triggered mode (the ``KW trig'' sample), bursts detected by KW in the waiting mode and simultaneously detected by \textit{Swift}/BAT (the ``KW \& BAT'' sample), and the events detected by \textit{Fermi}/GBM (the ``GBM only'' sample).

As the set of bursts detected by GBM during the time interval of interest overlaps with the first two subsamples, we distinguish two sets of the GBM bursts: all GRBs detected (``GBM all'') and the ones that are not included in the ``KW trig'' and ``KW \& BAT'' samples (``GBM only''). The former GBM subset is used when the GBM data are analysed separately from the data of other instruments to check for possible biases, and the latter GBM subsample is included in the ``Total'' sample.  

\begin{figure*}
\centering
\includegraphics[width=0.28\textwidth]{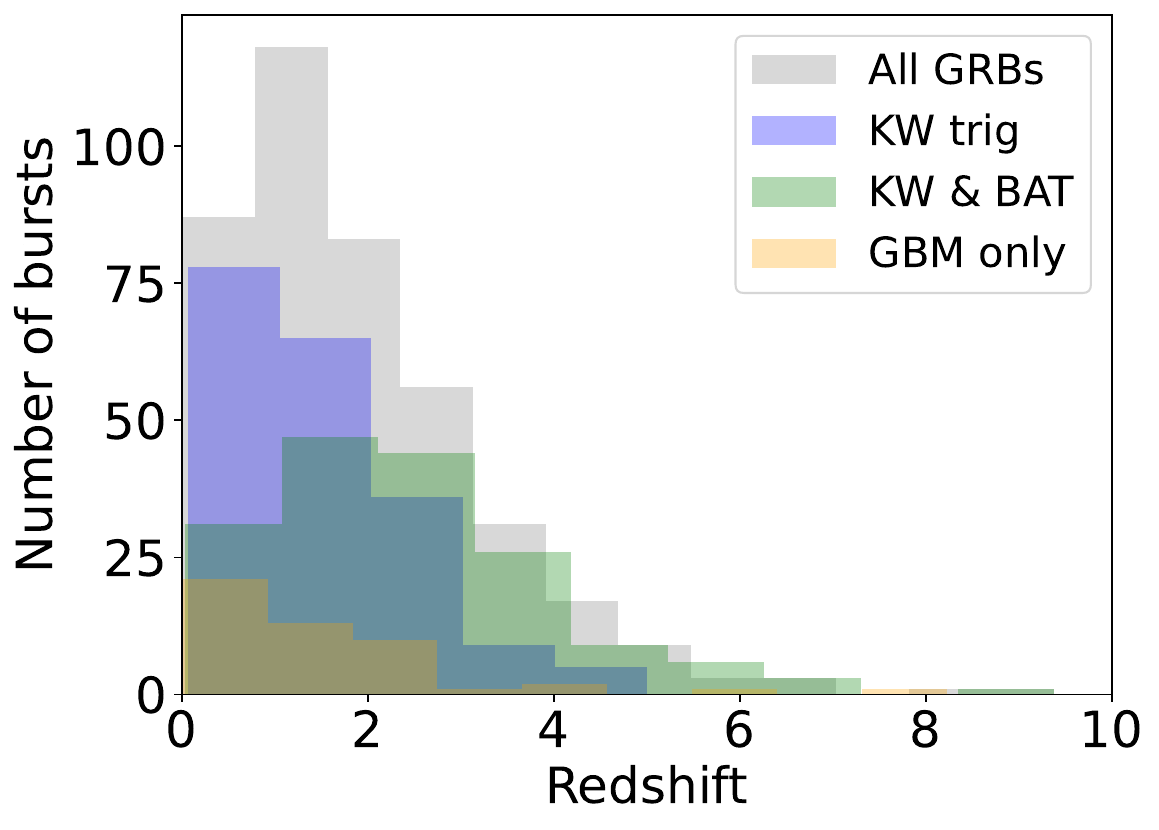}
\includegraphics[width=0.28\textwidth]{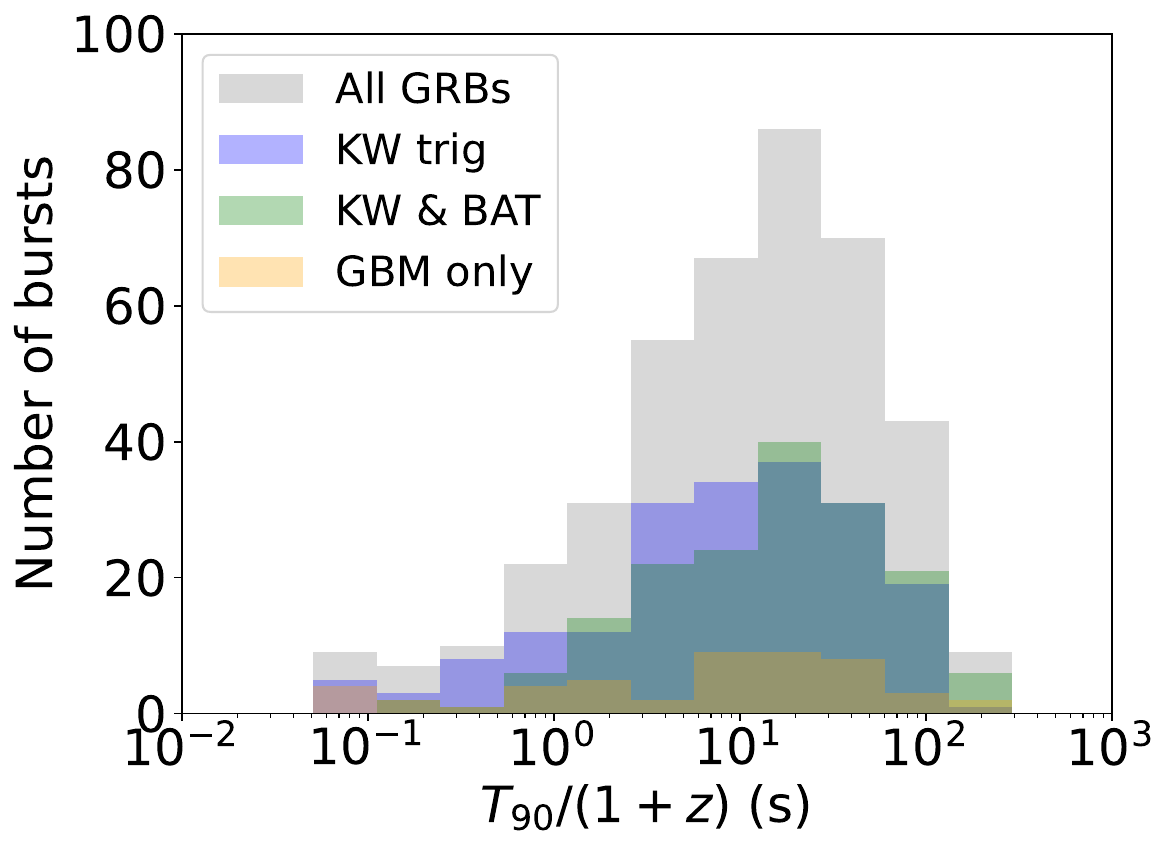}
\includegraphics[width=0.28\textwidth]{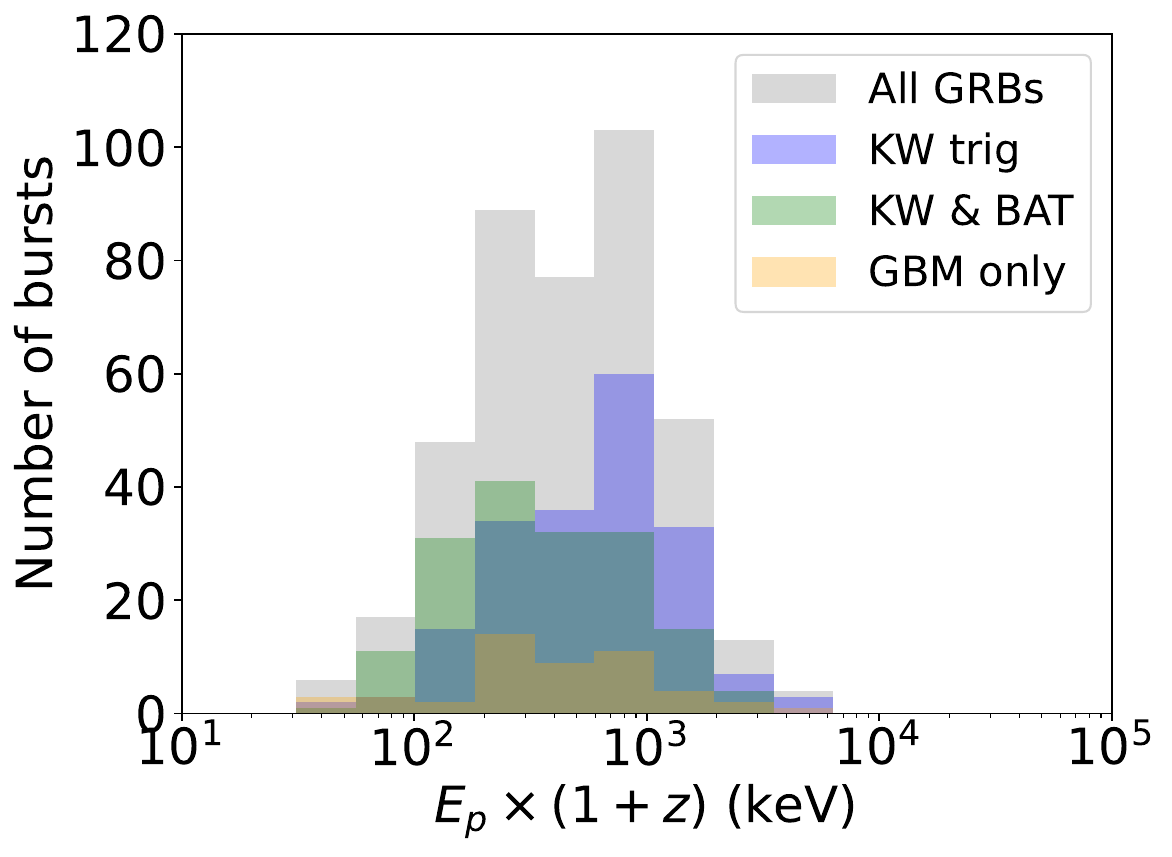}
\caption{Distributions of redshifts (left), rest-frame durations (middle), and rest-frame peak energies (right) of time-integrated spectra.}
\label{fig:sample}
\end{figure*}

The distribution of the bursts in each subsample is given in Table~\ref{tab:sample}.
Figure~\ref{fig:sample} presents the distributions of GRB properties used in this work: redshifts, rest-frame durations, and rest-frame peak energies.
All the subsamples under consideration comprise the peak energies of time-integrated spectra (which, typically, cover the total burst duration) and the burst $T_{90}$ durations.
The peak energies were obtained from spectral fits with the Band function \citep{Band1993}, unless the Band fit is unconstrained. In the latter case, the peak energy of the power law with exponential cutoff model was adopted instead.

\subsection{The ``KW trig'' sample}
The sample of triggered KW bursts with known redshifts covering the range $0.07 \lesssim z \lesssim 5.0$ from \citet{Tsvetkova2017}, updated with the recent GRBs, comprises 193 events, whose spectra are measured over the 20~keV -- 10 MeV range.
Following \citet{Minaev2021_err}, we use the updated or corrected redshifts for two GRBs from \citet{Tsvetkova2017}: $z = 1.9621$ \citep{Perley2017} for GRB~020819B, and $z \simeq 1.0$ \citep{Knust2017} for GRB~150424A.
GRB~050603 was removed from the sample as its redshift based on the alleged detection of a bright Lyman $\alpha$ line \citep{Berger2005} is highly unreliable given the deep non-detection of the host galaxy in follow-up observations \citep{Hjorth2012}. Finally, we excluded GRB~070508 from the sample as its redshift was withdrawn upon deeper analysis \citep{Fynbo2009}.
In this paper, we adopted the durations computed in the G2G3 ($\sim$80--1200~keV) energy window, unless otherwise specified.

\subsection{The ``KW \& BAT'' sample}
We do not use Swift/BAT spectral data alone as the limited spectral range of the instrument introduces bias to the spectral parameters and energetics \citep{Sakamoto2011_cc}. Accordingly, we adopted the sample from \citet{Tsvetkova2021} containing 167 GRBs with known redshifts ($0.04 \lesssim z \lesssim 9.4$), which were detected by KW in the waiting mode (20--1500~keV band) and triggered BAT.
For this sample, we, following \cite{Minaev2020}, corrected the redshift of GRB~050803 from $z = 0.422$ to $z = 4.3$ \citep{Perley2016}.
The afterglow spectrum of GRB~160327A is indicative of a Lyman-alpha drop at a redshift between 4.90 and 5.01, with a most probable value at $z = 4.99$, which we adopted as the redshift of this burst, although even lower redshifts could not be discarded for unusually large Hydrogen column densities. 
Due to the low S/N ratio of the spectrum, it was impossible to confirm any other absorption lines \citep{deUgartePostigo2016}.
Following \citep{Tsvetkova2021}, we adopted the durations computed in the 25--350~keV band, unless otherwise specified.

\subsection{The ``GBM all'' and ``GBM only'' samples}
These sets of bursts comprise the GBM events with known redshifts spanning the range $0.01\lesssim z \lesssim 8.2$ collected from \citet{Gruber2011, Atteia2017, Minaev2020, Minaev2021_err} and GCN circulars\footnote{\url{https://gcn.gsfc.nasa.gov}}.
The GRB $T_{90}$ durations and the peak energies of the time-integrated spectra were extracted either from \texttt{FERMIGBRST}--Fermi GBM Burst Catalogue\footnote{\url{https://heasarc.gsfc.nasa.gov/w3browse/fermi/fermigbrst.html}} or from the GCN circulars.
The ``GBM all'' sample contains 168 events, and, after removal of the overlap with the ``KW trig'' and ``KW \& BAT'' samples, 49 of them comprise the ``GBM only'' set.
Following \citet{Minaev2021_err}, we corrected redshift of GRB~150101B: the value of z = 0.093, taken from \citet{Zhang2018}, was based on an incorrect preliminary measurement by \citet{Castro-Tirado2015}, which was lately corrected in \citet{Levan2015} leading to $z = 0.1343$ \citep{Fong2016}.
The GBM durations are computed in the 50--300~keV energy range.

\begin{table}
\caption[]{Burst samples}
\label{tab:sample}
$$ 
 \begin{tabular}{lr}
    \hline
    \noalign{\smallskip}
    Sample & Number of bursts \\
    \noalign{\smallskip}
    \hline
    \noalign{\smallskip}
    ``KW trig'' & 193 \\
    ``KW \& BAT'' & 167 \\
    ``GBM only'' & 49 \\
    ``Total'' & 409 \\
    \hline
    ``GBM all'' & 168 \\
    \noalign{\smallskip}
    \hline
 \end{tabular}
$$ 
\end{table}

\section{Analysis and Results} \label{sec:results}
\begin{figure*}
   \centering
   \includegraphics[width=0.45\textwidth]{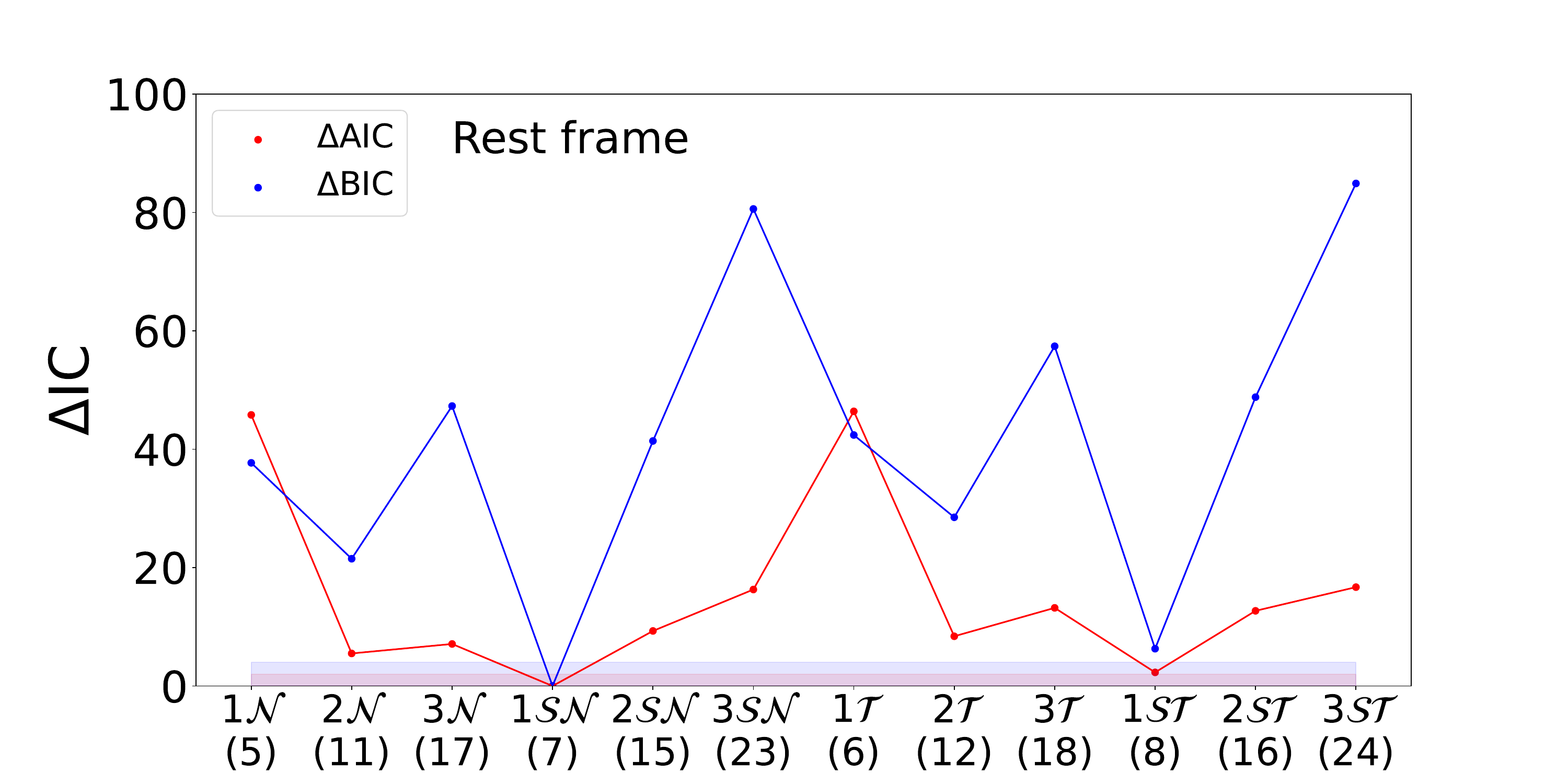}
   \includegraphics[width=0.45\textwidth]{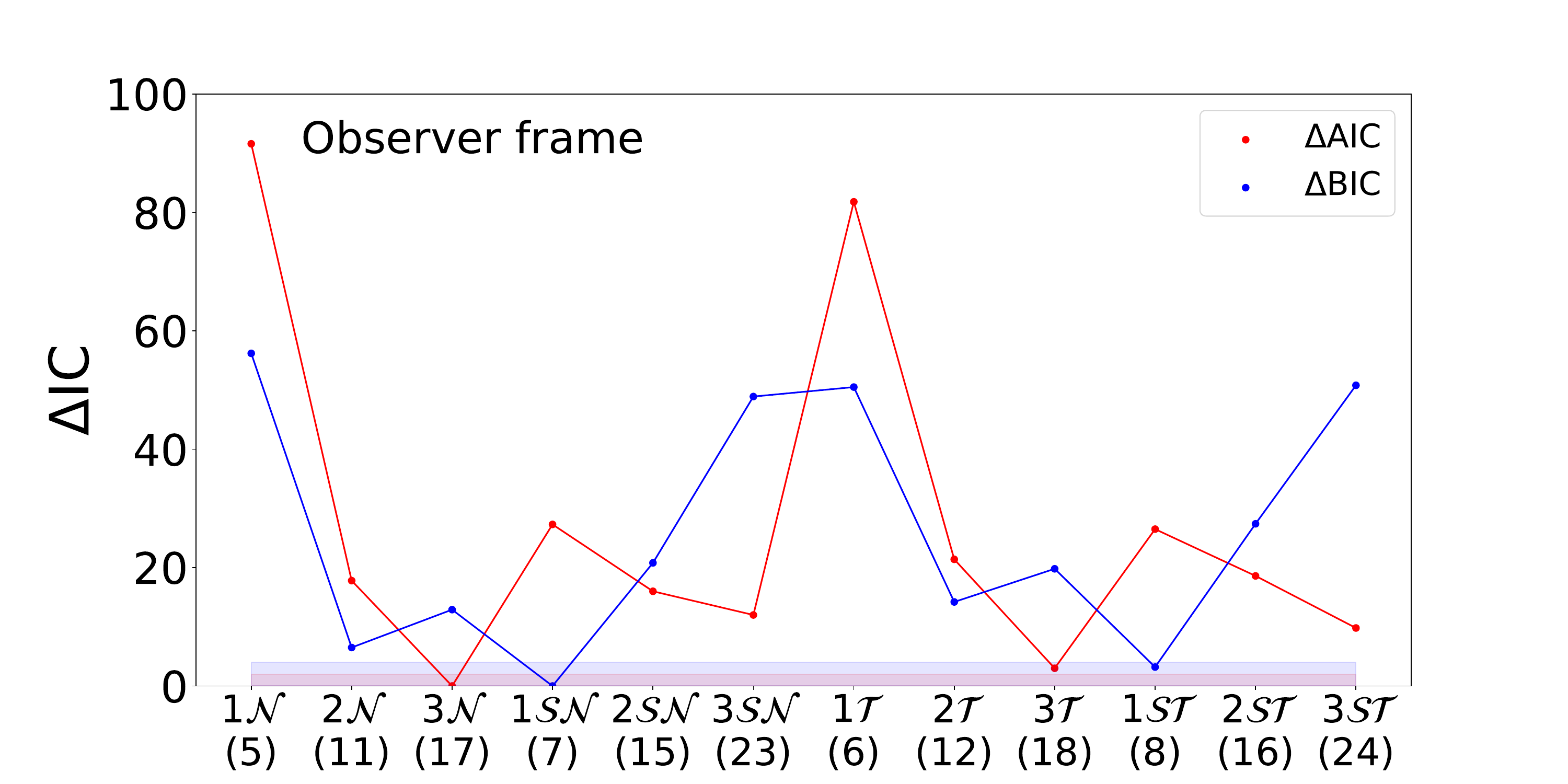}
   \caption{Information criteria scores for the ``Total'' sample. 
   The following designations are introduced for the models: $\mathcal{N}$ ($\mathcal{SN}$) and $\mathcal{T}$ ($\mathcal{ST}$) denote the symmetric (skewed) Gaussian and Student distributions, correspondingly; the number to the left of the model provides the amount of clusters; the number in parenthesis gives the amount of free parameters of the model.
   The red filled area corresponds to the $\Delta_{IC} < 2$, while the blue filled area marks the $2< \Delta_{IC} < 4$ area.
   }
\label{fig:ic}
\end{figure*}

We fitted the skewed and non-skewed Gaussian and Student distributions to the data on the $E_p$ vs $T_{90}$ plane for the ``Total'' sample in the observer and the source frames.
As we followed a common approximation procedure, we moved its description into the Appendix~\ref{sec:methods} where its details along with all designations can be found.
We also exploited the ``KW trig'', ``KW \& BAT'', ``GBM all'' subsamples to test how the instrumental biases affect the GRB classification.
The Akaike and Bayesian information criteria (IC) were used to estimate relative performance of models.
The $\Delta_i$ (difference between the IC of the $i$-th model and the minimum one within a set of models) for AIC and BIC yielded from this analysis, along with the corresponding model weights $w_i$, are summarised in Tables~\ref{tab:rest_total}, \ref{tab:rest_kw}, \ref{tab:rest_bat}, \ref{tab:rest_gbm} (rest frame) and Table~\ref{tab:obs_total}, \ref{tab:obs_kw}, \ref{tab:obs_bat}, \ref{tab:obs_gbm} (observer frame) and illustrated on Figure~\ref{fig:ic}.
In all the tables providing IC, the (equally) preferred models ($\Delta_\textrm{AIC}$ or $\Delta_\textrm{BIC} < 2$) are marked with the red colour, while the acceptable models with $2 < \Delta_\textrm{AIC} < 4$ are denoted with blue. 

In the rest frame, the AIC gives preference to the models based on the one-component skewed distributions for all but ``KW \& BAT'' samples, also allowing two (three) clusters with symmetric distributions for the ``KW'' (``GBM all'') subsamples, while the ``KW \& BAT'' sample can be nicely fitted only by a symmetric distribution with any number of clusters.
Meanwhile, BIC only supports the one-cluster distributions, skewed for the ``Total'', ``KW trig'', and ``GBM all'' samples and non-skewed for the ``KW \& BAT'' sample. 
It is noticeable that the three-component models can hardly be considered as acceptable to describe the GRBs on the hardness-duration plane in the rest frame.

In the observer frame, AIC allows three-component symmetric models for all samples, while in the BIC framework, the one-component skewed models are clearly preferred over others.
However, three-component skewed distribution also fits the ``KW \& BAT'' and ``GBM all'' samples.
The asymmetric one-component models are also admitted for the ``KW \& BAT'' sample which can be easily explained by the observational biases: the detection of the short hard GRBs is significantly suppressed by both the soft BAT energy band, whose channels prevail in number over three KW channels, and the coarse time resolution in the waiting mode of the KW experiment.

\section{Discussion} \label{sec:discussion}
It can be noted that, as expected, modelling the data with skewed distributions generally reduces the number of possible clusters, i.e. the distribution components, of GRBs compared to the fits with symmetric models.
In addition, AIC offers a systematically wider choice of acceptable models and allows higher number of clusters than BIC. 
Remarkably, BIC yields a one-cluster model, based on a skewed distribution, as the best one for almost all samples both in the observer and the rest frames, with the only one exception of the ``KW \& BAT'' sample considered in the rest frame, for which the non-skewed one-component model is preferred.
As one can see, a broader selection of models (based on AIC) is allowed for the individual subsamples than for the mixed ``Total'' sample, where the peculiarities of different gamma-detectors with distinct sensitivities and trigger algorithms are smoothed.
It is also noticeable that in many cases, similar results were obtained for the Gaussian and Student distributions in the AIC framework. 
This is not surprising, given that when the number of degrees of freedom (the sample size) is large, the Student distribution essentially coincides with the Gaussian due to the central limit theorem.

Thus, while BIC favours a single GRB cluster in both observer and rest frames, AIC gives the preference to the three-component models in the observer frame and one-component model in the rest frame.

\subsection{Comparison with other studies}
A significant part of studies considers the observer-frame burst durations and the ``count ratio'' hardnesses to draw conclusions on the GRB classification (see, e.g., the Introduction in \citet{Tarnopolski2019} and the Table~1 in \citet{Salmon2022b} for a comprehensive summary of efforts to classify GRBs).
Some of these studies found an evidence for the third class, while other works did not confirm this conclusion.
However, several studies, such as \citet{Zhang2009, Li2020, Luo2023}, which employed multiple criteria, including redshift, found no necessity to invoke the intermediate class of GRBs.
Moreover, a study of almost 300 \textit{Swift}/BAT bursts with known redshifts carried out using the GMM and EM algorithm showed that two components suffice in both observer and rest frames based on the BIC \citep{Yang2016}.

In this work, we found that, in most cases, a single cluster well describes the population of GRBs with known redshifts in their source frame on the hardness ($E_\textrm{p}$)--duration plane, which is in agreement with \cite{Tsvetkova2017}.
This can be explained by blurring of the cluster boundaries in the source frame due to a wider range of the long GRB redshifts and consequent decrease of contrast in hardnesses and durations of short bursts compared to the long ones. The difference in the redshift range of long and short GRBs may result from the selection effects, such as fainter afterglows of short bursts \citep{Berger2014}, and the merger delay with respect to star formation \citep{Zhang2009}.

\subsection{Selection effects and other distortions}
\label{sec:biases}
GRB observations are subject to selection effects as the observed sample may not reflect the true, underlying population.
This observational distortions play a crucial role for GRBs \citep{Turpin2016, Dainotti2017}, which are particularly affected by the Malmquist bias that favours the brightest objects against faint objects at large distances leading to the sample incompleteness.
For the sample of GRBs with known redshifts, the selection effects can be divided into two categories: the instrument-specific effects, caused by its trigger sensitivity to the burst prompt emission parameters; and the ``external'' biases originating from the process of GRB localisation and redshift measurement.
The selection effects may affect the peak energy, the isotropic energy release, the peak or the isotropic luminosity (see, e.g., \citealt{Dainotti2018}).

The prerequisite of a reliable redshift estimate introduces an additional bias:
the ``KW trig'' and ``GBM all'' subsamples include around 12\% of short bursts against 16\% present in full samples of all  bursts which triggered the instruments, while the sample of BAT GRBs contains twice less short events compared to the aforementioned subsamples, both with measured redshifts and in total, and only 2\% in the ``KW \& BAT'' subsample.

The GRB durations are subject to three redshift-driven effects: 
(1) cosmological time dilation by the factor of $(1 + z)$; 
(2) a dependence of sharpness of the light curve peaks on the energy bandpass where it is collected yields narrower temporal structures in the source frame, according to the narrowing of pulses with energy \citep{Norris1996, Fenimore1995}, as any given energy window in the observer frame, corresponds to a harder rest-frame band;
(3) the ``tip of the iceberg'' effect consisting in the decrease of the GRB light curve S/N ratio with larger redshifts, so that the ``wings'' of the pulses drop below the background level, which leaves only the brightest part of the burst being detectable \citep{Kocevski2013}.
The first effect can be easily corrected if the redshift is known: $T_\textrm{90, z} = T_\textrm{90, obs}/(1 + z)$.
The second effect, in principle, may be addressed by applying the empirical equation from \citet{Fenimore1995} which states that the duration of individual GRB emission episodes decreases with the increase of photon energy: $T \propto E^{-0.4}$.
However, in practice, this property should be used with caution as it is only valid on average. 
It is known that the power-law index 0.4 can vary wildly on individual GRBs (see, e.g.,  \citealt{Borgonovo2007}).
The third effect is hard to be corrected for.

To assess the instrument-specific selection effects, we conducted analysis for three subsamples (``KW trig'', ``KW \& BAT'', and ``GBM all'') separately, in addition to the ``Total'' sample.
It turns out that in the rest frame, the claim of the one-component skewed distribution as the best or, at least, suitable model for the ``Total'' sample remains valid for the ``KW trig'' and ``GBM all'' subsamples, leaving non-skewed models as preferable for the ``KW \& BAT'' subsample, which is the one the most subject to biases. 
Meanwhile, in the observer frame, the best model differs depending on the IC applied.
However, for each IC, the suitable model is in agreement within all subsamples and the ``Total'' sample, despite allowing some other models as suitable too.

To evaluate the effect of the Malmquist bias, which leads to the insecure detection of the events with energetics being near the trigger threshold, we excluded 10\% of the faintest busts (based on their S/N) in each subsample and in the ``Total'' sample and conducted the clustering analysis on the hardness-duration plane in the rest frame.
The results, provided in Table~\ref{tab:snr}, are in a good agreement with the ones claimed from the original sample and the subsamples only evoking additional models based on the symmetric distributions for the ``KW trig'' (three clusters) and ``GBM all'' (one and two clusters) samples and retracting a three-component symmetric distribution for the ``KW \& BAT'' sample.

\subsection{$T_{90}$ vs $T_{50}$ duration}
\label{sec:T90_vs_T50}
\begin{figure*}
   \centering
   \includegraphics[width=0.45\textwidth]{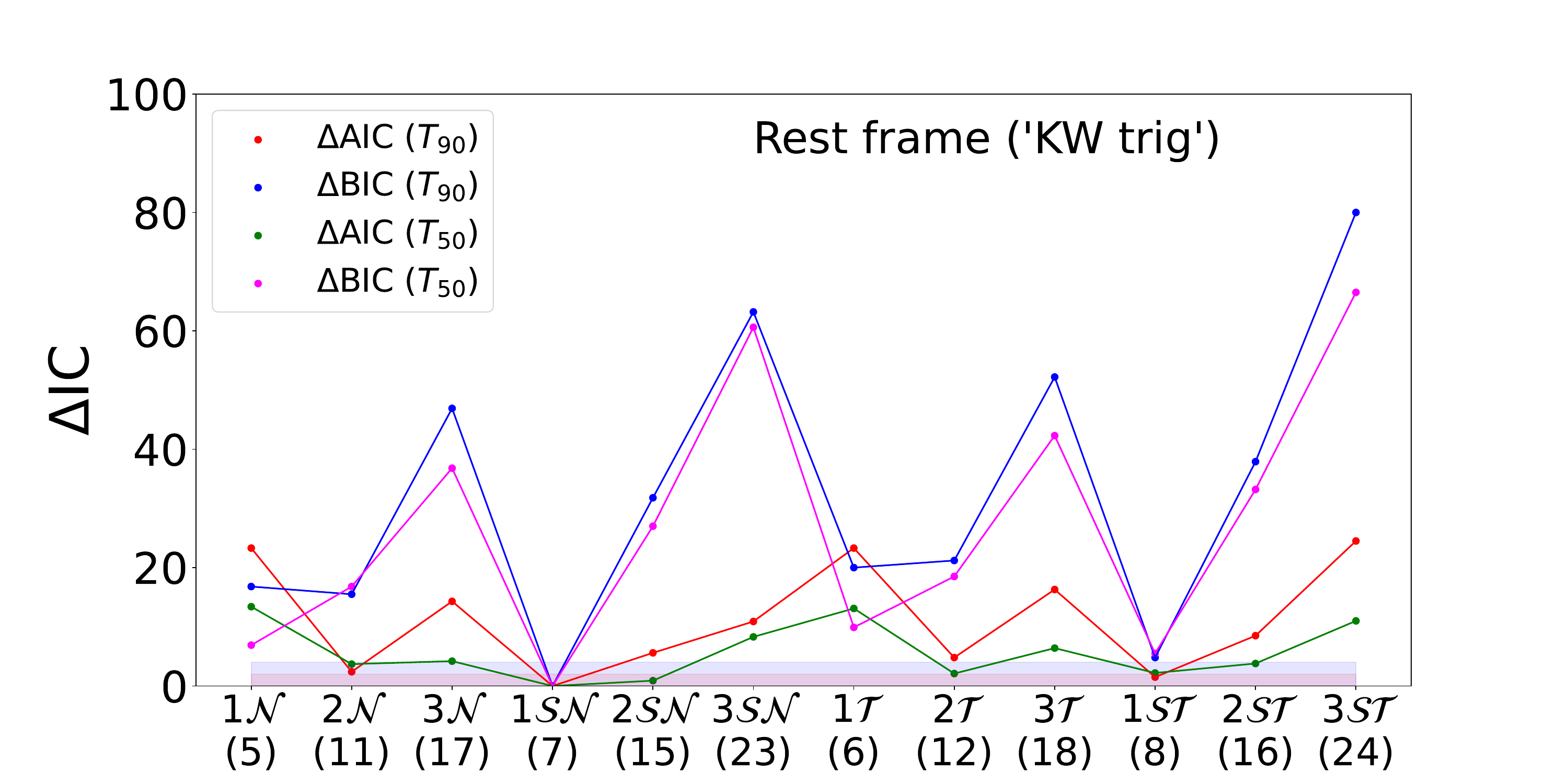}
    \includegraphics[width=0.45\textwidth]{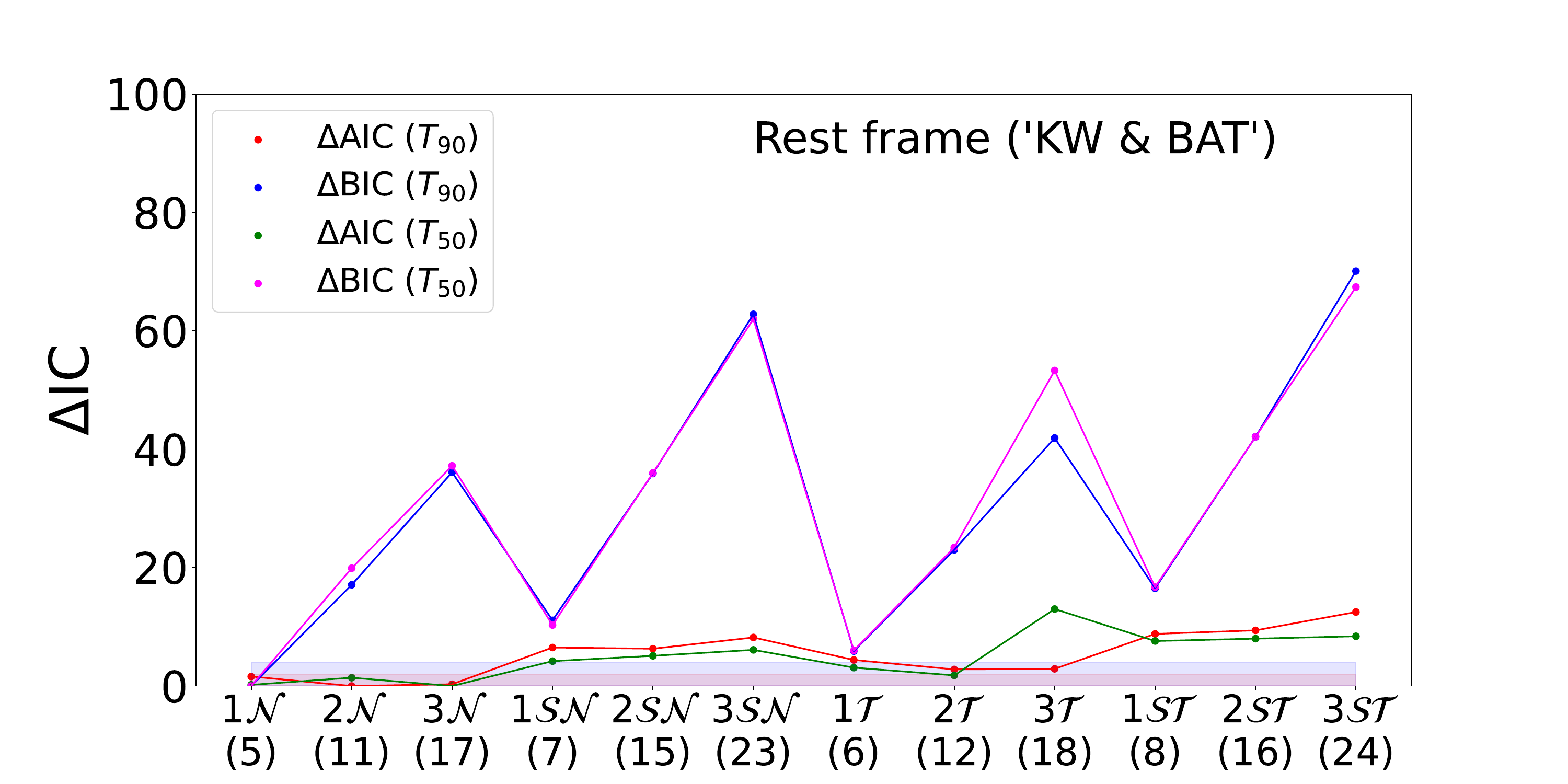}
    
   \caption{Comparison of information criteria scores computed using the $T_{50}$ and $T_{90}$ durations for the ``KW trig'' and ``KW \& BAT'' samples. 
   The following designations are introduced for the models: $\mathcal{N}$ ($\mathcal{SN}$) and $\mathcal{T}$ ($\mathcal{ST}$) denote the symmetric (skewed) Gaussian and Student distributions, correspondingly; the number to the left of the model provides the amount of clusters; the number in parenthesis gives the amount of free parameters of the model.
   The red filled area corresponds to the $\Delta_{IC} < 2$, while the blue filled area marks the $2< \Delta_{IC} < 4$ area.
   }
\label{fig:t50}
\end{figure*}

As \citet{Svinkin2019} argued that $T_{50}$ is a more robust\footnote{The robustness may be a consequence of $T_{50}$ being less subject to the ``tip of the iceberg'' effect.} duration measure than $T_{90}$, we studied the GRB clustering on the $E_p$--$T_{50}$ plane in the GRB source frame applied to the ``KW trig'' and ``KW \& BAT'' samples, for which we had all the $T_{50}$ durations available.
Table~\ref{tab:T50kw} and \ref{tab:T50bat} along with the Figure~\ref{fig:t50} provide the yielded ICs.
One can see that the results are in good agreement with the ones obtained using the $T_{90}$ duration, except that two-component distributions are allowed, in addition to others,  in the AIC framework, for the ``KW trig'' sample.

\subsection{Observer- vs rest-frame durations}
\label{sec:obs_vs_rest}
A recent study of GRB minimum variability timescale \citep{Camisasca2023} showed that dependence of the GRB duration on its redshift may be weaker than expected (if present at all).
This can be explained by the fact that three different effects (see Section~\ref{sec:biases}) can reduce each other decreasing the contribution of the cosmological time dilation, as also found by previous investigations \citep{Kouveliotou1993, Golkhou2014, Golkhou2015, Littlejohns2014}. 
Therefore, we examined the ``Total'' sample on the plane of the rest-frame peak energies and the observer-frame $T_{90}$ durations.
The ICs, presented in Table~\ref{tab:obsT90}, show that, based on BIC, the accepted models are in a nice agreement with the ones yielded from the modelling of the ``Total'' sample in both the rest and the observer frames. 
Nevertheless, additional multi-component models emerged in the AIC framework, which, however, tends to overfit the data.

\subsection{Assessment of the uncertainties on IC}
\label{sec:resampling}
To assess performance of the applied method, we carry out data resampling using the Bootstrap \citep{Efron1979} and Jackknife \citep{Quenouille1949, Quenouille1956, Tukey1958} techniques.
Applying Bootstrap, we drew 100 samples from the original ``Total'' one keeping the same sample size. 
When applying the Jackknife approach, we generated two sets of new samples, 100 samples each, excluding 10\% and 30\% of the original data.
Table~\ref{tab:sim} provides the mean values and standard deviations, which we used as proxy to estimate the IC uncertainties, for each set of samples.
Remarkably, the adopted uncertainties are significantly higher than the thresholds on $\Delta_\textrm{IC}$ for accepting a model as the best or, at least, a suitable one.
The same conclusion is valid when the IC uncertainties are estimated using a set of 100 samples generated from the $1 \sigma$ statistical errors of the rest-frame $T_{90}$ and $E_p$ (Table~\ref{tab:sim_err}).
Thus, the results based on the applied technique should be treated with caution.

\subsection{Validation of the technique reliability}
\label{sec:validation}
To test the robustness of the technique applied in this work, we first generated a mock sample of 400 points on the hardness-duration plane, which corresponds in size to the original ``Total'' sample, and a mock 10,000-point sample from the two-dimensional two-component symmetric Gaussian distribution obtained within the previous analysis of the ``Total'' sample.
After application of the technique described in Section~\ref{sec:methods} to these mock samples, we found that the original model is firmly selected as the preferred one (the separation between the IC of the models was well above the best model selection criteria, even including the large uncertainties assessed in the previous subsection) for the larger mock sample, while for the smaller sample the one- and three-component non-skewed models were preferred in addition to the original one.
Thus, the size of the sample under investigation is crucial to ensure the robustness of the technique.

\subsection{Rest- vs observer-frame light curve energy bands}
\begin{figure}
\centering
\includegraphics[width=0.95\linewidth]{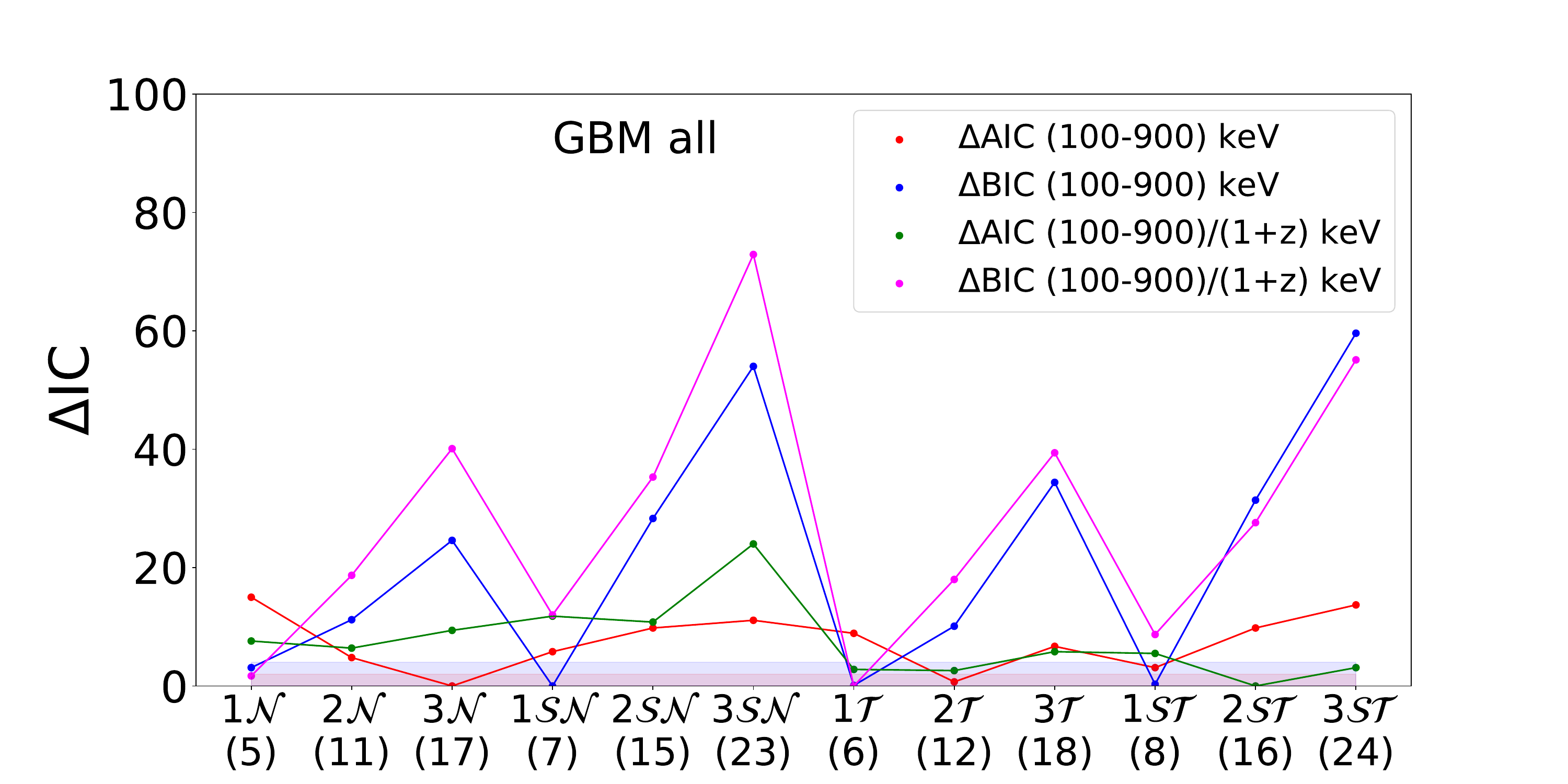}
\caption{Comparison of information criteria scores computed for the rest-frame  hardness-duration distributions of 155 GRBs from the ``GBM all'' sample.
    The durations were computed in the observer- and rest-frame energy windows $100-900$~keV, which in the latter case corresponds to the $\frac{100}{1+z}-\frac{900}{1+z}$~keV instrumental bandpass. 
    In both cases, the durations were corrected for the cosmological time dilation.
    The following designations are introduced for the models: $\mathcal{N}$ ($\mathcal{SN}$) and $\mathcal{T}$ ($\mathcal{ST}$) denote the symmetric (skewed) Gaussian and Student distributions, correspondingly; the number to the left of the model provides the amount of clusters; the number in parenthesis gives the amount of free parameters of the model.
   The red filled area corresponds to the $\Delta_{IC} < 2$, while the blue filled area marks the $2< \Delta_{IC} < 4$ area.
   }
\label{fig:gbm_t90}
\end{figure}

To address the problem of GRB pulse narrowing with increasing energy which results in the duration bias, one can compute durations based on the GRB time histories in two energy windows: one is fixed in the observer frame, while other is set constant in the rest frame.
However, as KW light curves are collected in three fixed energy windows, and the energy band of BAT is too narrow to provide multiple energy windows for light curves, this kind of bias assessment appears to be feasible only for the GBM data.
Thus, we computed $T_{90}$ for the ``GBM all'' sample based on the light curves in two energy ranges: 100--900~keV in the observer and rest frames. 
In the latter case, the bandpass was transformed to $100/(1+z)-900/(1+z)$~keV interval.
The durations were computed using the techniques similar to the ones described in Sections~\ref{app:kw} and \ref{app:bat}.
We excluded some relatively dim bursts from the sample, as the shift of the 50--300~keV band, conventionally used by the GBM team to evaluate durations, towards the harder 100-900~keV range resulted in the S/N decrease and consequent absence of a reliable $T_{90}$ estimate for these events, thereby leaving a set of 155 GRBs under consideration.
Then we applied the clustering procedure to two data sets which comprise the rest-frame hardnesses and cosmological time dilation-corrected durations assessed in two aforementioned energy bands.
Figure~\ref{fig:gbm_t90} and Table~\ref{tab:rest_900keV} present the comparison of the results obtained in these two cases.
For both data sets, as well as for the one which includes the rest-frame $T_{90}$ estimated in the standard 50--300~keV band, BIC shows clear preference to single-component distributions.
AIC, expectedly, offered a wider choice of the amount of clusters, though, allowing a one-component distribution as a suitable model in all three cases.

Remarkably, the ``classic'' estimate of hardness, which is the ratio of counts collected in two different energy ranges, is subject to a similar bias, thus, making the spectral peak energy a more robust proxy for the GRB hardness.

\section{Conclusions} \label{sec:conclusions}
As it was suggested that the redshift distribution may play a crucial role in the GRB classification (e.g., \citealt{Tarnopolski2019}), we studied clustering of GRBs on the hardness-duration, namely, $E_p$--$T_{90}$ plane in both the observer and the rest frame using the data obtained by three different GRB detectors.
The results obtained from the sample of 409 GRBs with known redshifts detected in a wide energy range, one of the largest to date, in terms of the informational criteria, do not show  evidence of the third class of GRBs in the rest-frame\footnote{However, the results drawn from a limited-size sample should be treated with caution (see Section~\ref{sec:validation}).}.
In addition, the results of our study confirm the conclusions made in \citet{Tsvetkova2017} for a sample of 150 KW GRBs with known redshifts detected in the triggered mode: a quite prominent boundary between the type~I and type~II GRBs becomes blurred when the data is transformed from the observer frame to the rest-frame.
An obvious dependence of robustness of the adopted technique on the sample size highlights importance of the prospective missions aimed at substantial increase of reliable GRB redshift estimates, e.g., \textit{THESEUS} \citep{Amati2018, Amati2021}.

\begin{acknowledgements}
We thank the anonymous referee for their valuable comments and suggestions that led to an overall improvement of this work.
We thank Adam Goldstein for giving clarification on the GBM duration calculation procedure.
We also gratefully acknowledge discussions with Dmitry Svinkin.
A.T. acknowledges support from the HERMES Pathfinder–Operazioni 2022-25-HH.0 grant.
L.A. acknowledges support from INAF grant programme 2022.
\end{acknowledgements}

\bibliography{clustering_2d}{}

\begin{thebibliography}{108}
\expandafter\ifx\csname natexlab\endcsname\relax\def\natexlab#1{#1}\fi

\bibitem[{{Acuner} \& {Ryde}(2018)}]{Acuner2018}
{Acuner}, Z. \& {Ryde}, F. 2018, \mnras, 475, 1708

\bibitem[{{Ahumada} {et~al.}(2021){Ahumada}, {Singer}, {Anand}, {Coughlin}, {Kasliwal}, {Ryan}, {Andreoni}, {Cenko}, {Fremling}, {Kumar}, {Pang}, {Burns}, {Cunningham}, {Dichiara}, {Dietrich}, {Svinkin}, {Almualla}, {Castro-Tirado}, {De}, {Dunwoody}, {Gatkine}, {Hammerstein}, {Iyyani}, {Mangan}, {Perley}, {Purkayastha}, {Bellm}, {Bhalerao}, {Bolin}, {Bulla}, {Cannella}, {Chandra}, {Duev}, {Frederiks}, {Gal-Yam}, {Graham}, {Ho}, {Hurley}, {Karambelkar}, {Kool}, {Kulkarni}, {Mahabal}, {Masci}, {McBreen}, {Pandey}, {Reusch}, {Ridnaia}, {Rosnet}, {Rusholme}, {Carracedo}, {Smith}, {Soumagnac}, {Stein}, {Troja}, {Tsvetkova}, {Walters}, \& {Valeev}}]{Ahumada2021}
{Ahumada}, T., {Singer}, L.~P., {Anand}, S., {et~al.} 2021, Nature Astronomy, 5, 917

\bibitem[{{Akaike}(1974)}]{Akaike1974}
{Akaike}, H. 1974, IEEE Transactions on Automatic Control, 19, 716

\bibitem[{{Amati} {et~al.}(2018){Amati}, {O'Brien}, {G{\"o}tz}, {Bozzo}, {Tenzer}, {Frontera}, {Ghirlanda}, {Labanti}, {Osborne}, {Stratta}, {Tanvir}, {Willingale}, {Attina}, {Campana}, {Castro-Tirado}, {Contini}, {Fuschino}, {Gomboc}, {Hudec}, {Orleanski}, {Renotte}, {Rodic}, {Bagoly}, {Blain}, {Callanan}, {Covino}, {Ferrara}, {Le Floch}, {Marisaldi}, {Mereghetti}, {Rosati}, {Vacchi}, {D'Avanzo}, {Giommi}, {Piranomonte}, {Piro}, {Reglero}, {Rossi}, {Santangelo}, {Salvaterra}, {Tagliaferri}, {Vergani}, {Vinciguerra}, {Briggs}, {Campolongo}, {Ciolfi}, {Connaughton}, {Cordier}, {Morelli}, {Orlandini}, {Adami}, {Argan}, {Atteia}, {Auricchio}, {Balazs}, {Baldazzi}, {Basa}, {Basak}, {Bellutti}, {Bernardini}, {Bertuccio}, {Braga}, {Branchesi}, {Brandt}, {Brocato}, {Budtz-Jorgensen}, {Bulgarelli}, {Burderi}, {Camp}, {Capozziello}, {Caruana}, {Casella}, {Cenko}, {Chardonnet}, {Ciardi}, {Colafrancesco}, {Dainotti}, {D'Elia}, {De Martino}, {De Pasquale}, {Del Monte}, {Della Valle}, {Drago}, {Evangelista}, {Feroci},
  {Finelli}, {Fiorini}, {Fynbo}, {Gal-Yam}, {Gendre}, {Ghisellini}, {Grado}, {Guidorzi}, {Hafizi}, {Hanlon}, {Hjorth}, {Izzo}, {Kiss}, {Kumar}, {Kuvvetli}, {Lavagna}, {Li}, {Longo}, {Lyutikov}, {Maio}, {Maiorano}, {Malcovati}, {Malesani}, {Margutti}, {Martin-Carrillo}, {Masetti}, {McBreen}, {Mignani}, {Morgante}, {Mundell}, {Nargaard-Nielsen}, {Nicastro}, {Palazzi}, {Paltani}, {Panessa}, {Pareschi}, {Pe'er}, {Penacchioni}, {Pian}, {Piedipalumbo}, {Piran}, {Rauw}, {Razzano}, {Read}, {Rezzolla}, {Romano}, {Ruffini}, {Savaglio}, {Sguera}, {Schady}, {Skidmore}, {Song}, {Stanway}, {Starling}, {Topinka}, {Troja}, {van Putten}, {Vanzella}, {Vercellone}, {Wilson-Hodge}, {Yonetoku}, {Zampa}, {Zampa}, {Zhang}, {Zhang}, {Zhang}, {Zhang}, {Antonelli}, {Bianco}, {Boci}, {Boer}, {Botticella}, {Boulade}, {Butler}, {Campana}, {Capitanio}, {Celotti}, {Chen}, {Colpi}, {Comastri}, {Cuby}, {Dadina}, {De Luca}, {Dong}, {Ettori}, {Gandhi}, {Geza}, {Greiner}, {Guiriec}, {Harms}, {Hernanz}, {Hornstrup}, {Hutchinson}, {Israel},
  {Jonker}, {Kaneko}, {Kawai}, {Wiersema}, {Korpela}, {Lebrun}, {Lu}, {MacFadyen}, {Malaguti}, {Maraschi}, {Melandri}, {Modjaz}, {Morris}, {Omodei}, {Paizis}, {P{\'a}ta}, {Petrosian}, {Rachevski}, {Rhoads}, {Ryde}, {Sabau-Graziati}, {Shigehiro}, {Sims}, {Soomin}, {Sz{\'e}csi}, {Urata}, {Uslenghi}, {Valenziano}, {Vianello}, {Vojtech}, {Watson}, \& {Zicha}}]{Amati2018}
{Amati}, L., {O'Brien}, P., {G{\"o}tz}, D., {et~al.} 2018, Advances in Space Research, 62, 191

\bibitem[{{Amati} {et~al.}(2021){Amati}, {O'Brien}, {G{\"o}tz}, {Bozzo}, {Santangelo}, {Tanvir}, {Frontera}, {Mereghetti}, {Osborne}, {Blain}, {Basa}, {Branchesi}, {Burderi}, {Caballero-Garc{\'\i}a}, {Castro-Tirado}, {Christensen}, {Ciolfi}, {De Rosa}, {Doroshenko}, {Ferrara}, {Ghirlanda}, {Hanlon}, {Heddermann}, {Hutchinson}, {Labanti}, {Le Floch}, {Lerman}, {Paltani}, {Reglero}, {Rezzolla}, {Rosati}, {Salvaterra}, {Stratta}, {Tenzer}, \& {Theseus Consortium}}]{Amati2021}
{Amati}, L., {O'Brien}, P.~T., {G{\"o}tz}, D., {et~al.} 2021, Experimental Astronomy, 52, 183

\bibitem[{{Aptekar} {et~al.}(1995){Aptekar}, {Frederiks}, {Golenetskii}, {Ilynskii}, {Mazets}, {Panov}, {Sokolova}, {Terekhov}, {Sheshin}, {Cline}, \& {Stilwell}}]{Aptekar1995}
{Aptekar}, R.~L., {Frederiks}, D.~D., {Golenetskii}, S.~V., {et~al.} 1995, \ssr, 71, 265

\bibitem[{{Atteia} {et~al.}(2017){Atteia}, {Heussaff}, {Dezalay}, {Klotz}, {Turpin}, {Tsvetkova}, {Frederiks}, {Zolnierowski}, {Daigne}, \& {Mochkovitch}}]{Atteia2017}
{Atteia}, J.~L., {Heussaff}, V., {Dezalay}, J.~P., {et~al.} 2017, \apj, 837, 119

\bibitem[{Azzalini \& Capitanio(2002)}]{Azzalini2002}
Azzalini, A. \& Capitanio, A. 2002, Journal of the Royal Statistical Society Series B: Statistical Methodology, 61, 579

\bibitem[{Azzalini \& Capitanio(2003)}]{Azzalini2003}
Azzalini, A. \& Capitanio, A. 2003, Journal of the Royal Statistical Society Series B: Statistical Methodology, 65, 367

\bibitem[{{Band} {et~al.}(1993){Band}, {Matteson}, {Ford}, {Schaefer}, {Palmer}, {Teegarden}, {Cline}, {Briggs}, {Paciesas}, {Pendleton}, {Fishman}, {Kouveliotou}, {Meegan}, {Wilson}, \& {Lestrade}}]{Band1993}
{Band}, D., {Matteson}, J., {Ford}, L., {et~al.} 1993, \apj, 413, 281

\bibitem[{{Barthelmy} {et~al.}(2005){Barthelmy}, {Barbier}, {Cummings}, {Fenimore}, {Gehrels}, {Hullinger}, {Krimm}, {Markwardt}, {Palmer}, {Parsons}, {Sato}, {Suzuki}, {Takahashi}, {Tashiro}, \& {Tueller}}]{Barthelmy2005}
{Barthelmy}, S.~D., {Barbier}, L.~M., {Cummings}, J.~R., {et~al.} 2005, \ssr, 120, 143

\bibitem[{Basso {et~al.}(2010)Basso, Lachos, Cabral, \& Ghosh}]{Basso2010}
Basso, R.~M., Lachos, V.~H., Cabral, C. R.~B., \& Ghosh, P. 2010, Computational Statistics \& Data Analysis, 54, 2926

\bibitem[{{Berger}(2014)}]{Berger2014}
{Berger}, E. 2014, \araa, 52, 43

\bibitem[{{Berger} \& {Becker}(2005)}]{Berger2005}
{Berger}, E. \& {Becker}, G. 2005, GRB Coordinates Network, 3520, 1

\bibitem[{{Biesiada}(2007)}]{Biesiada2007}
{Biesiada}, M. 2007, \jcap, 2007, 003

\bibitem[{{Blinnikov} {et~al.}(1984){Blinnikov}, {Novikov}, {Perevodchikova}, \& {Polnarev}}]{Blinnikov1984}
{Blinnikov}, S.~I., {Novikov}, I.~D., {Perevodchikova}, T.~V., \& {Polnarev}, A.~G. 1984, Soviet Astronomy Letters, 10, 177

\bibitem[{{Borgonovo} {et~al.}(2007){Borgonovo}, {Frontera}, {Guidorzi}, {Montanari}, {Vetere}, \& {Soffitta}}]{Borgonovo2007}
{Borgonovo}, L., {Frontera}, F., {Guidorzi}, C., {et~al.} 2007, \aap, 465, 765

\bibitem[{Burnham \& Anderson(2004)}]{Burnham2004}
Burnham, K.~P. \& Anderson, D.~R. 2004, Sociological Methods \& Research, 33, 261

\bibitem[{Cabral {et~al.}(2012)Cabral, Lachos, \& Prates}]{Cabral2012}
Cabral, C. R.~B., Lachos, V.~H., \& Prates, M.~O. 2012, Computational Statistics \& Data Analysis, 56, 126

\bibitem[{{Camisasca} {et~al.}(2023){Camisasca}, {Guidorzi}, {Amati}, {Frontera}, {Song}, {Xiao}, {Xiong}, {Zhang}, {Margutti}, {Kobayashi}, {Mundell}, {Ge}, {Gomboc}, {Jia}, {Jordana-Mitjans}, {Li}, {Li}, {Maccary}, {Shrestha}, {Xue}, \& {Zhang}}]{Camisasca2023}
{Camisasca}, A.~E., {Guidorzi}, C., {Amati}, L., {et~al.} 2023, \aap, 671, A112

\bibitem[{{Castro-Tirado} {et~al.}(2015){Castro-Tirado}, {Sanchez-Ramirez}, {Gorosabel}, \& {Scarpa}}]{Castro-Tirado2015}
{Castro-Tirado}, A.~J., {Sanchez-Ramirez}, R., {Gorosabel}, J., \& {Scarpa}, R. 2015, GRB Coordinates Network, 17278, 1

\bibitem[{{Dainotti} \& {Amati}(2018)}]{Dainotti2018}
{Dainotti}, M.~G. \& {Amati}, L. 2018, \pasp, 130, 051001

\bibitem[{{Dainotti} \& {Del Vecchio}(2017)}]{Dainotti2017}
{Dainotti}, M.~G. \& {Del Vecchio}, R. 2017, \nar, 77, 23

\bibitem[{{de Ugarte Postigo} {et~al.}(2011){de Ugarte Postigo}, {Horv{\'a}th}, {Veres}, {Bagoly}, {Kann}, {Th{\"o}ne}, {Balazs}, {D'Avanzo}, {Aloy}, {Foley}, {Campana}, {Mao}, {Jakobsson}, {Covino}, {Fynbo}, {Gorosabel}, {Castro-Tirado}, {Amati}, \& {Nardini}}]{deUgartePostigo2011}
{de Ugarte Postigo}, A., {Horv{\'a}th}, I., {Veres}, P., {et~al.} 2011, \aap, 525, A109

\bibitem[{{de Ugarte Postigo} {et~al.}(2016){de Ugarte Postigo}, {Tanvir}, {Cano}, {Izzo}, {Fynbo}, {Sanchez-Ramirez}, {Thoene}, \& {Pesev}}]{deUgartePostigo2016}
{de Ugarte Postigo}, A., {Tanvir}, N.~R., {Cano}, Z., {et~al.} 2016, GRB Coordinates Network, 19245, 1

\bibitem[{{Dezalay} {et~al.}(1991){Dezalay}, {Barat}, {Talon}, {Sunyaev}, {Terekhov}, \& {Kuznetsov}}]{Dezalay1991}
{Dezalay}, J.~P., {Barat}, C., {Talon}, R., {et~al.} 1991, in American Institute of Physics Conference Series, Vol. 265, Gamma-ray Bursts, 304--309

\bibitem[{{Dichiara} {et~al.}(2021){Dichiara}, {Troja}, {Beniamini}, {O'Connor}, {Moss}, {Lien}, {Ricci}, {Amati}, {Ryan}, \& {Sakamoto}}]{Dichiara2021}
{Dichiara}, S., {Troja}, E., {Beniamini}, P., {et~al.} 2021, \apjl, 911, L28

\bibitem[{{Dimple} {et~al.}(2023){Dimple}, {Misra}, \& {Arun}}]{Dimple2023}
{Dimple}, {Misra}, K., \& {Arun}, K.~G. 2023, \apjl, 949, L22

\bibitem[{{Dimple} {et~al.}(2024){Dimple}, {Misra}, \& {Arun}}]{Dimple2024}
{Dimple}, {Misra}, K., \& {Arun}, K.~G. 2024, \apj, 974, 55

\bibitem[{Efron(1979)}]{Efron1979}
Efron, B. 1979, The Annals of Statistics, 7, 1

\bibitem[{{Eichler} {et~al.}(1989){Eichler}, {Livio}, {Piran}, \& {Schramm}}]{Eichler1989}
{Eichler}, D., {Livio}, M., {Piran}, T., \& {Schramm}, D.~N. 1989, \nat, 340, 126

\bibitem[{{Fenimore} {et~al.}(1995){Fenimore}, {in 't Zand}, {Norris}, {Bonnell}, \& {Nemiroff}}]{Fenimore1995}
{Fenimore}, E.~E., {in 't Zand}, J.~J.~M., {Norris}, J.~P., {Bonnell}, J.~T., \& {Nemiroff}, R.~J. 1995, \apjl, 448, L101

\bibitem[{{Fong} {et~al.}(2016){Fong}, {Margutti}, {Chornock}, {Berger}, {Shappee}, {Levan}, {Tanvir}, {Smith}, {Milne}, {Laskar}, {Fox}, {Lunnan}, {Blanchard}, {Hjorth}, {Wiersema}, {van der Horst}, \& {Zaritsky}}]{Fong2016}
{Fong}, W., {Margutti}, R., {Chornock}, R., {et~al.} 2016, \apj, 833, 151

\bibitem[{{Fynbo} {et~al.}(2009){Fynbo}, {Jakobsson}, {Prochaska}, {Malesani}, {Ledoux}, {de Ugarte Postigo}, {Nardini}, {Vreeswijk}, {Wiersema}, {Hjorth}, {Sollerman}, {Chen}, {Th{\"o}ne}, {Bj{\"o}rnsson}, {Bloom}, {Castro-Tirado}, {Christensen}, {De Cia}, {Fruchter}, {Gorosabel}, {Graham}, {Jaunsen}, {Jensen}, {Kann}, {Kouveliotou}, {Levan}, {Maund}, {Masetti}, {Milvang-Jensen}, {Palazzi}, {Perley}, {Pian}, {Rol}, {Schady}, {Starling}, {Tanvir}, {Watson}, {Xu}, {Augusteijn}, {Grundahl}, {Telting}, \& {Quirion}}]{Fynbo2009}
{Fynbo}, J.~P.~U., {Jakobsson}, P., {Prochaska}, J.~X., {et~al.} 2009, \apjs, 185, 526

\bibitem[{{Gehrels} {et~al.}(2004){Gehrels}, {Chincarini}, {Giommi}, {Mason}, {Nousek}, {Wells}, {White}, {Barthelmy}, {Burrows}, {Cominsky}, {Hurley}, {Marshall}, {M{\'e}sz{\'a}ros}, {Roming}, {Angelini}, {Barbier}, {Belloni}, {Campana}, {Caraveo}, {Chester}, {Citterio}, {Cline}, {Cropper}, {Cummings}, {Dean}, {Feigelson}, {Fenimore}, {Frail}, {Fruchter}, {Garmire}, {Gendreau}, {Ghisellini}, {Greiner}, {Hill}, {Hunsberger}, {Krimm}, {Kulkarni}, {Kumar}, {Lebrun}, {Lloyd-Ronning}, {Markwardt}, {Mattson}, {Mushotzky}, {Norris}, {Osborne}, {Paczynski}, {Palmer}, {Park}, {Parsons}, {Paul}, {Rees}, {Reynolds}, {Rhoads}, {Sasseen}, {Schaefer}, {Short}, {Smale}, {Smith}, {Stella}, {Tagliaferri}, {Takahashi}, {Tashiro}, {Townsley}, {Tueller}, {Turner}, {Vietri}, {Voges}, {Ward}, {Willingale}, {Zerbi}, \& {Zhang}}]{Gehrels2004}
{Gehrels}, N., {Chincarini}, G., {Giommi}, P., {et~al.} 2004, \apj, 611, 1005

\bibitem[{{Gillanders} {et~al.}(2023){Gillanders}, {Troja}, {Fryer}, {Ristic}, {O'Connor}, {Fontes}, {Yang}, {Domoto}, {Rahmouni}, {Tanaka}, {Fox}, \& {Dichiara}}]{Gillanders2023}
{Gillanders}, J.~H., {Troja}, E., {Fryer}, C.~L., {et~al.} 2023, arXiv e-prints, arXiv:2308.00633

\bibitem[{{Golkhou} \& {Butler}(2014)}]{Golkhou2014}
{Golkhou}, V.~Z. \& {Butler}, N.~R. 2014, \apj, 787, 90

\bibitem[{{Golkhou} {et~al.}(2015){Golkhou}, {Butler}, \& {Littlejohns}}]{Golkhou2015}
{Golkhou}, V.~Z., {Butler}, N.~R., \& {Littlejohns}, O.~M. 2015, \apj, 811, 93

\bibitem[{{Gruber} {et~al.}(2011){Gruber}, {Greiner}, {von Kienlin}, {Rau}, {Briggs}, {Connaughton}, {Goldstein}, {van der Horst}, {Nardini}, {Bhat}, {Bissaldi}, {Burgess}, {Chaplin}, {Diehl}, {Fishman}, {Fitzpatrick}, {Foley}, {Gibby}, {Giles}, {Guiriec}, {Kippen}, {Kouveliotou}, {Lin}, {McBreen}, {Meegan}, {Olivares E.}, {Paciesas}, {Preece}, {Tierney}, \& {Wilson-Hodge}}]{Gruber2011}
{Gruber}, D., {Greiner}, J., {von Kienlin}, A., {et~al.} 2011, \aap, 531, A20

\bibitem[{{Hjorth} {et~al.}(2012){Hjorth}, {Malesani}, {Jakobsson}, {Jaunsen}, {Fynbo}, {Gorosabel}, {Kr{\"u}hler}, {Levan}, {Micha{\l}owski}, {Milvang-Jensen}, {M{\o}ller}, {Schulze}, {Tanvir}, \& {Watson}}]{Hjorth2012}
{Hjorth}, J., {Malesani}, D., {Jakobsson}, P., {et~al.} 2012, \apj, 756, 187

\bibitem[{{Horv{\'a}th}(1998)}]{Horvath1998}
{Horv{\'a}th}, I. 1998, \apj, 508, 757

\bibitem[{{Huja} {et~al.}(2009){Huja}, {M{\'e}sz{\'a}ros}, \& {{\v{R}}{\'\i}pa}}]{Huja2009}
{Huja}, D., {M{\'e}sz{\'a}ros}, A., \& {{\v{R}}{\'\i}pa}, J. 2009, \aap, 504, 67

\bibitem[{{Jespersen} {et~al.}(2020){Jespersen}, {Severin}, {Steinhardt}, {Vinther}, {Fynbo}, {Selsing}, \& {Watson}}]{Jespersen2020}
{Jespersen}, C.~K., {Severin}, J.~B., {Steinhardt}, C.~L., {et~al.} 2020, \apjl, 896, L20

\bibitem[{Kass \& Raftery(1995)}]{Kass1995}
Kass, R.~E. \& Raftery, A.~E. 1995, Journal of the American Statistical Association, 90, 773

\bibitem[{{Knust} {et~al.}(2017){Knust}, {Greiner}, {van Eerten}, {Schady}, {Kann}, {Chen}, {Delvaux}, {Graham}, {Klose}, {Kr{\"u}hler}, {McConnell}, {Nicuesa Guelbenzu}, {Perley}, {Schmidl}, {Schweyer}, {Tanga}, \& {Varela}}]{Knust2017}
{Knust}, F., {Greiner}, J., {van Eerten}, H.~J., {et~al.} 2017, \aap, 607, A84

\bibitem[{{Kocevski} \& {Petrosian}(2013)}]{Kocevski2013}
{Kocevski}, D. \& {Petrosian}, V. 2013, \apj, 765, 116

\bibitem[{{Koen} \& {Bere}(2012)}]{Koen2012}
{Koen}, C. \& {Bere}, A. 2012, \mnras, 420, 405

\bibitem[{{Kouveliotou} {et~al.}(1993){Kouveliotou}, {Meegan}, {Fishman}, {Bhat}, {Briggs}, {Koshut}, {Paciesas}, \& {Pendleton}}]{Kouveliotou1993}
{Kouveliotou}, C., {Meegan}, C.~A., {Fishman}, G.~J., {et~al.} 1993, \apjl, 413, L101

\bibitem[{{Kulkarni} \& {Desai}(2017)}]{Kulkarni2017}
{Kulkarni}, S. \& {Desai}, S. 2017, \apss, 362, 70

\bibitem[{{Kwong} \& {Nadarajah}(2018)}]{Kwong2018}
{Kwong}, H.~S. \& {Nadarajah}, S. 2018, \mnras, 473, 625

\bibitem[{{Levan} {et~al.}(2015){Levan}, {Hjorth}, {Wiersema}, \& {Tanvir}}]{Levan2015}
{Levan}, A., {Hjorth}, J., {Wiersema}, K., \& {Tanvir}, N. 2015, The Astronomer's Telegram, 6873, 1

\bibitem[{{Levan} {et~al.}(2024){Levan}, {Gompertz}, {Salafia}, {Bulla}, {Burns}, {Hotokezaka}, {Izzo}, {Lamb}, {Malesani}, {Oates}, {Ravasio}, {Rouco Escorial}, {Schneider}, {Sarin}, {Schulze}, {Tanvir}, {Ackley}, {Anderson}, {Brammer}, {Christensen}, {Dhillon}, {Evans}, {Fausnaugh}, {Fong}, {Fruchter}, {Fryer}, {Fynbo}, {Gaspari}, {Heintz}, {Hjorth}, {Kennea}, {Kennedy}, {Laskar}, {Leloudas}, {Mandel}, {Martin-Carrillo}, {Metzger}, {Nicholl}, {Nugent}, {Palmerio}, {Pugliese}, {Rastinejad}, {Rhodes}, {Rossi}, {Saccardi}, {Smartt}, {Stevance}, {Tohuvavohu}, {van der Horst}, {Vergani}, {Watson}, {Barclay}, {Bhirombhakdi}, {Breedt}, {Breeveld}, {Brown}, {Campana}, {Chrimes}, {D'Avanzo}, {D'Elia}, {De Pasquale}, {Dyer}, {Galloway}, {Garbutt}, {Green}, {Hartmann}, {Jakobsson}, {Kerry}, {Kouveliotou}, {Langeroodi}, {Le Floc'h}, {Leung}, {Littlefair}, {Munday}, {O'Brien}, {Parsons}, {Pelisoli}, {Sahman}, {Salvaterra}, {Sbarufatti}, {Steeghs}, {Tagliaferri}, {Th{\"o}ne}, {de Ugarte Postigo}, \& {Kann}}]{Levan2024}
{Levan}, A.~J., {Gompertz}, B.~P., {Salafia}, O.~S., {et~al.} 2024, \nat, 626, 737

\bibitem[{{Li} {et~al.}(2020){Li}, {Zhang}, \& {Yuan}}]{Li2020}
{Li}, Y., {Zhang}, B., \& {Yuan}, Q. 2020, \apj, 897, 154

\bibitem[{{Liddle}(2007)}]{Liddle2007}
{Liddle}, A.~R. 2007, \mnras, 377, L74

\bibitem[{{Lien} {et~al.}(2016){Lien}, {Sakamoto}, {Barthelmy}, {Baumgartner}, {Cannizzo}, {Chen}, {Collins}, {Cummings}, {Gehrels}, {Krimm}, {Markwardt}, {Palmer}, {Stamatikos}, {Troja}, \& {Ukwatta}}]{Lien2016}
{Lien}, A., {Sakamoto}, T., {Barthelmy}, S.~D., {et~al.} 2016, \apj, 829, 7

\bibitem[{{Littlejohns} \& {Butler}(2014)}]{Littlejohns2014}
{Littlejohns}, O.~M. \& {Butler}, N.~R. 2014, \mnras, 444, 3948

\bibitem[{{Luo} {et~al.}(2023){Luo}, {Wang}, {Zhu-Ge}, {Li}, {Zou}, \& {Zhang}}]{Luo2023}
{Luo}, J.-W., {Wang}, F.-F., {Zhu-Ge}, J.-M., {et~al.} 2023, \apj, 959, 44

\bibitem[{{MacFadyen} \& {Woosley}(1999)}]{MacFadyen1999}
{MacFadyen}, A.~I. \& {Woosley}, S.~E. 1999, \apj, 524, 262

\bibitem[{{Mazets} {et~al.}(1981){Mazets}, {Golenetskii}, {Ilinskii}, {Panov}, {Aptekar}, {Gurian}, {Proskura}, {Sokolov}, {Sokolova}, \& {Kharitonova}}]{Mazets1981}
{Mazets}, E.~P., {Golenetskii}, S.~V., {Ilinskii}, V.~N., {et~al.} 1981, \apss, 80, 3

\bibitem[{{Meegan} {et~al.}(2009){Meegan}, {Lichti}, {Bhat}, {Bissaldi}, {Briggs}, {Connaughton}, {Diehl}, {Fishman}, {Greiner}, {Hoover}, {van der Horst}, {von Kienlin}, {Kippen}, {Kouveliotou}, {McBreen}, {Paciesas}, {Preece}, {Steinle}, {Wallace}, {Wilson}, \& {Wilson-Hodge}}]{Meegan2009}
{Meegan}, C., {Lichti}, G., {Bhat}, P.~N., {et~al.} 2009, \apj, 702, 791

\bibitem[{{Minaev} \& {Pozanenko}(2020)}]{Minaev2020}
{Minaev}, P.~Y. \& {Pozanenko}, A.~S. 2020, \mnras, 492, 1919

\bibitem[{{Minaev} \& {Pozanenko}(2021)}]{Minaev2021_err}
{Minaev}, P.~Y. \& {Pozanenko}, A.~S. 2021, \mnras, 504, 926

\bibitem[{{Narayan} {et~al.}(1992){Narayan}, {Paczynski}, \& {Piran}}]{Narayan1992}
{Narayan}, R., {Paczynski}, B., \& {Piran}, T. 1992, \apjl, 395, L83

\bibitem[{{Norris} \& {Bonnell}(2006)}]{Norris2006}
{Norris}, J.~P. \& {Bonnell}, J.~T. 2006, \apj, 643, 266

\bibitem[{{Norris} {et~al.}(1984){Norris}, {Cline}, {Desai}, \& {Teegarden}}]{Norris1984}
{Norris}, J.~P., {Cline}, T.~L., {Desai}, U.~D., \& {Teegarden}, B.~J. 1984, \nat, 308, 434

\bibitem[{{Norris} {et~al.}(1996){Norris}, {Nemiroff}, {Bonnell}, {Scargle}, {Kouveliotou}, {Paciesas}, {Meegan}, \& {Fishman}}]{Norris1996}
{Norris}, J.~P., {Nemiroff}, R.~J., {Bonnell}, J.~T., {et~al.} 1996, \apj, 459, 393

\bibitem[{{Paczynski}(1986)}]{Paczynski1986}
{Paczynski}, B. 1986, \apjl, 308, L43

\bibitem[{{Paczynski}(1991)}]{Paczynski1991}
{Paczynski}, B. 1991, \actaa, 41, 257

\bibitem[{{Paczy{\'n}ski}(1998)}]{Paczynski1998}
{Paczy{\'n}ski}, B. 1998, \apjl, 494, L45

\bibitem[{{Perley} {et~al.}(2017){Perley}, {Kr{\"u}hler}, {Schady}, {Micha{\l}owski}, {Th{\"o}ne}, {Petry}, {Graham}, {Greiner}, {Klose}, {Schulze}, \& {Kim}}]{Perley2017}
{Perley}, D.~A., {Kr{\"u}hler}, T., {Schady}, P., {et~al.} 2017, \mnras, 465, L89

\bibitem[{{Perley} {et~al.}(2016){Perley}, {Tanvir}, {Hjorth}, {Laskar}, {Berger}, {Chary}, {de Ugarte Postigo}, {Fynbo}, {Kr{\"u}hler}, {Levan}, {Micha{\l}owski}, \& {Schulze}}]{Perley2016}
{Perley}, D.~A., {Tanvir}, N.~R., {Hjorth}, J., {et~al.} 2016, \apj, 817, 8

\bibitem[{Prates {et~al.}(2013)Prates, Lachos, \& Barbosa~Cabral}]{Prates2013}
Prates, M.~O., Lachos, V.~H., \& Barbosa~Cabral, C.~R. 2013, Journal of Statistical Software, 54, 1–20

\bibitem[{Quenouille(1949)}]{Quenouille1949}
Quenouille, M.~H. 1949, The Annals of Mathematical Statistics, 20, 355

\bibitem[{Quenouille(1956)}]{Quenouille1956}
Quenouille, M.~H. 1956, Biometrika, 43, 353

\bibitem[{{Rastinejad} {et~al.}(2022){Rastinejad}, {Gompertz}, {Levan}, {Fong}, {Nicholl}, {Lamb}, {Malesani}, {Nugent}, {Oates}, {Tanvir}, {de Ugarte Postigo}, {Kilpatrick}, {Moore}, {Metzger}, {Ravasio}, {Rossi}, {Schroeder}, {Jencson}, {Sand}, {Smith}, {Ag{\"u}{\'\i} Fern{\'a}ndez}, {Berger}, {Blanchard}, {Chornock}, {Cobb}, {De Pasquale}, {Fynbo}, {Izzo}, {Kann}, {Laskar}, {Marini}, {Paterson}, {Escorial}, {Sears}, \& {Th{\"o}ne}}]{Rastinejad2022}
{Rastinejad}, J.~C., {Gompertz}, B.~P., {Levan}, A.~J., {et~al.} 2022, \nat, 612, 223

\bibitem[{{Rossi} {et~al.}(2022){Rossi}, {Rothberg}, {Palazzi}, {Kann}, {D'Avanzo}, {Amati}, {Klose}, {Perego}, {Pian}, {Guidorzi}, {Pozanenko}, {Savaglio}, {Stratta}, {Agapito}, {Covino}, {Cusano}, {D'Elia}, {De Pasquale}, {Della Valle}, {Kuhn}, {Izzo}, {Loffredo}, {Masetti}, {Melandri}, {Minaev}, {Guelbenzu}, {Paris}, {Paiano}, {Plantet}, {Rossi}, {Salvaterra}, {Schulze}, {Veillet}, \& {Volnova}}]{Rossi2022}
{Rossi}, A., {Rothberg}, B., {Palazzi}, E., {et~al.} 2022, \apj, 932, 1

\bibitem[{{Sakamoto} {et~al.}(2008){Sakamoto}, {Barthelmy}, {Barbier}, {Cummings}, {Fenimore}, {Gehrels}, {Hullinger}, {Krimm}, {Markwardt}, {Palmer}, {Parsons}, {Sato}, {Stamatikos}, {Tueller}, {Ukwatta}, \& {Zhang}}]{Sakamoto2008}
{Sakamoto}, T., {Barthelmy}, S.~D., {Barbier}, L., {et~al.} 2008, \apjs, 175, 179

\bibitem[{{Sakamoto} {et~al.}(2011{\natexlab{a}}){Sakamoto}, {Barthelmy}, {Baumgartner}, {Cummings}, {Fenimore}, {Gehrels}, {Krimm}, {Markwardt}, {Palmer}, {Parsons}, {Sato}, {Stamatikos}, {Tueller}, {Ukwatta}, \& {Zhang}}]{Sakamoto2011}
{Sakamoto}, T., {Barthelmy}, S.~D., {Baumgartner}, W.~H., {et~al.} 2011{\natexlab{a}}, \apjs, 195, 2

\bibitem[{{Sakamoto} {et~al.}(2011{\natexlab{b}}){Sakamoto}, {Pal'Shin}, {Yamaoka}, {Ohno}, {Sato}, {Aptekar}, {Barthelmy}, {Baumgartner}, {Cummings}, {Fenimore}, {Frederiks}, {Gehrels}, {Golenetskii}, {Krimm}, {Markwardt}, {Onda}, {Palmer}, {Parsons}, {Stamatikos}, {Sugita}, {Tashiro}, {Tueller}, \& {Ukwatta}}]{Sakamoto2011_cc}
{Sakamoto}, T., {Pal'Shin}, V., {Yamaoka}, K., {et~al.} 2011{\natexlab{b}}, \pasj, 63, 215

\bibitem[{{Salmon} {et~al.}(2022{\natexlab{a}}){Salmon}, {Hanlon}, \& {Martin-Carrillo}}]{Salmon2022a}
{Salmon}, L., {Hanlon}, L., \& {Martin-Carrillo}, A. 2022{\natexlab{a}}, Galaxies, 10, 78

\bibitem[{{Salmon} {et~al.}(2022{\natexlab{b}}){Salmon}, {Hanlon}, \& {Martin-Carrillo}}]{Salmon2022b}
{Salmon}, L., {Hanlon}, L., \& {Martin-Carrillo}, A. 2022{\natexlab{b}}, Galaxies, 10, 77

\bibitem[{{Scargle} {et~al.}(2013){Scargle}, {Norris}, {Jackson}, \& {Chiang}}]{Scargle2013}
{Scargle}, J.~D., {Norris}, J.~P., {Jackson}, B., \& {Chiang}, J. 2013, \apj, 764, 167

\bibitem[{{Schwarz}(1978)}]{Schwarz1978}
{Schwarz}, G. 1978, Annals of Statistics, 6, 461

\bibitem[{{Svinkin} {et~al.}(2019){Svinkin}, {Aptekar}, {Golenetskii}, {Frederiks}, {Ulanov}, \& {Tsvetkova}}]{Svinkin2019}
{Svinkin}, D.~S., {Aptekar}, R.~L., {Golenetskii}, S.~V., {et~al.} 2019, in Journal of Physics Conference Series, Vol. 1400, Journal of Physics Conference Series, 022010

\bibitem[{{Svinkin} {et~al.}(2016){Svinkin}, {Frederiks}, {Aptekar}, {Golenetskii}, {Pal'shin}, {Oleynik}, {Tsvetkova}, {Ulanov}, {Cline}, \& {Hurley}}]{Svinkin2016}
{Svinkin}, D.~S., {Frederiks}, D.~D., {Aptekar}, R.~L., {et~al.} 2016, \apjs, 224, 10

\bibitem[{{Tarnopolski}(2015)}]{Tarnopolski2015}
{Tarnopolski}, M. 2015, \aap, 581, A29

\bibitem[{{Tarnopolski}(2016{\natexlab{a}})}]{Tarnopolski2016b}
{Tarnopolski}, M. 2016{\natexlab{a}}, \mnras, 458, 2024

\bibitem[{{Tarnopolski}(2016{\natexlab{b}})}]{Tarnopolski2016c}
{Tarnopolski}, M. 2016{\natexlab{b}}, \apss, 361, 125

\bibitem[{{Tarnopolski}(2016{\natexlab{c}})}]{Tarnopolski2016a}
{Tarnopolski}, M. 2016{\natexlab{c}}, \na, 46, 54

\bibitem[{{Tarnopolski}(2019{\natexlab{a}})}]{Tarnopolski2019}
{Tarnopolski}, M. 2019{\natexlab{a}}, \apj, 870, 105

\bibitem[{{Tarnopolski}(2019{\natexlab{b}})}]{Tarnopolski2019a}
{Tarnopolski}, M. 2019{\natexlab{b}}, \apj, 887, 97

\bibitem[{{Troja} {et~al.}(2022){Troja}, {Fryer}, {O'Connor}, {Ryan}, {Dichiara}, {Kumar}, {Ito}, {Gupta}, {Wollaeger}, {Norris}, {Kawai}, {Butler}, {Aryan}, {Misra}, {Hosokawa}, {Murata}, {Niwano}, {Pandey}, {Kutyrev}, {van Eerten}, {Chase}, {Hu}, {Caballero-Garcia}, \& {Castro-Tirado}}]{Troja2022}
{Troja}, E., {Fryer}, C.~L., {O'Connor}, B., {et~al.} 2022, \nat, 612, 228

\bibitem[{{Tsvetkova} {et~al.}(2017){Tsvetkova}, {Frederiks}, {Golenetskii}, {Lysenko}, {Oleynik}, {Pal'shin}, {Svinkin}, {Ulanov}, {Cline}, {Hurley}, \& {Aptekar}}]{Tsvetkova2017}
{Tsvetkova}, A., {Frederiks}, D., {Golenetskii}, S., {et~al.} 2017, \apj, 850, 161

\bibitem[{{Tsvetkova} {et~al.}(2021){Tsvetkova}, {Frederiks}, {Svinkin}, {Aptekar}, {Cline}, {Golenetskii}, {Hurley}, {Lysenko}, {Ridnaia}, \& {Ulanov}}]{Tsvetkova2021}
{Tsvetkova}, A., {Frederiks}, D., {Svinkin}, D., {et~al.} 2021, \apj, 908, 83

\bibitem[{{Tsvetkova} {et~al.}(2022){Tsvetkova}, {Svinkin}, {Karpov}, \& {Frederiks}}]{Tsvetkova2022}
{Tsvetkova}, A., {Svinkin}, D., {Karpov}, S., \& {Frederiks}, D. 2022, Universe, 8, 373

\bibitem[{Tukey(1958)}]{Tukey1958}
Tukey, J.~W. 1958, The Annals of Mathematical Statistics, 29, 614

\bibitem[{{Turpin} {et~al.}(2016){Turpin}, {Heussaff}, {Dezalay}, {Atteia}, {Klotz}, \& {Dornic}}]{Turpin2016}
{Turpin}, D., {Heussaff}, V., {Dezalay}, J.~P., {et~al.} 2016, \apj, 831, 28

\bibitem[{{von Kienlin} {et~al.}(2014){von Kienlin}, {Meegan}, {Paciesas}, {Bhat}, {Bissaldi}, {Briggs}, {Burgess}, {Byrne}, {Chaplin}, {Cleveland}, {Connaughton}, {Collazzi}, {Fitzpatrick}, {Foley}, {Gibby}, {Giles}, {Goldstein}, {Greiner}, {Gruber}, {Guiriec}, {van der Horst}, {Kouveliotou}, {Layden}, {McBreen}, {McGlynn}, {Pelassa}, {Preece}, {Rau}, {Tierney}, {Wilson-Hodge}, {Xiong}, {Younes}, \& {Yu}}]{vonKienlin2014}
{von Kienlin}, A., {Meegan}, C.~A., {Paciesas}, W.~S., {et~al.} 2014, \apjs, 211, 13

\bibitem[{{Wagenmakers} \& {Farrell}(2004)}]{Wagenmakers2004}
{Wagenmakers}, E.-J. \& {Farrell}, S. 2004, Psychonomic Bulletin \& Review, 11, 192

\bibitem[{{Woosley}(1993)}]{Woosley1993}
{Woosley}, S.~E. 1993, \apj, 405, 273

\bibitem[{{Woosley} \& {Bloom}(2006)}]{Woosley2006}
{Woosley}, S.~E. \& {Bloom}, J.~S. 2006, \araa, 44, 507

\bibitem[{{Yang} {et~al.}(2016){Yang}, {Zhang}, \& {Jiang}}]{Yang2016}
{Yang}, E.~B., {Zhang}, Z.~B., \& {Jiang}, X.~X. 2016, \apss, 361, 257

\bibitem[{{Yang} {et~al.}(2022){Yang}, {Ai}, {Zhang}, {Zhang}, {Liu}, {Wang}, {Yang}, {Yin}, {Li}, \& {L{\"u}}}]{Yang2022}
{Yang}, J., {Ai}, S., {Zhang}, B.-B., {et~al.} 2022, \nat, 612, 232

\bibitem[{{Yang} {et~al.}(2024){Yang}, {Troja}, {O'Connor}, {Fryer}, {Im}, {Durbak}, {Paek}, {Ricci}, {Bom}, {Gillanders}, {Castro-Tirado}, {Peng}, {Dichiara}, {Ryan}, {van Eerten}, {Dai}, {Chang}, {Choi}, {De}, {Hu}, {Kilpatrick}, {Kutyrev}, {Jeong}, {Lee}, {Makler}, {Navarete}, \& {P{\'e}rez-Garc{\'\i}a}}]{Yang2024}
{Yang}, Y.-H., {Troja}, E., {O'Connor}, B., {et~al.} 2024, \nat, 626, 742

\bibitem[{{Zhang} {et~al.}(2009){Zhang}, {Zhang}, {Virgili}, {Liang}, {Kann}, {Wu}, {Proga}, {Lv}, {Toma}, {M{\'e}sz{\'a}ros}, {Burrows}, {Roming}, \& {Gehrels}}]{Zhang2009}
{Zhang}, B., {Zhang}, B.-B., {Virgili}, F.~J., {et~al.} 2009, \apj, 703, 1696

\bibitem[{{Zhang} {et~al.}(2016){Zhang}, {Yang}, {Choi}, \& {Chang}}]{Zhang2016}
{Zhang}, Z.-B., {Yang}, E.-B., {Choi}, C.-S., \& {Chang}, H.-Y. 2016, \mnras, 462, 3243

\bibitem[{{Zhang} {et~al.}(2018){Zhang}, {Zhang}, {Zhao}, {Luo}, {Jiang}, {Wang}, {Han}, \& {Terheide}}]{Zhang2018}
{Zhang}, Z.~B., {Zhang}, C.~T., {Zhao}, Y.~X., {et~al.} 2018, \pasp, 130, 054202

\bibitem[{{Zitouni} {et~al.}(2015){Zitouni}, {Guessoum}, {Azzam}, \& {Mochkovitch}}]{Zitouni2015}
{Zitouni}, H., {Guessoum}, N., {Azzam}, W.~J., \& {Mochkovitch}, R. 2015, \apss, 357, 7

\end{thebibliography}
\bibliographystyle{aa}

\begin{appendix}
\section{Instrumentation}
\label{sec:instrumentation}
\subsection{Konus-\textit{Wind}} \label{app:kw}
The Konus-Wind (hereafter KW; \citealt{Aptekar1995}) experiment has been operating since 1994 November continuously observes the full sky by two omnidirectional NaI detectors (S1 and S2) with high temporal resolution (down to 2~ms for light curves and 64~ms for spectra) in the wide energy range of $\sim$10~keV--10~MeV, nominally, which is now $\sim$20~keV--20~MeV.
The instrument has two operational modes: waiting and triggered. 
In the triggered mode, the light curves are recorded in three energy windows: G1~($\sim$13--50~keV), G2~($\sim$50--200~keV), and G3~($\sim$200--760~keV) with the time resolution varying from 2~ms to 256~ms.
The detector sensitivity\footnote{The KW sensitivity to GRBs is approximately defined as the lowest energy flux in the sample.} can be estimated as 1--$ 2\times 10^{-6}$~erg~cm$^{-2}$~s$^{-1}$ and $5 \times 10^{-7}$~erg~cm$^{-2}$~s$^{-1}$  in the triggered and waiting modes, correspondingly.
Further details on the triggered and waiting mode data can be found in \citet{Tsvetkova2021}.
As for June 2024, the instrument has detected $\sim$ 3800 triggered GRBs.

\subsection{\textit{Swift}/BAT} \label{app:bat}
The Neil Gehrels  \textit{Swift} Observatory, dedicated primarily to GRB afterglow studies, was launched on 2004~November~20~\citep{Gehrels2004}.
BAT is a highly sensitive, large field-of-view (FOV;  1.4~sr for $>50\%$ coded FOV and 2.2~sr for $>10\%$ coded FOV) coded-aperture telescope onboard \textit{Swift} that detects and localises GRBs in real time.
The BAT energy range spans 14--150~keV in imaging mode.
The detector sensitivity is around $1 \times 10^{-8}$ ~erg~cm$^{-2}$~s$^{-1}$.
However, the effective limiting flux of the sample is dominated by the lower the KW waiting-mode sensitivity.
Further details of the BAT instrument, including the in-orbit calibrations, can be found in \cite{Barthelmy2005} and the BAT GRB catalogues (\citealt{Sakamoto2008, Sakamoto2011, Lien2016}).
Up to June 2024, BAT has detected $\sim$1600 GRBs, of which $\sim$430 events have measured redshifts.

\subsection{\textit{Fermi}/GBM} \label{app:gbm}
The \textit{Fermi} Gamma-ray Space Observatory, dedicated to study transient Gamma-ray sources, was launched in June~2008.
The Gamma-ray Burst Monitor (GBM; \citealt{Meegan2009}) instrument onboard it observes the whole sky not occulted by the Earth ($>8$~sr) covering the energy range from 8~keV to 40~MeV and is composed of twelve NaI(Tl) detectors and two bismuth-germanate (BGO) scintillation detectors.
The instrument sensitivity is around $5 \times 10^{-8}$~erg~cm$^{-2}$~s$^{-1}$.
More details on the design and performance of the facilities detecting GRBs may be found in \citet{Tsvetkova2022}.
So far, GBM has detected $\sim$3800 triggered GRBs.

\section{Remarks on the duration computation}
\label{sec:durations}
Despite $T_{90}$, the time interval that contains 5\% to 95\% of the total burst count fluence, is defined unambiguously, the methodology of the computation of the total GRB duration ($T_{100}$), which $T_{90}$ is connected with, differs in these three samples. 
(1) To compute $T_{100}$ durations for the triggered KW sample, a concatenation of waiting-mode and triggered-mode light curves in the $\sim$80--1200~keV energy range was used. 
The burst's start and end times were determined at 5$\sigma$ excess above background on the timescales from 2~ms to 2.944~s in the interval from $T_0-200$~s to $T_0+240$~s (the end of the KW triggered mode record). 
In some cases, the search interval was narrowed to exclude a non-GRB event overlapping with a target event.
The background was approximated by a constant, using, typically, the interval from $T-1200$~s to $T-200$~s\footnote{After the triggered-mode measurements are finished, KW switches into the data-readout mode for $\sim$1~hr and no measurements are available for this time interval.}, during which count rates in all energy ranges of both KW detectors are consistent with being Poisson distributed.
Further details can be found in Section 4.1.1. in \citet{Tsvetkova2017}.
(2) The $T_{100}$ durations for the ``KW \& BAT'' sample were determined using the Bayesian block decomposition \citep{Scargle2013} of the BAT light curve collected with the 64-ms time binning in the 25--350~keV energy band. 
This energy band was selected as, first, being common to both instruments and, second, being less sensitive to weak precursors or soft extended tails. 
(3) The standard practise for the GBM $T_{100}$ calculation is to generally identify, by eye, what are the pre-burst and post-burst time intervals for background assessment.  
After background is fit to pre- and post-burst regions, the light curve is split into 64-ms bins and each bin is fit to retrieve the deconvolved photon flux.  
That flux is then cumulatively added, which generates an S-shaped curve for a single-pulse GRB (see Fig 3 in \citealt{vonKienlin2014}). The time at which the curve deviates from the plateau indicates the start and stop times of the burst. 

We compared the $T_{90}$ durations of the bursts belonging to the intersection of the ``GBM all'' and two other samples (the ``KW trig'' and ``KW \& BAT'' samples obviously do not overlap).
Overall, the median value of the $T_{90}$ ratios is around unity, despite for some outliers, especially for the weak bursts, the discrepancy between the durations may reach 1.5 orders of magnitude.

\section{Methods} \label{sec:methods}
In this Section, we follow the notations in \citet{Tarnopolski2019}.

\subsection{Distributions}
We approximated the data on the hardness-duration plane with the Normal and Student distributions, both skewed and non-skewed. 
The Student-$t$ distribution is symmetric, similarly to the normal distribution, but has a wider spread and a more slender shape.
For each of the four distributions, we fitted the data with a mixture of one to three components, therefore implying one, two, or three clusters of GRBs.
The probability density function (PDF) of a mixture of $n$ components with individual PDFs $f_i(\mathbf{x}, \theta^{(i)})$ is 
$$f(\mathbf{x}, \theta) = \sum^n_{i=1}A_i f_i(\mathbf{x}, \theta^{(i)}),$$
where $A_i$ are the weights: $\sum^n_{i=1} A_i = 1$, and $\theta = \cup^n_{i=1} \theta^{(i)}$ are the distribution parameters.

The $k$-dimensional Gaussian (Normal) distribution is described by the PDF:
$$f^{(N)}_k(\mathbf{x, \mu, \Sigma}) = \frac{1}{\sqrt{(2 \pi)^k |\Sigma|}} \exp \left[-\frac{1}{2}(\mathbf{x} - \mathbf{\mu})^\top \Sigma^{-1} (\mathbf{x} - \mathbf{\mu})\right],$$
where $\mathbf{\mu}$ is the location vector, which in the case of a non-skewed distribution is also the mean, $\Sigma$ is the covariance matrix, and $|\Sigma| = \det \Sigma$. 
A mixture of $n$ components has $p =  6n - 1$ free parameters for $\mathbf{k=2}$.

The multivariate skew-normal distribution \citep{Azzalini2002, Prates2013} is defined by
$$f_k^{(SN)}(\mathbf{x, \mu, \Sigma, \lambda}) = 2f^{(N)}_k(\mathbf{x, \mu, \Sigma}) \Phi(\lambda^\top \Sigma^{-1/2}(\mathbf{x} - \mathbf{\mu})),$$
where $\Phi$ is the Cumulative Distribution Function (CDF) of a univariate standard normal distribution, and $\mathbf{\lambda}$ is the skewness parameter vector. 
Indeed, the skewed Normal distribution reduces to the Gaussian one at $\mathbf{\lambda} = 0$. 
The mean and the covariance of the skewed-Normal distribution are given by 
$\mathbf{m_\mathrm{skewed}} = \mathbf{\mu} + \sqrt{\frac{2}{\pi}} \frac{\mathbf{\Sigma}\mathbf{\lambda}}{\sqrt{1+\mathbf{\lambda}^\top\mathbf{\Sigma}\mathbf{\lambda}}}$ and $\mathbf{\Sigma_\mathrm{skewed}} = \mathbf{\Sigma} - (\mathbf{m} - \mathbf{\mu})(\mathbf{m} - \mathbf{\mu})^\top$, 
where $^\top$ denotes the transpose of a matrix, correspondingly.
For this distribution, the number of free parameters defining mixture of $n$ components is $p = 8n - 1$.

The multivariate Student distribution \citep{Basso2010, Cabral2012, Prates2013} with $\nu$ degrees of freedom (d.o.f.) is described by
\begin{equation}
  \begin{aligned}
        f^{(T)}_k(\mathbf{x, \mu, \Sigma}, \nu) & = \frac{1}{\sqrt{(\pi\nu)^k |\Sigma|}} \frac{\Gamma\left(\frac{\nu+k}{2}\right)}{\Gamma\left(\frac{\nu}{2}\right)} \\
        & \times \left(1 + \frac{1}{\nu}(\mathbf{x} - \mathbf{\mu})^\top\mathbf{\Sigma}^{-1}(\mathbf{x} - \mathbf{\mu})\right)^{-\frac{\nu+k}{2}},
  \end{aligned}
\end{equation}
where $\Gamma$ is the gamma function, $\mu$ is the mean of the distribution for $\nu>1$, while the covariance matrix (for $\nu>2$) is $\frac{\nu}{\nu-2}\mathbf{\Sigma}$.
The Student distribution approaches the Gaussian in the limit of $\nu \rightarrow \infty$.
A mixture of $n$ components has $p = 6n$ free parameters.

The multivariate skew-Student distribution \citep{Cabral2012, Prates2013} is defined as
\begin{equation}
  \begin{aligned}
    f^{(ST)}_k(\mathbf{x, \mu, \Sigma}, \nu, \lambda) & = 2f_k^{(T)}(\mathbf{x, \mu, \Sigma}, \nu) T_{\nu+k}\\
    & \times \left(\sqrt{\frac{\nu+k}{\nu+(\mathbf{x}-\mathbf{\mu})^\top\Sigma^{-1}(\mathbf{x}-\mathbf{\mu})}} \mathbf{\lambda}^\top\mathbf{\Sigma}^{-1/2}(\mathbf{x}-\mathbf{\mu})\right),
  \end{aligned}
\end{equation}
where $T_{\nu+k}$ is the CDF of the standard univariate Student distribution with $\nu+k$ d.o.f., and $\lambda$ defines the skewness parameter vector. 
The skewed Student distribution reduces to the non-skewed one when $\lambda=0$, meanwhile it approaches the skewed Normal distribution in the limit of $\nu \rightarrow \infty$. 
The mean for $\nu > 1$ is given by $m_\mathrm{skewed} = \mu + \omega \xi$, while its covariance for $\nu > 2$ is $\Sigma_\mathrm{skewed} = \frac{\nu}{\nu-2}\mathbf{\Sigma} - (\mathbf{m} - \mathbf{\mu})(\mathbf{m} - \mathbf{\mu})^\top$, where $\xi = \sqrt{\frac{\nu}{\pi(1+\mathbf{\lambda^\top \Sigma \lambda})}} \frac{\Gamma\left(\frac{\nu-1}{2}\right)}{\Gamma\left(\frac{\nu}{2}\right)}\mathbf{\Sigma\lambda}$ and $\omega = \mathrm{diag}(\Sigma_{11}, ..., \Sigma_{kk})^{1/2}$ \citep{Azzalini2003}.
A mixture of $n$ components is described by $p = 8n$ free parameters and has a non-zero skewness unless $\lambda = 0$ for $n > 3$.

\subsection{Model comparison using Information Criteria}
For a multidimensional distribution with a PDF $f = f(\mathbf{x}, \theta)$, possibly a mixture, where $\theta = {\theta_i}^p_{i=1}$ is a set of $p$ parameters, the log-likelihood function is $$\mathcal{L}_p(\theta) = \sum^{N}_{i=1} \ln f(x_i, \theta),$$
where $\{x_i\}^N_{i=1}$ are $N$ data points from the sample to which a distribution is fitted. 

To fit a model to the data,  we searched the set of parameters $\hat{\theta}$ that yields the highest value of likelihood ($\mathcal{L}_\textrm{p,max}$) using the package \texttt{mixsmsn}\footnote{\url{https://cran.r-project.org/web/packages/mixsmsn/index.html}, \url{https://cran.r-project.org/web/packages/mixsmsn/mixsmsn.pdf}} \citep{Prates2013} developed for the R language\footnote{\url{http://www.R-project.org/}}.
This package implements routines for maximum likelihood estimation via an expectation maximization EM-type algorithm \citep{Basso2010, Cabral2012}.
The initial values for the distribution parameters were obtained using a combination of the \texttt{R} function \texttt{kmeans} and the method of moments.

For both nested and non-nested models, the Akaike (AIC) and Bayesian (BIC) information criteria (IC) may be applied \citep{Akaike1974, Schwarz1978, Burnham2004, Biesiada2007, Liddle2007, Tarnopolski2016a, Tarnopolski2016b} to estimate of the model performance.
They are defined as 
$$\textrm{AIC} = 2p - 2\mathcal{L}_\textrm{p,max}$$
and 
$$\textrm{BIC} = p \ln N - 2\mathcal{L}_\textrm{p,max}.$$

The favoured model is selected based on the minimum of AIC or BIC. 
Both IC comprise two competing terms: the first measuring the model complexity (number of free parameters) and the second measuring the goodness of fit (or more precisely, its lack). 
The IC penalise the use of an excessive number of parameters: the preference is given to the models with fewer parameters, as long as the others do not yield an essentially better fit. 
For BIC, the penalisation term is greater than the corresponding term for the AIC: $p \ln N > 2 p$, for $N \geq 8$. 
Consequently, the BIC provides much more strict penalisation of additional parameters than AIC, especially for large samples.
Hence, BIC tends to underfit the data giving preference to an excessively simple model, while AIC is biased towards overfitting, i.e. favouring models with excessive parameters. 
This may lead to selection of different best-fit models using two criteria.
Essentially, these two IC answer different questions:
AIC selects a model that better describes the data (the model being a real description of the data is never considered), while BIC intends finding the true model among the set of candidates \citep{Tarnopolski2019}. 
\citet{Tarnopolski2019a} found that BIC is more accurate then AIC when applied to the non-skewed distributions, while both criteria are consistent with each other for the skewed distributions.

The suitable models of fit can be assessed using the difference, $\mathbf{\Delta_i = \textrm{AIC}_i - \textrm{AIC}_\textrm{min}}$, where $\textrm{AIC}_\textrm{min}$ is the minimum value of AIC over the set of 12 models, i.e. four distributions with different number of components, ranging from one to three. 
If $\Delta_i < 2$, then there is strong support for the $i$-th model (or the evidence against it is worth only a small mention), and it is highly probable that this model is a proper description of the data. 
If $2 < \Delta_i < 4$, then there is significant support for the $i$-th model. 
When $4 < \Delta_i < 7$, there is considerably less support, and models with $ \Delta_i > 10$ have essentially no support \citep{Burnham2004, Biesiada2007}. 
It is worth noting that when two models with similar $\mathcal{L}_\textrm{max}$ are considered, the $\Delta_i$ only depends on the number of parameters due to the $2p$ term. 
Thus, $\Delta_i / (2\Delta p) < 1$, where $\Delta p$ is the difference in the model parameter number, evidences that the relative improvement is owing to the fit improvement, not an increase of the number of parameters only.

Similarly for BIC, $\Delta_i = \mathrm{BIC}_i - \mathrm{BIC}_\textrm{min}$, where $\mathrm{BIC}_\textrm{min}$ corresponds to the minimum BIC value within the set of 12 models, defines the degree of preference of the i-th model: if $\Delta_i < 2$, then there is substantial support for the i-th model. 
For $2 < \Delta_i< 6$, there is positive evidence against the i-th model. 
If $6 < \Delta_i < 10$, the evidence is strong, and models with $\Delta_i > 10$ yield very strong evidence against the i-th model, i.e. essentially no support \citep{Kass1995}.

Additionally, we estimated the Akaike weights, which can be directly interpreted as conditional probabilities for each model (see \citealt{Burnham2004, Wagenmakers2004} and references therein):
$$w_i = \frac{\exp(-\Delta_i/2)}{\sum_{r=1}^{R} \exp(-\Delta_r/2)},$$
where the denominator represents a sum over all models.
The Akaike weight can be interpreted as the probability of the model to be the best given the data and the set of models.
Thus, the strength of evidence in favour of one model relative to another can be assessed using the ratio of their weights $w_i/w_j$ or normalised probability $w_i/(w_i + w_j)$.
Similar values for BIC, called the ``Schwarz'' weights, can be obtained by replacing the AIC values by the BIC values in the equation for the Akaike weights.

\section{Tables with parameters of fits of GRB hardness-duration distributions.}

\onecolumn
\begin{table}
\caption[]{Parameters of the fits with different models to the rest-frame hardness-duration distributions of the ``Total'' sample.}
\label{tab:rest_total}
$$
\begin{tabular}{llrrrrcccc}
\hline
\noalign{\smallskip}
Distr. &  Cl. & $\Delta_\textrm{AIC}$ & $w_i$ (AIC) & $\Delta_\textrm{BIC}$ & \
            $w_i$ (BIC) & $\mathcal{L}_\textrm{max}$ & $\mu$ & $\Sigma$ & $\lambda$ \\
\noalign{\smallskip}
\hline
\noalign{\smallskip}
$\mathcal{N}$ & 1 & 45.8 & $<0.001$ & 37.7 & $<0.001$ & -659.3 & (0.99, 2.64) & $\left( \begin{array}{cc}0.73 & -0.01 \\ -0.01 & 0.40 \end{array} \right)$ & \ldots \\
\hline
$\mathcal{N}$ & 2 & 5.5 & 0.044 & 21.5 & $<0.001$ & -633.1 & (0.34, 2.55) & $\left( \begin{array}{cc}0.71 & -0.06 \\ -0.06 & 0.43 \end{array} \right)$ & \ldots \\
 &  &  &  &  &  & -633.1 & (1.34, 2.70) & $\left( \begin{array}{cc}0.71 & -0.03 \\ -0.06 & 0.37 \end{array} \right)$ & \ldots \\
\hline
$\mathcal{N}$ & 3 & 7.1 & 0.020 & 47.3 & $<0.001$ & -628.0 & (1.37, 2.34) & $\left( \begin{array}{cc}0.49 & -0.02 \\ -0.02 & 0.25 \end{array} \right)$ & \ldots \\
 &  &  &  &  &  & -628.0 & (0.19, 2.56) & $\left( \begin{array}{cc}0.49 & -0.07 \\ -0.02 & 0.45 \end{array} \right)$ & \ldots \\
 &  &  &  &  &  & -628.0 & (1.25, 2.90) & $\left( \begin{array}{cc}0.49 & 0.01 \\ -0.02 & 0.25 \end{array} \right)$ & \ldots \\
\hline
$\mathcal{SN}$ & 1 & \textcolor{red}{0.0} & 0.696 & \textcolor{red}{0.0} & 0.959 & -634.4 & (1.83, 2.70) & $\left( \begin{array}{cc}1.11 & 0.02 \\ 0.02 & 0.41 \end{array} \right)$ & (-3.35, -0.38) \\
\hline
$\mathcal{SN}$ & 2 & 9.3 & 0.007 & 41.4 & $<0.001$ & -631.1 & (0.81, 2.71) & $\left( \begin{array}{cc}0.82 & -0.00 \\ -0.00 & 0.46 \end{array} \right)$ & (-1.17, -0.71) \\
 &  &  &  &  &  & -631.1 & (1.11, 2.97) & $\left( \begin{array}{cc}0.82 & -0.10 \\ -0.00 & 0.45 \end{array} \right)$ & (0.91, -1.14) \\
\hline
$\mathcal{SN}$ & 3 & 16.3 & $<0.001$ & 80.6 & $<0.001$ & -626.6 & (1.31, 2.96) & $\left( \begin{array}{cc}0.40 & -0.10 \\ -0.10 & 0.48 \end{array} \right)$ & (0.93, -1.27) \\
 &  &  &  &  &  & -626.6 & (1.02, 2.84) & $\left( \begin{array}{cc}0.40 & 0.11 \\ -0.10 & 0.45 \end{array} \right)$ & (-1.12, -0.95) \\
 &  &  &  &  &  & -626.6 & (0.04, 2.94) & $\left( \begin{array}{cc}0.40 & 0.05 \\ -0.10 & 0.48 \end{array} \right)$ & (-0.81, -0.65) \\
\hline
$\mathcal{T}$ & 1 & 46.4 & $<0.001$ & 42.4 & $<0.001$ & -658.6 & (1.01, 2.65) & $\left( \begin{array}{cc}0.70 & -0.01 \\ -0.01 & 0.39 \end{array} \right)$ & \ldots \\
\hline
$\mathcal{T}$ & 2 & 8.4 & 0.010 & 28.5 & $<0.001$ & -633.6 & (1.33, 2.69) & $\left( \begin{array}{cc}0.45 & -0.03 \\ -0.03 & 0.37 \end{array} \right)$ & \ldots \\
 &  &  &  &  &  & -633.6 & (0.30, 2.55) & $\left( \begin{array}{cc}0.45 & -0.07 \\ -0.03 & 0.43 \end{array} \right)$ & \ldots \\
\hline
$\mathcal{T}$ & 3 & 13.2 & 0.001 & 57.4 & $<0.001$ & -630.0 & (0.04, 2.59) & $\left( \begin{array}{cc}0.64 & -0.06 \\ -0.06 & 0.46 \end{array} \right)$ & \ldots \\
 &  &  &  &  &  & -630.0 & (0.94, 2.67) & $\left( \begin{array}{cc}0.64 & 0.01 \\ -0.06 & 0.36 \end{array} \right)$ & \ldots \\
 &  &  &  &  &  & -630.0 & (1.61, 2.65) & $\left( \begin{array}{cc}0.64 & -0.04 \\ -0.06 & 0.39 \end{array} \right)$ & \ldots \\
\hline
$\mathcal{ST}$ & 1 & \textcolor{blue}{2.3} & 0.220 & 6.3 & 0.041 & -634.6 & (1.81, 2.70) & $\left( \begin{array}{cc}1.06 & 0.02 \\ 0.02 & 0.40 \end{array} \right)$ & (-3.14, -0.37) \\
\hline
$\mathcal{ST}$ & 2 & 12.7 & 0.001 & 48.8 & $<0.001$ & -631.7 & (0.74, 2.68) & $\left( \begin{array}{cc}0.79 & -0.01 \\ -0.01 & 0.45 \end{array} \right)$ & (-1.09, -0.60) \\
 &  &  &  &  &  & -631.7 & (1.12, 2.95) & $\left( \begin{array}{cc}0.79 & -0.09 \\ -0.01 & 0.44 \end{array} \right)$ & (0.81, -1.02) \\
\hline
$\mathcal{ST}$ & 3 & 16.7 & $<0.001$ & 84.9 & $<0.001$ & -625.8 & (1.06, 2.84) & $\left( \begin{array}{cc}0.55 & 0.11 \\ 0.11 & 0.45 \end{array} \right)$ & (-1.11, -0.88) \\
 &  &  &  &  &  & -625.8 & (-0.40, 3.13) & $\left( \begin{array}{cc}0.55 & 0.09 \\ 0.11 & 0.45 \end{array} \right)$ & (-0.34, -0.89) \\
 &  &  &  &  &  & -625.8 & (1.35, 2.93) & $\left( \begin{array}{cc}0.55 & -0.09 \\ 0.11 & 0.47 \end{array} \right)$ & (0.88, -1.14) \\
\hline
\noalign{\smallskip}
\end{tabular}
$$
\tablefoot{The acceptable models are shown in red ($\Delta \textrm{IC} < 2$) and blue ($2 < \Delta \textrm{AIC} < 4$) colour.
Cl. provides the number of GRB classes, $\mathcal{N}$ ($\mathcal{SN}$) and $\mathcal{T}$ ($\mathcal{ST}$) denote the symmetric (skewed) Gaussian and Student distributions, correspondingly.}
\end{table}

\twocolumn
\onecolumn
\begin{table}
\caption[]{Parameters of the fits with different models to the rest-frame hardness-duration distributions of the ``KW trig'' sample.}
\label{tab:rest_kw}
$$
\begin{tabular}{llrrrrcccc}
\hline
\noalign{\smallskip}
Distr. &  Cl. & $\Delta_\textrm{AIC}$ & $w_i$ (AIC) & $\Delta_\textrm{BIC}$ & \
            $w_i$ (BIC) & $\mathcal{L}_\textrm{max}$ & $\mu$ & $\Sigma$ & $\lambda$ \\
\noalign{\smallskip}
\hline
\noalign{\smallskip}
$\mathcal{N}$ & 1 & 23.3 & $<0.001$ & 16.8 & $<0.001$ & -298.5 & (0.91, 2.75) & $\left( \begin{array}{cc}0.74 & -0.01 \\ -0.01 & 0.37 \end{array} \right)$ & \ldots \\
\hline
$\mathcal{N}$ & 2 & \textcolor{blue}{2.4} & 0.155 & 15.5 & $<0.001$ & -282.0 & (0.38, 2.62) & $\left( \begin{array}{cc}0.78 & -0.11 \\ -0.11 & 0.43 \end{array} \right)$ & \ldots \\
 &  &  &  &  &  & -282.0 & (1.25, 2.84) & $\left( \begin{array}{cc}0.78 & -0.01 \\ -0.11 & 0.29 \end{array} \right)$ & \ldots \\
\hline
$\mathcal{N}$ & 3 & 14.3 & $<0.001$ & 46.9 & $<0.001$ & -282.0 & (-0.24, 2.83) & $\left( \begin{array}{cc}0.56 & -0.01 \\ -0.01 & 0.43 \end{array} \right)$ & \ldots \\
 &  &  &  &  &  & -282.0 & (1.50, 2.71) & $\left( \begin{array}{cc}0.56 & 0.01 \\ -0.01 & 0.36 \end{array} \right)$ & \ldots \\
 &  &  &  &  &  & -282.0 & (0.78, 2.77) & $\left( \begin{array}{cc}0.56 & 0.05 \\ -0.01 & 0.34 \end{array} \right)$ & \ldots \\
\hline
$\mathcal{SN}$ & 1 & \textcolor{red}{0.0} & 0.514 & \textcolor{red}{0.0} & 0.916 & -284.8 & (1.79, 2.73) & $\left( \begin{array}{cc}1.14 & -0.02 \\ -0.02 & 0.37 \end{array} \right)$ & (-3.78, 0.12) \\
\hline
$\mathcal{SN}$ & 2 & 5.6 & 0.031 & 31.8 & $<0.001$ & -279.7 & (1.11, 3.12) & $\left( \begin{array}{cc}0.47 & -0.06 \\ -0.06 & 0.48 \end{array} \right)$ & (0.59, -2.21) \\
 &  &  &  &  &  & -279.7 & (0.59, 2.77) & $\left( \begin{array}{cc}0.47 & -0.04 \\ -0.06 & 0.45 \end{array} \right)$ & (-1.60, -0.45) \\
\hline
$\mathcal{SN}$ & 3 & 10.9 & 0.002 & 63.2 & $<0.001$ & -274.3 & (-0.33, 3.00) & $\left( \begin{array}{cc}0.50 & 0.08 \\ 0.08 & 0.37 \end{array} \right)$ & (-0.70, 0.15) \\
 &  &  &  &  &  & -274.3 & (1.48, 3.01) & $\left( \begin{array}{cc}0.50 & -0.00 \\ 0.08 & 0.51 \end{array} \right)$ & (1.11, -3.46) \\
 &  &  &  &  &  & -274.3 & (1.04, 3.18) & $\left( \begin{array}{cc}0.50 & 0.16 \\ 0.08 & 0.52 \end{array} \right)$ & (-1.06, -2.86) \\
\hline
$\mathcal{T}$ & 1 & 23.3 & $<0.001$ & 20.0 & $<0.001$ & -297.5 & (0.95, 2.76) & $\left( \begin{array}{cc}0.68 & -0.01 \\ -0.01 & 0.34 \end{array} \right)$ & \ldots \\
\hline
$\mathcal{T}$ & 2 & 4.8 & 0.047 & 21.2 & $<0.001$ & -282.3 & (1.25, 2.83) & $\left( \begin{array}{cc}0.44 & -0.01 \\ -0.01 & 0.28 \end{array} \right)$ & \ldots \\
 &  &  &  &  &  & -282.3 & (0.37, 2.62) & $\left( \begin{array}{cc}0.44 & -0.11 \\ -0.01 & 0.43 \end{array} \right)$ & \ldots \\
\hline
$\mathcal{T}$ & 3 & 16.3 & $<0.001$ & 52.2 & $<0.001$ & -282.0 & (1.51, 2.71) & $\left( \begin{array}{cc}0.32 & 0.01 \\ 0.01 & 0.36 \end{array} \right)$ & \ldots \\
 &  &  &  &  &  & -282.0 & (-0.28, 2.84) & $\left( \begin{array}{cc}0.32 & -0.00 \\ 0.01 & 0.43 \end{array} \right)$ & \ldots \\
 &  &  &  &  &  & -282.0 & (0.78, 2.77) & $\left( \begin{array}{cc}0.32 & 0.05 \\ 0.01 & 0.34 \end{array} \right)$ & \ldots \\
\hline
$\mathcal{ST}$ & 1 & \textcolor{red}{1.5} & 0.243 & 4.8 & 0.083 & -284.6 & (1.74, 2.76) & $\left( \begin{array}{cc}1.04 & -0.00 \\ -0.00 & 0.35 \end{array} \right)$ & (-3.28, -0.08) \\
\hline
$\mathcal{ST}$ & 2 & 8.5 & 0.007 & 37.9 & $<0.001$ & -280.1 & (0.79, 2.73) & $\left( \begin{array}{cc}0.89 & -0.05 \\ -0.05 & 0.44 \end{array} \right)$ & (-1.44, -0.53) \\
 &  &  &  &  &  & -280.1 & (1.14, 3.11) & $\left( \begin{array}{cc}0.89 & -0.06 \\ -0.05 & 0.43 \end{array} \right)$ & (0.56, -1.83) \\
\hline
$\mathcal{ST}$ & 3 & 24.5 & $<0.001$ & 80.0 & $<0.001$ & -280.1 & (1.01, 2.80) & $\left( \begin{array}{cc}0.49 & 0.04 \\ 0.04 & 0.25 \end{array} \right)$ & (0.90, 0.93) \\
 &  &  &  &  &  & -280.1 & (0.52, 2.55) & $\left( \begin{array}{cc}0.49 & -0.10 \\ 0.04 & 0.44 \end{array} \right)$ & (-1.05, 0.70) \\
 &  &  &  &  &  & -280.1 & (1.05, 2.58) & $\left( \begin{array}{cc}0.49 & -0.05 \\ 0.04 & 0.34 \end{array} \right)$ & (0.66, -1.16) \\
\hline
\noalign{\smallskip}
\end{tabular}
$$
\tablefoot{The acceptable models are shown in red ($\Delta \textrm{IC} < 2$) and blue ($2 < \Delta \textrm{AIC} < 4$) colour.
Cl. provides the number of GRB classes, $\mathcal{N}$ ($\mathcal{SN}$) and $\mathcal{T}$ ($\mathcal{ST}$) denote the symmetric (skewed) Gaussian and Student distributions, correspondingly.}
\end{table}

\twocolumn
\onecolumn
\begin{table}
\caption[]{Parameters of the fits with different models to the rest-frame hardness-duration distributions of the ``KW \& BAT'' sample.}
\label{tab:rest_bat}
$$
\begin{tabular}{llrrrrcccc}
\hline
\noalign{\smallskip}
Distr. &  Cl. & $\Delta_\textrm{AIC}$ & $w_i$ (AIC) & $\Delta_\textrm{BIC}$ & \
            $w_i$ (BIC) & $\mathcal{L}_\textrm{max}$ & $\mu$ & $\Sigma$ & $\lambda$ \\
\noalign{\smallskip}
\hline
\noalign{\smallskip}
$\mathcal{N}$ & 1 & \textcolor{red}{1.6} & 0.149 & \textcolor{red}{0.0} & 0.946 & -235.6 & (1.14, 2.54) & $\left( \begin{array}{cc}0.63 & 0.02 \\ 0.02 & 0.38 \end{array} \right)$ & \ldots \\
\hline
$\mathcal{N}$ & 2 & \textcolor{red}{0.0} & 0.331 & 17.1 & $<0.001$ & -228.7 & (1.47, 2.60) & $\left( \begin{array}{cc}0.41 & -0.04 \\ -0.04 & 0.37 \end{array} \right)$ & \ldots \\
 &  &  &  &  &  & -228.7 & (0.56, 2.45) & $\left( \begin{array}{cc}0.41 & 0.03 \\ -0.04 & 0.38 \end{array} \right)$ & \ldots \\
\hline
$\mathcal{N}$ & 3 & \textcolor{red}{0.3} & 0.285 & 36.1 & $<0.001$ & -222.9 & (0.46, 2.27) & $\left( \begin{array}{cc}0.50 & -0.03 \\ -0.03 & 0.29 \end{array} \right)$ & \ldots \\
 &  &  &  &  &  & -222.9 & (1.11, 2.56) & $\left( \begin{array}{cc}0.50 & -0.13 \\ -0.03 & 0.32 \end{array} \right)$ & \ldots \\
 &  &  &  &  &  & -222.9 & (1.72, 2.74) & $\left( \begin{array}{cc}0.50 & -0.11 \\ -0.03 & 0.35 \end{array} \right)$ & \ldots \\
\hline
$\mathcal{SN}$ & 1 & 6.5 & 0.013 & 11.1 & 0.004 & -236.0 & (1.48, 2.34) & $\left( \begin{array}{cc}0.72 & -0.04 \\ -0.04 & 0.43 \end{array} \right)$ & (-0.88, 0.82) \\
\hline
$\mathcal{SN}$ & 2 & 6.3 & 0.014 & 35.9 & $<0.001$ & -227.9 & (0.88, 2.27) & $\left( \begin{array}{cc}0.60 & -0.03 \\ -0.03 & 0.42 \end{array} \right)$ & (-1.67, 1.17) \\
 &  &  &  &  &  & -227.9 & (1.35, 2.39) & $\left( \begin{array}{cc}0.60 & -0.01 \\ -0.03 & 0.43 \end{array} \right)$ & (0.46, 0.83) \\
\hline
$\mathcal{SN}$ & 3 & 8.2 & 0.005 & 62.8 & $<0.001$ & -220.9 & (1.59, 2.59) & $\left( \begin{array}{cc}0.32 & -0.08 \\ -0.08 & 0.38 \end{array} \right)$ & (0.66, 0.69) \\
 &  &  &  &  &  & -220.9 & (1.25, 2.60) & $\left( \begin{array}{cc}0.32 & -0.09 \\ -0.08 & 0.35 \end{array} \right)$ & (-1.88, -1.13) \\
 &  &  &  &  &  & -220.9 & (0.68, 2.03) & $\left( \begin{array}{cc}0.32 & -0.12 \\ -0.08 & 0.33 \end{array} \right)$ & (-2.79, 1.93) \\
\hline
$\mathcal{T}$ & 1 & 4.4 & 0.037 & 5.9 & 0.050 & -236.0 & (1.14, 2.54) & $\left( \begin{array}{cc}0.62 & 0.02 \\ 0.02 & 0.38 \end{array} \right)$ & \ldots \\
\hline
$\mathcal{T}$ & 2 & \textcolor{blue}{2.8} & 0.082 & 23.0 & $<0.001$ & -229.2 & (0.60, 2.47) & $\left( \begin{array}{cc}0.51 & 0.04 \\ 0.04 & 0.38 \end{array} \right)$ & \ldots \\
 &  &  &  &  &  & -229.2 & (1.49, 2.59) & $\left( \begin{array}{cc}0.51 & -0.04 \\ 0.04 & 0.37 \end{array} \right)$ & \ldots \\
\hline
$\mathcal{T}$ & 3 & \textcolor{blue}{2.9} & 0.078 & 41.9 & $<0.001$ & -223.2 & (1.71, 2.72) & $\left( \begin{array}{cc}0.29 & -0.10 \\ -0.10 & 0.35 \end{array} \right)$ & \ldots \\
 &  &  &  &  &  & -223.2 & (1.09, 2.55) & $\left( \begin{array}{cc}0.29 & -0.13 \\ -0.10 & 0.32 \end{array} \right)$ & \ldots \\
 &  &  &  &  &  & -223.2 & (0.40, 2.27) & $\left( \begin{array}{cc}0.29 & -0.03 \\ -0.10 & 0.29 \end{array} \right)$ & \ldots \\
\hline
$\mathcal{ST}$ & 1 & 8.8 & 0.004 & 16.5 & $<0.001$ & -236.1 & (1.46, 2.36) & $\left( \begin{array}{cc}0.70 & -0.03 \\ -0.03 & 0.42 \end{array} \right)$ & (-0.80, 0.71) \\
\hline
$\mathcal{ST}$ & 2 & 9.4 & 0.003 & 42.1 & $<0.001$ & -228.4 & (0.86, 2.27) & $\left( \begin{array}{cc}0.58 & -0.02 \\ -0.02 & 0.42 \end{array} \right)$ & (-1.47, 1.06) \\
 &  &  &  &  &  & -228.4 & (1.35, 2.41) & $\left( \begin{array}{cc}0.58 & -0.01 \\ -0.02 & 0.42 \end{array} \right)$ & (0.44, 0.75) \\
\hline
$\mathcal{ST}$ & 3 & 12.5 & 0.001 & 70.1 & $<0.001$ & -222.0 & (1.26, 2.75) & $\left( \begin{array}{cc}0.45 & -0.01 \\ -0.01 & 0.29 \end{array} \right)$ & (0.24, 0.27) \\
 &  &  &  &  &  & -222.0 & (1.65, 2.05) & $\left( \begin{array}{cc}0.45 & -0.04 \\ -0.01 & 0.25 \end{array} \right)$ & (-0.67, 1.54) \\
 &  &  &  &  &  & -222.0 & (0.77, 2.14) & $\left( \begin{array}{cc}0.45 & -0.07 \\ -0.01 & 0.41 \end{array} \right)$ & (-3.18, 2.62) \\
\hline
\noalign{\smallskip}
\end{tabular}
$$
\tablefoot{The acceptable models are shown in red ($\Delta \textrm{IC} < 2$) and blue ($2 < \Delta \textrm{AIC} < 4$) colour.
Cl. provides the number of GRB classes, $\mathcal{N}$ ($\mathcal{SN}$) and $\mathcal{T}$ ($\mathcal{ST}$) denote the symmetric (skewed) Gaussian and Student distributions, correspondingly.}
\end{table}

\twocolumn
\onecolumn
\begin{table}
\caption[]{Parameters of the fits with different models to the rest-frame hardness-duration distributions of the ``GBM all'' sample.}
\label{tab:rest_gbm}
$$
\begin{tabular}{llrrrrcccc}
\hline
\noalign{\smallskip}
Distr. &  Cl. & $\Delta_\textrm{AIC}$ & $w_i$ (AIC) & $\Delta_\textrm{BIC}$ & \
            $w_i$ (BIC) & $\mathcal{L}_\textrm{max}$ & $\mu$ & $\Sigma$ & $\lambda$ \\
\noalign{\smallskip}
\hline
\noalign{\smallskip}
$\mathcal{N}$ & 1 & 21.9 & $<0.001$ & 12.1 & 0.002 & -299.5 & (0.98, 2.67) & $\left( \begin{array}{cc}0.72 & 0.02 \\ 0.02 & 0.48 \end{array} \right)$ & \ldots \\
\hline
$\mathcal{N}$ & 2 & 8.6 & 0.005 & 17.6 & $<0.001$ & -286.9 & (0.33, 2.59) & $\left( \begin{array}{cc}0.81 & -0.05 \\ -0.05 & 0.48 \end{array} \right)$ & \ldots \\
 &  &  &  &  &  & -286.9 & (1.26, 2.70) & $\left( \begin{array}{cc}0.81 & 0.03 \\ -0.05 & 0.48 \end{array} \right)$ & \ldots \\
\hline
$\mathcal{N}$ & 3 & \textcolor{red}{1.2} & 0.215 & 28.9 & $<0.001$ & -277.2 & (0.94, 2.74) & $\left( \begin{array}{cc}0.62 & 0.13 \\ 0.13 & 0.51 \end{array} \right)$ & \ldots \\
 &  &  &  &  &  & -277.2 & (-0.81, 2.85) & $\left( \begin{array}{cc}0.62 & 0.11 \\ 0.13 & 0.52 \end{array} \right)$ & \ldots \\
 &  &  &  &  &  & -277.2 & (1.30, 2.55) & $\left( \begin{array}{cc}0.62 & -0.06 \\ 0.13 & 0.38 \end{array} \right)$ & \ldots \\
\hline
$\mathcal{SN}$ & 1 & 5.8 & 0.022 & 2.3 & 0.216 & -289.5 & (1.77, 2.66) & $\left( \begin{array}{cc}1.06 & 0.01 \\ 0.01 & 0.48 \end{array} \right)$ & (-2.76, 0.10) \\
\hline
$\mathcal{SN}$ & 2 & 9.7 & 0.003 & 31.2 & $<0.001$ & -283.5 & (0.78, 2.24) & $\left( \begin{array}{cc}0.90 & -0.15 \\ -0.15 & 0.56 \end{array} \right)$ & (-1.35, 1.12) \\
 &  &  &  &  &  & -283.5 & (1.04, 2.31) & $\left( \begin{array}{cc}0.90 & 0.09 \\ -0.15 & 0.61 \end{array} \right)$ & (1.07, 1.70) \\
\hline
$\mathcal{SN}$ & 3 & 22.9 & $<0.001$ & 69.4 & $<0.001$ & -282.1 & (1.02, 2.51) & $\left( \begin{array}{cc}0.52 & 0.11 \\ 0.11 & 0.59 \end{array} \right)$ & (0.96, 2.26) \\
 &  &  &  &  &  & -282.1 & (0.17, 2.84) & $\left( \begin{array}{cc}0.52 & 0.02 \\ 0.11 & 0.59 \end{array} \right)$ & (-0.81, -0.81) \\
 &  &  &  &  &  & -282.1 & (0.93, 2.67) & $\left( \begin{array}{cc}0.52 & -0.10 \\ 0.11 & 0.43 \end{array} \right)$ & (1.09, -1.62) \\
\hline
$\mathcal{T}$ & 1 & 10.4 & 0.002 & 3.8 & 0.102 & -292.8 & (1.04, 2.65) & $\left( \begin{array}{cc}0.61 & 0.01 \\ 0.01 & 0.41 \end{array} \right)$ & \ldots \\
\hline
$\mathcal{T}$ & 2 & 4.7 & 0.037 & 16.8 & $<0.001$ & -283.9 & (1.25, 2.69) & $\left( \begin{array}{cc}0.42 & 0.01 \\ 0.01 & 0.41 \end{array} \right)$ & \ldots \\
 &  &  &  &  &  & -283.9 & (0.28, 2.56) & $\left( \begin{array}{cc}0.42 & -0.06 \\ 0.01 & 0.40 \end{array} \right)$ & \ldots \\
\hline
$\mathcal{T}$ & 3 & \textcolor{red}{0.0} & 0.392 & 30.8 & $<0.001$ & -275.6 & (0.94, 2.71) & $\left( \begin{array}{cc}0.57 & 0.10 \\ 0.10 & 0.45 \end{array} \right)$ & \ldots \\
 &  &  &  &  &  & -275.6 & (-0.81, 2.90) & $\left( \begin{array}{cc}0.57 & 0.09 \\ 0.10 & 0.36 \end{array} \right)$ & \ldots \\
 &  &  &  &  &  & -275.6 & (1.39, 2.52) & $\left( \begin{array}{cc}0.57 & -0.07 \\ 0.10 & 0.36 \end{array} \right)$ & \ldots \\
\hline
$\mathcal{ST}$ & 1 & \textcolor{red}{0.4} & 0.321 & \textcolor{red}{0.0} & 0.681 & -285.8 & (1.63, 2.61) & $\left( \begin{array}{cc}0.84 & -0.01 \\ -0.01 & 0.42 \end{array} \right)$ & (-1.89, 0.20) \\
\hline
$\mathcal{ST}$ & 2 & 9.9 & 0.003 & 34.4 & $<0.001$ & -282.5 & (0.98, 2.32) & $\left( \begin{array}{cc}0.84 & -0.12 \\ -0.12 & 0.44 \end{array} \right)$ & (-1.54, 0.71) \\
 &  &  &  &  &  & -282.5 & (1.12, 2.45) & $\left( \begin{array}{cc}0.84 & 0.04 \\ -0.12 & 0.50 \end{array} \right)$ & (1.06, 1.24) \\
\hline
$\mathcal{ST}$ & 3 & 13.5 & $<0.001$ & 63.1 & $<0.001$ & -276.4 & (1.08, 2.56) & $\left( \begin{array}{cc}0.48 & 0.03 \\ 0.03 & 0.46 \end{array} \right)$ & (0.83, 2.76) \\
 &  &  &  &  &  & -276.4 & (0.94, 2.47) & $\left( \begin{array}{cc}0.48 & -0.09 \\ 0.03 & 0.14 \end{array} \right)$ & (3.65, -3.48) \\
 &  &  &  &  &  & -276.4 & (0.57, 2.50) & $\left( \begin{array}{cc}0.48 & -0.14 \\ 0.03 & 0.44 \end{array} \right)$ & (-4.76, -5.61) \\
\hline
\noalign{\smallskip}
\end{tabular}
$$
\tablefoot{The acceptable models are shown in red ($\Delta \textrm{IC} < 2$) and blue ($2 < \Delta \textrm{AIC} < 4$) colour.
Cl. provides the number of GRB classes, $\mathcal{N}$ ($\mathcal{SN}$) and $\mathcal{T}$ ($\mathcal{ST}$) denote the symmetric (skewed) Gaussian and Student distributions, correspondingly.}
\end{table}

\twocolumn
\onecolumn
\begin{table}
\caption[]{Parameters of the fits with different models to the observer-frame hardness-duration distributions of the ``Total'' sample.}
\label{tab:obs_total}
$$
\begin{tabular}{llrrrrcccc}
\hline
\noalign{\smallskip}
Distr. &  Cl. & $\Delta_\textrm{AIC}$ & $w_i$ (AIC) & $\Delta_\textrm{BIC}$ & \
            $w_i$ (BIC) & $\mathcal{L}_\textrm{max}$ & $\mu$ & $\Sigma$ & $\lambda$ \\
\noalign{\smallskip}
\hline
\noalign{\smallskip}
$\mathcal{N}$ & 1 & 91.6 & $<0.001$ & 56.2 & $<0.001$ & -648.6 & (1.40, 2.23) & $\left( \begin{array}{cc}0.75 & -0.02 \\ -0.02 & 0.38 \end{array} \right)$ & \ldots \\
\hline
$\mathcal{N}$ & 2 & 17.8 & $<0.001$ & 6.5 & 0.031 & -605.7 & (1.67, 2.23) & $\left( \begin{array}{cc}0.48 & 0.01 \\ 0.01 & 0.35 \end{array} \right)$ & \ldots \\
 &  &  &  &  &  & -605.7 & (0.36, 2.21) & $\left( \begin{array}{cc}0.48 & -0.11 \\ 0.01 & 0.46 \end{array} \right)$ & \ldots \\
\hline
$\mathcal{N}$ & 3 & \textcolor{red}{0.0} & 0.811 & 12.9 & 0.001 & -590.8 & (1.95, 2.16) & $\left( \begin{array}{cc}0.35 & 0.00 \\ 0.00 & 0.33 \end{array} \right)$ & \ldots \\
 &  &  &  &  &  & -590.8 & (1.20, 2.23) & $\left( \begin{array}{cc}0.35 & 0.09 \\ 0.00 & 0.37 \end{array} \right)$ & \ldots \\
 &  &  &  &  &  & -590.8 & (-0.42, 2.67) & $\left( \begin{array}{cc}0.35 & 0.10 \\ 0.00 & 0.39 \end{array} \right)$ & \ldots \\
\hline
$\mathcal{SN}$ & 1 & 27.3 & $<0.001$ & \textcolor{red}{0.0} & 0.804 & -614.5 & (2.28, 2.15) & $\left( \begin{array}{cc}1.15 & -0.05 \\ -0.05 & 0.38 \end{array} \right)$ & (-3.82, 0.43) \\
\hline
$\mathcal{SN}$ & 2 & 16.0 & $<0.001$ & 20.8 & $<0.001$ & -600.8 & (1.27, 1.68) & $\left( \begin{array}{cc}0.93 & -0.27 \\ -0.27 & 0.57 \end{array} \right)$ & (-2.45, 2.89) \\
 &  &  &  &  &  & -600.8 & (1.59, 2.08) & $\left( \begin{array}{cc}0.93 & 0.02 \\ -0.27 & 0.39 \end{array} \right)$ & (0.54, 0.79) \\
\hline
$\mathcal{SN}$ & 3 & 12.0 & 0.002 & 48.9 & $<0.001$ & -590.8 & (1.79, 2.02) & $\left( \begin{array}{cc}0.39 & 0.03 \\ 0.03 & 0.37 \end{array} \right)$ & (0.51, 0.67) \\
 &  &  &  &  &  & -590.8 & (-0.24, 2.91) & $\left( \begin{array}{cc}0.39 & 0.15 \\ 0.03 & 0.42 \end{array} \right)$ & (-0.53, -0.68) \\
 &  &  &  &  &  & -590.8 & (1.49, 2.18) & $\left( \begin{array}{cc}0.39 & 0.07 \\ 0.03 & 0.38 \end{array} \right)$ & (-1.18, 0.38) \\
\hline
$\mathcal{T}$ & 1 & 81.8 & $<0.001$ & 50.5 & $<0.001$ & -642.7 & (1.47, 2.22) & $\left( \begin{array}{cc}0.66 & -0.00 \\ -0.00 & 0.35 \end{array} \right)$ & \ldots \\
\hline
$\mathcal{T}$ & 2 & 21.4 & $<0.001$ & 14.2 & 0.001 & -606.5 & (1.68, 2.24) & $\left( \begin{array}{cc}0.47 & 0.00 \\ 0.00 & 0.35 \end{array} \right)$ & \ldots \\
 &  &  &  &  &  & -606.5 & (0.40, 2.20) & $\left( \begin{array}{cc}0.47 & -0.12 \\ 0.00 & 0.45 \end{array} \right)$ & \ldots \\
\hline
$\mathcal{T}$ & 3 & \textcolor{blue}{3.0} & 0.181 & 19.8 & $<0.001$ & -591.3 & (-0.42, 2.67) & $\left( \begin{array}{cc}0.38 & 0.10 \\ 0.10 & 0.39 \end{array} \right)$ & \ldots \\
 &  &  &  &  &  & -591.3 & (1.23, 2.23) & $\left( \begin{array}{cc}0.38 & 0.09 \\ 0.10 & 0.36 \end{array} \right)$ & \ldots \\
 &  &  &  &  &  & -591.3 & (1.99, 2.15) & $\left( \begin{array}{cc}0.38 & 0.00 \\ 0.10 & 0.32 \end{array} \right)$ & \ldots \\
\hline
$\mathcal{ST}$ & 1 & 26.5 & $<0.001$ & 3.2 & 0.162 & -613.0 & (2.24, 2.17) & $\left( \begin{array}{cc}1.04 & -0.04 \\ -0.04 & 0.36 \end{array} \right)$ & (-3.19, 0.31) \\
\hline
$\mathcal{ST}$ & 2 & 18.6 & $<0.001$ & 27.4 & $<0.001$ & -601.1 & (1.60, 2.09) & $\left( \begin{array}{cc}0.47 & 0.01 \\ 0.01 & 0.38 \end{array} \right)$ & (0.37, 0.64) \\
 &  &  &  &  &  & -601.1 & (1.12, 1.66) & $\left( \begin{array}{cc}0.47 & -0.29 \\ 0.01 & 0.60 \end{array} \right)$ & (-2.58, 3.22) \\
\hline
$\mathcal{ST}$ & 3 & 9.8 & 0.006 & 50.8 & $<0.001$ & -588.7 & (-0.42, 3.03) & $\left( \begin{array}{cc}0.37 & 0.10 \\ 0.10 & 0.49 \end{array} \right)$ & (0.41, -2.07) \\
 &  &  &  &  &  & -588.7 & (1.63, 2.25) & $\left( \begin{array}{cc}0.37 & 0.08 \\ 0.10 & 0.37 \end{array} \right)$ & (-1.45, 0.15) \\
 &  &  &  &  &  & -588.7 & (1.93, 2.05) & $\left( \begin{array}{cc}0.37 & 0.02 \\ 0.10 & 0.34 \end{array} \right)$ & (0.30, 0.41) \\
\hline
\noalign{\smallskip}
\end{tabular}
$$
\tablefoot{The acceptable models are shown in red ($\Delta \textrm{IC} < 2$) and blue ($2 < \Delta \textrm{AIC} < 4$) colour.
Cl. provides the number of GRB classes, $\mathcal{N}$ ($\mathcal{SN}$) and $\mathcal{T}$ ($\mathcal{ST}$) denote the symmetric (skewed) Gaussian and Student distributions, correspondingly.}
\end{table}

\twocolumn
\onecolumn
\begin{table}
\caption[]{Parameters of the fits with different models to the observer-frame hardness-duration distributions of the ``KW trig'' sample.}
\label{tab:obs_kw}
$$
\begin{tabular}{llrrrrcccc}
\hline
\noalign{\smallskip}
Distr. &  Cl. & $\Delta_\textrm{AIC}$ & $w_i$ (AIC) & $\Delta_\textrm{BIC}$ & \
            $w_i$ (BIC) & $\mathcal{L}_\textrm{max}$ & $\mu$ & $\Sigma$ & $\lambda$ \\
\noalign{\smallskip}
\hline
\noalign{\smallskip}
$\mathcal{N}$ & 1 & 58.8 & $<0.001$ & 33.4 & $<0.001$ & -286.5 & (1.27, 2.39) & $\left( \begin{array}{cc}0.76 & -0.01 \\ -0.01 & 0.34 \end{array} \right)$ & \ldots \\
\hline
$\mathcal{N}$ & 2 & 13.6 & 0.001 & 7.8 & 0.014 & -257.9 & (1.51, 2.37) & $\left( \begin{array}{cc}0.49 & 0.03 \\ 0.03 & 0.30 \end{array} \right)$ & \ldots \\
 &  &  &  &  &  & -257.9 & (-0.05, 2.51) & $\left( \begin{array}{cc}0.49 & -0.06 \\ 0.03 & 0.47 \end{array} \right)$ & \ldots \\
\hline
$\mathcal{N}$ & 3 & \textcolor{red}{0.0} & 0.717 & 13.8 & 0.001 & -245.1 & (1.79, 2.29) & $\left( \begin{array}{cc}0.35 & 0.05 \\ 0.05 & 0.26 \end{array} \right)$ & \ldots \\
 &  &  &  &  &  & -245.1 & (1.06, 2.42) & $\left( \begin{array}{cc}0.35 & 0.14 \\ 0.05 & 0.32 \end{array} \right)$ & \ldots \\
 &  &  &  &  &  & -245.1 & (-0.44, 2.82) & $\left( \begin{array}{cc}0.35 & 0.15 \\ 0.05 & 0.31 \end{array} \right)$ & \ldots \\
\hline
$\mathcal{SN}$ & 1 & 18.9 & $<0.001$ & \textcolor{red}{0.0} & 0.710 & -264.5 & (2.15, 2.27) & $\left( \begin{array}{cc}1.15 & -0.08 \\ -0.08 & 0.35 \end{array} \right)$ & (-5.00, 1.15) \\
\hline
$\mathcal{SN}$ & 2 & 23.0 & $<0.001$ & 30.2 & $<0.001$ & -258.6 & (1.26, 1.94) & $\left( \begin{array}{cc}0.98 & -0.24 \\ -0.24 & 0.52 \end{array} \right)$ & (-2.17, 2.07) \\
 &  &  &  &  &  & -258.6 & (1.46, 2.22) & $\left( \begin{array}{cc}0.98 & 0.05 \\ -0.24 & 0.35 \end{array} \right)$ & (0.67, 0.94) \\
\hline
$\mathcal{SN}$ & 3 & 10.1 & 0.005 & 43.5 & $<0.001$ & -244.2 & (-0.14, 2.86) & $\left( \begin{array}{cc}0.51 & 0.15 \\ 0.15 & 0.31 \end{array} \right)$ & (-1.68, 0.45) \\
 &  &  &  &  &  & -244.2 & (1.69, 2.53) & $\left( \begin{array}{cc}0.51 & 0.15 \\ 0.15 & 0.33 \end{array} \right)$ & (-2.29, -0.68) \\
 &  &  &  &  &  & -244.2 & (1.82, 2.09) & $\left( \begin{array}{cc}0.51 & 0.11 \\ 0.15 & 0.37 \end{array} \right)$ & (0.88, 1.27) \\
\hline
$\mathcal{T}$ & 1 & 50.7 & $<0.001$ & 28.5 & $<0.001$ & -281.4 & (1.36, 2.38) & $\left( \begin{array}{cc}0.64 & -0.00 \\ -0.00 & 0.30 \end{array} \right)$ & \ldots \\
\hline
$\mathcal{T}$ & 2 & 16.2 & $<0.001$ & 13.7 & 0.001 & -258.2 & (-0.06, 2.52) & $\left( \begin{array}{cc}0.59 & -0.06 \\ -0.06 & 0.47 \end{array} \right)$ & \ldots \\
 &  &  &  &  &  & -258.2 & (1.51, 2.37) & $\left( \begin{array}{cc}0.59 & 0.03 \\ -0.06 & 0.30 \end{array} \right)$ & \ldots \\
\hline
$\mathcal{T}$ & 3 & \textcolor{red}{1.9} & 0.277 & 18.9 & $<0.001$ & -245.0 & (1.12, 2.40) & $\left( \begin{array}{cc}0.49 & 0.12 \\ 0.12 & 0.31 \end{array} \right)$ & \ldots \\
 &  &  &  &  &  & -245.0 & (-0.44, 2.82) & $\left( \begin{array}{cc}0.49 & 0.15 \\ 0.12 & 0.30 \end{array} \right)$ & \ldots \\
 &  &  &  &  &  & -245.0 & (1.85, 2.28) & $\left( \begin{array}{cc}0.49 & 0.05 \\ 0.12 & 0.25 \end{array} \right)$ & \ldots \\
\hline
$\mathcal{ST}$ & 1 & 17.5 & $<0.001$ & \textcolor{red}{1.9} & 0.274 & -262.9 & (2.10, 2.29) & $\left( \begin{array}{cc}0.99 & -0.06 \\ -0.06 & 0.32 \end{array} \right)$ & (-3.73, 0.80) \\
\hline
$\mathcal{ST}$ & 2 & 25.5 & $<0.001$ & 36.0 & $<0.001$ & -258.9 & (1.22, 1.96) & $\left( \begin{array}{cc}0.97 & -0.23 \\ -0.23 & 0.51 \end{array} \right)$ & (-2.09, 1.90) \\
 &  &  &  &  &  & -258.9 & (1.47, 2.24) & $\left( \begin{array}{cc}0.97 & 0.04 \\ -0.23 & 0.34 \end{array} \right)$ & (0.58, 0.82) \\
\hline
$\mathcal{ST}$ & 3 & 19.5 & $<0.001$ & 56.1 & $<0.001$ & -247.9 & (1.57, 2.29) & $\left( \begin{array}{cc}0.40 & 0.02 \\ 0.02 & 0.32 \end{array} \right)$ & (1.05, 1.20) \\
 &  &  &  &  &  & -247.9 & (-0.29, 3.01) & $\left( \begin{array}{cc}0.40 & 0.18 \\ 0.02 & 0.34 \end{array} \right)$ & (-0.32, -0.64) \\
 &  &  &  &  &  & -247.9 & (1.33, 2.34) & $\left( \begin{array}{cc}0.40 & 0.06 \\ 0.02 & 0.32 \end{array} \right)$ & (-1.42, -0.89) \\
\hline
\noalign{\smallskip}
\end{tabular}
$$
\tablefoot{The acceptable models are shown in red ($\Delta \textrm{IC} < 2$) and blue ($2 < \Delta \textrm{AIC} < 4$) colour.
Cl. provides the number of GRB classes, $\mathcal{N}$ ($\mathcal{SN}$) and $\mathcal{T}$ ($\mathcal{ST}$) denote the symmetric (skewed) Gaussian and Student distributions, correspondingly.}
\end{table}

\twocolumn
\onecolumn
\begin{table}
\caption[]{Parameters of the fits with different models to the observer-frame hardness-duration distributions of the ``KW \& BAT'' sample.}
\label{tab:obs_bat}
$$
\begin{tabular}{llrrrrcccc}
\hline
\noalign{\smallskip}
Distr. &  Cl. & $\Delta_\textrm{AIC}$ & $w_i$ (AIC) & $\Delta_\textrm{BIC}$ & \
            $w_i$ (BIC) & $\mathcal{L}_\textrm{max}$ & $\mu$ & $\Sigma$ & $\lambda$ \\
\noalign{\smallskip}
\hline
\noalign{\smallskip}
$\mathcal{N}$ & 1 & 13.6 & 0.001 & 7.4 & 0.023 & -196.7 & (1.63, 2.04) & $\left( \begin{array}{cc}0.62 & 0.05 \\ 0.05 & 0.31 \end{array} \right)$ & \ldots \\
\hline
$\mathcal{N}$ & 2 & 7.9 & 0.012 & 20.4 & $<0.001$ & -187.9 & (1.30, 2.03) & $\left( \begin{array}{cc}0.55 & 0.07 \\ 0.07 & 0.31 \end{array} \right)$ & \ldots \\
 &  &  &  &  &  & -187.9 & (2.14, 2.06) & $\left( \begin{array}{cc}0.55 & 0.03 \\ 0.07 & 0.31 \end{array} \right)$ & \ldots \\
\hline
$\mathcal{N}$ & 3 & 4.4 & 0.067 & 35.6 & $<0.001$ & -180.1 & (1.94, 1.68) & $\left( \begin{array}{cc}0.40 & 0.03 \\ 0.03 & 0.10 \end{array} \right)$ & \ldots \\
 &  &  &  &  &  & -180.1 & (2.23, 2.21) & $\left( \begin{array}{cc}0.40 & 0.00 \\ 0.03 & 0.23 \end{array} \right)$ & \ldots \\
 &  &  &  &  &  & -180.1 & (1.32, 2.07) & $\left( \begin{array}{cc}0.40 & 0.09 \\ 0.03 & 0.29 \end{array} \right)$ & \ldots \\
\hline
$\mathcal{SN}$ & 1 & \textcolor{red}{0.0} & 0.609 & \textcolor{red}{0.0} & 0.932 & -188.0 & (2.40, 2.17) & $\left( \begin{array}{cc}0.98 & 0.11 \\ 0.11 & 0.32 \end{array} \right)$ & (-3.95, -0.67) \\
\hline
$\mathcal{SN}$ & 2 & 13.1 & 0.001 & 38.1 & $<0.001$ & -186.5 & (1.67, 2.01) & $\left( \begin{array}{cc}0.67 & 0.06 \\ 0.06 & 0.32 \end{array} \right)$ & (-1.80, 0.80) \\
 &  &  &  &  &  & -186.5 & (1.99, 2.12) & $\left( \begin{array}{cc}0.67 & 0.02 \\ 0.06 & 0.32 \end{array} \right)$ & (1.07, -0.72) \\
\hline
$\mathcal{SN}$ & 3 & 5.0 & 0.050 & 54.9 & $<0.001$ & -174.4 & (1.36, 2.09) & $\left( \begin{array}{cc}0.32 & -0.02 \\ -0.02 & 0.26 \end{array} \right)$ & (0.60, 0.84) \\
 &  &  &  &  &  & -174.4 & (2.18, 2.43) & $\left( \begin{array}{cc}0.32 & -0.02 \\ -0.02 & 0.32 \end{array} \right)$ & (0.15, -1.24) \\
 &  &  &  &  &  & -174.4 & (1.67, 1.55) & $\left( \begin{array}{cc}0.32 & -0.10 \\ -0.02 & 0.26 \end{array} \right)$ & (-2.32, 3.66) \\
\hline
$\mathcal{T}$ & 1 & 16.3 & $<0.001$ & 13.2 & 0.001 & -197.1 & (1.64, 2.04) & $\left( \begin{array}{cc}0.61 & 0.05 \\ 0.05 & 0.31 \end{array} \right)$ & \ldots \\
\hline
$\mathcal{T}$ & 2 & 10.2 & 0.004 & 25.9 & $<0.001$ & -188.1 & (2.14, 2.06) & $\left( \begin{array}{cc}0.28 & 0.03 \\ 0.03 & 0.31 \end{array} \right)$ & \ldots \\
 &  &  &  &  &  & -188.1 & (1.30, 2.04) & $\left( \begin{array}{cc}0.28 & 0.08 \\ 0.03 & 0.31 \end{array} \right)$ & \ldots \\
\hline
$\mathcal{T}$ & 3 & \textcolor{blue}{3.5} & 0.106 & 37.8 & $<0.001$ & -178.7 & (2.19, 2.20) & $\left( \begin{array}{cc}0.24 & -0.00 \\ -0.00 & 0.27 \end{array} \right)$ & \ldots \\
 &  &  &  &  &  & -178.7 & (1.24, 1.75) & $\left( \begin{array}{cc}0.24 & -0.02 \\ -0.00 & 0.17 \end{array} \right)$ & \ldots \\
 &  &  &  &  &  & -178.7 & (1.43, 2.22) & $\left( \begin{array}{cc}0.24 & -0.03 \\ -0.00 & 0.23 \end{array} \right)$ & \ldots \\
\hline
$\mathcal{ST}$ & 1 & \textcolor{blue}{3.0} & 0.136 & 6.1 & 0.044 & -188.4 & (2.40, 2.17) & $\left( \begin{array}{cc}0.97 & 0.11 \\ 0.11 & 0.32 \end{array} \right)$ & (-3.87, -0.66) \\
\hline
$\mathcal{ST}$ & 2 & 15.8 & $<0.001$ & 43.9 & $<0.001$ & -186.8 & (1.66, 2.01) & $\left( \begin{array}{cc}0.65 & 0.06 \\ 0.06 & 0.32 \end{array} \right)$ & (-1.63, 0.75) \\
 &  &  &  &  &  & -186.8 & (2.00, 2.12) & $\left( \begin{array}{cc}0.65 & 0.02 \\ 0.06 & 0.31 \end{array} \right)$ & (0.95, -0.65) \\
\hline
$\mathcal{ST}$ & 3 & 7.5 & 0.014 & 60.5 & $<0.001$ & -174.7 & (1.38, 2.08) & $\left( \begin{array}{cc}0.29 & -0.02 \\ -0.02 & 0.26 \end{array} \right)$ & (0.48, 0.96) \\
 &  &  &  &  &  & -174.7 & (2.17, 2.42) & $\left( \begin{array}{cc}0.29 & -0.01 \\ -0.02 & 0.31 \end{array} \right)$ & (0.10, -1.18) \\
 &  &  &  &  &  & -174.7 & (1.66, 1.55) & $\left( \begin{array}{cc}0.29 & -0.10 \\ -0.02 & 0.26 \end{array} \right)$ & (-2.19, 3.57) \\
\hline
\noalign{\smallskip}
\end{tabular}
$$
\tablefoot{The acceptable models are shown in red ($\Delta \textrm{IC} < 2$) and blue ($2 < \Delta \textrm{AIC} < 4$) colour.
Cl. provides the number of GRB classes, $\mathcal{N}$ ($\mathcal{SN}$) and $\mathcal{T}$ ($\mathcal{ST}$) denote the symmetric (skewed) Gaussian and Student distributions, correspondingly.}
\end{table}

\twocolumn
\onecolumn
\begin{table}
\caption[]{Parameters of the fits with different models to the observer-frame hardness-duration distributions of the ``GBM all'' sample.}
\label{tab:obs_gbm}
$$
\begin{tabular}{llrrrrcccc}
\hline
\noalign{\smallskip}
Distr. &  Cl. & $\Delta_\textrm{AIC}$ & $w_i$ (AIC) & $\Delta_\textrm{BIC}$ & \
            $w_i$ (BIC) & $\mathcal{L}_\textrm{max}$ & $\mu$ & $\Sigma$ & $\lambda$ \\
\noalign{\smallskip}
\hline
\noalign{\smallskip}
$\mathcal{N}$ & 1 & 57.1 & $<0.001$ & 40.1 & $<0.001$ & -300.0 & (1.37, 2.28) & $\left( \begin{array}{cc}0.75 & 0.01 \\ 0.01 & 0.47 \end{array} \right)$ & \ldots \\
\hline
$\mathcal{N}$ & 2 & 11.2 & 0.001 & 12.9 & 0.002 & -271.0 & (1.59, 2.26) & $\left( \begin{array}{cc}0.46 & 0.05 \\ 0.05 & 0.45 \end{array} \right)$ & \ldots \\
 &  &  &  &  &  & -271.0 & (-0.14, 2.43) & $\left( \begin{array}{cc}0.46 & 0.02 \\ 0.05 & 0.53 \end{array} \right)$ & \ldots \\
\hline
$\mathcal{N}$ & 3 & \textcolor{red}{0.0} & 0.395 & 20.4 & $<0.001$ & -259.4 & (1.50, 2.23) & $\left( \begin{array}{cc}0.45 & 0.05 \\ 0.05 & 0.36 \end{array} \right)$ & \ldots \\
 &  &  &  &  &  & -259.4 & (-0.21, 2.46) & $\left( \begin{array}{cc}0.45 & 0.04 \\ 0.05 & 0.54 \end{array} \right)$ & \ldots \\
 &  &  &  &  &  & -259.4 & (2.09, 2.45) & $\left( \begin{array}{cc}0.45 & 0.01 \\ 0.05 & 0.83 \end{array} \right)$ & \ldots \\
\hline
$\mathcal{SN}$ & 1 & 24.4 & $<0.001$ & 13.5 & 0.001 & -281.6 & (2.19, 2.23) & $\left( \begin{array}{cc}1.11 & -0.02 \\ -0.02 & 0.47 \end{array} \right)$ & (-3.79, 0.42) \\
\hline
$\mathcal{SN}$ & 2 & 8.1 & 0.007 & 22.3 & $<0.001$ & -265.5 & (1.41, 1.83) & $\left( \begin{array}{cc}0.48 & 0.11 \\ 0.11 & 0.62 \end{array} \right)$ & (0.62, 1.87) \\
 &  &  &  &  &  & -265.5 & (-0.10, 2.05) & $\left( \begin{array}{cc}0.48 & -0.00 \\ 0.11 & 0.65 \end{array} \right)$ & (-0.09, 1.01) \\
\hline
$\mathcal{SN}$ & 3 & 9.1 & 0.004 & 48.3 & $<0.001$ & -258.0 & (1.79, 2.33) & $\left( \begin{array}{cc}0.68 & 0.11 \\ 0.11 & 0.35 \end{array} \right)$ & (-1.90, -0.88) \\
 &  &  &  &  &  & -258.0 & (1.72, 1.92) & $\left( \begin{array}{cc}0.68 & 0.09 \\ 0.11 & 0.78 \end{array} \right)$ & (1.25, 2.10) \\
 &  &  &  &  &  & -258.0 & (-0.41, 2.50) & $\left( \begin{array}{cc}0.68 & 0.17 \\ 0.11 & 0.51 \end{array} \right)$ & (-0.19, 0.41) \\
\hline
$\mathcal{T}$ & 1 & 26.4 & $<0.001$ & 12.5 & 0.002 & -283.6 & (1.48, 2.25) & $\left( \begin{array}{cc}0.55 & 0.02 \\ 0.02 & 0.35 \end{array} \right)$ & \ldots \\
\hline
$\mathcal{T}$ & 2 & \textcolor{red}{1.9} & 0.153 & 6.7 & 0.034 & -265.3 & (-0.20, 2.44) & $\left( \begin{array}{cc}0.48 & -0.02 \\ -0.02 & 0.45 \end{array} \right)$ & \ldots \\
 &  &  &  &  &  & -265.3 & (1.57, 2.25) & $\left( \begin{array}{cc}0.48 & 0.04 \\ -0.02 & 0.37 \end{array} \right)$ & \ldots \\
\hline
$\mathcal{T}$ & 3 & \textcolor{red}{2.0} & 0.145 & 25.5 & $<0.001$ & -259.4 & (2.08, 2.42) & $\left( \begin{array}{cc}0.20 & 0.01 \\ 0.01 & 0.77 \end{array} \right)$ & \ldots \\
 &  &  &  &  &  & -259.4 & (-0.23, 2.47) & $\left( \begin{array}{cc}0.20 & 0.04 \\ 0.01 & 0.52 \end{array} \right)$ & \ldots \\
 &  &  &  &  &  & -259.4 & (1.50, 2.23) & $\left( \begin{array}{cc}0.20 & 0.05 \\ 0.01 & 0.35 \end{array} \right)$ & \ldots \\
\hline
$\mathcal{ST}$ & 1 & 7.7 & 0.008 & \textcolor{red}{0.0} & 0.962 & -272.2 & (2.00, 2.18) & $\left( \begin{array}{cc}0.76 & -0.02 \\ -0.02 & 0.37 \end{array} \right)$ & (-2.06, 0.47) \\
\hline
$\mathcal{ST}$ & 2 & 7.6 & 0.009 & 24.9 & $<0.001$ & -264.2 & (1.43, 1.96) & $\left( \begin{array}{cc}0.45 & 0.08 \\ 0.08 & 0.47 \end{array} \right)$ & (0.42, 1.09) \\
 &  &  &  &  &  & -264.2 & (-0.08, 2.16) & $\left( \begin{array}{cc}0.45 & -0.04 \\ 0.08 & 0.53 \end{array} \right)$ & (-0.20, 0.68) \\
\hline
$\mathcal{ST}$ & 3 & \textcolor{red}{0.7} & 0.278 & 43.0 & $<0.001$ & -252.8 & (1.80, 2.25) & $\left( \begin{array}{cc}0.60 & 0.05 \\ 0.05 & 0.32 \end{array} \right)$ & (-2.24, 0.01) \\
 &  &  &  &  &  & -252.8 & (2.03, 1.87) & $\left( \begin{array}{cc}0.60 & 0.04 \\ 0.05 & 0.68 \end{array} \right)$ & (0.40, 1.46) \\
 &  &  &  &  &  & -252.8 & (-0.74, 2.65) & $\left( \begin{array}{cc}0.60 & 0.12 \\ 0.05 & 0.41 \end{array} \right)$ & (13.15, -3.73) \\
\hline
\noalign{\smallskip}
\end{tabular}
$$
\tablefoot{The acceptable models are shown in red ($\Delta \textrm{IC} < 2$) and blue ($2 < \Delta \textrm{AIC} < 4$) colour.
Cl. provides the number of GRB classes, $\mathcal{N}$ ($\mathcal{SN}$) and $\mathcal{T}$ ($\mathcal{ST}$) denote the symmetric (skewed) Gaussian and Student distributions, correspondingly.}
\end{table}

\twocolumn
\onecolumn

\begin{table}
\caption[]{Information criteria for the rest-frame distributions with 10\% of the dimmest bursts excluded.}
\label{tab:snr}
$$ 
 \begin{tabular}{llrrrrr}
 
    \hline
    \noalign{\smallskip}
    Statistics & Cl. & Sample & $\Delta_\textrm{AIC}$ & $w_i$ (AIC) & $\Delta_\textrm{BIC}$ & $w_i$ (BIC) \\
    \noalign{\smallskip}
    \hline
    \noalign{\smallskip}
    
$\mathcal{N}$ & 1 & Total & 44.7 & $<0.001$ & 36.8 & $<0.001$ \\
$\mathcal{N}$ & 2 & Total & 5.7 & 0.040 & 21.3 & $<0.001$ \\
$\mathcal{N}$ & 3 & Total & 8.9 & 0.008 & 47.9 & $<0.001$ \\
\hline
$\mathcal{SN}$ & 1 & Total & \textcolor{red}{0.0} & 0.692 & \textcolor{red}{0.0} & 0.953 \\
$\mathcal{SN}$ & 2 & Total & 10.3 & 0.004 & 41.5 & $<0.001$ \\
$\mathcal{SN}$ & 3 & Total & 16.8 & $<0.001$ & 79.2 & $<0.001$ \\
\hline
$\mathcal{T}$ & 1 & Total & 45.4 & $<0.001$ & 41.5 & $<0.001$ \\
$\mathcal{T}$ & 2 & Total & 8.4 & 0.010 & 27.9 & $<0.001$ \\
$\mathcal{T}$ & 3 & Total & 10.7 & 0.003 & 53.6 & $<0.001$ \\
\hline
$\mathcal{ST}$ & 1 & Total & \textcolor{blue}{2.1} & 0.242 & 6.0 & 0.047 \\
$\mathcal{ST}$ & 2 & Total & 14.6 & $<0.001$ & 49.8 & $<0.001$ \\
$\mathcal{ST}$ & 3 & Total & 19.2 & $<0.001$ & 85.6 & $<0.001$ \\
\hline
\hline
$\mathcal{N}$ & 1 & KW trig & 20.2 & $<0.001$ & 13.9 & 0.001 \\
$\mathcal{N}$ & 2 & KW trig & \textcolor{red}{0.8} & 0.252 & 13.4 & 0.001 \\
$\mathcal{N}$ & 3 & KW trig & \textcolor{blue}{3.7} & 0.059 & 35.3 & $<0.001$ \\
\hline
$\mathcal{SN}$ & 1 & KW trig & \textcolor{red}{0.0} & 0.375 & \textcolor{red}{0.0} & 0.915 \\
$\mathcal{SN}$ & 2 & KW trig & 4.6 & 0.038 & 29.9 & $<0.001$ \\
$\mathcal{SN}$ & 3 & KW trig & 7.0 & 0.011 & 57.6 & $<0.001$ \\
\hline
$\mathcal{T}$ & 1 & KW trig & 19.8 & $<0.001$ & 16.7 & $<0.001$ \\
$\mathcal{T}$ & 2 & KW trig & \textcolor{blue}{3.2} & 0.076 & 19.0 & $<0.001$ \\
$\mathcal{T}$ & 3 & KW trig & 6.2 & 0.017 & 41.0 & $<0.001$ \\
\hline
$\mathcal{ST}$ & 1 & KW trig & \textcolor{red}{1.7} & 0.160 & 4.8 & 0.083 \\
$\mathcal{ST}$ & 2 & KW trig & 7.2 & 0.010 & 35.7 & $<0.001$ \\
$\mathcal{ST}$ & 3 & KW trig & 11.0 & 0.002 & 64.7 & $<0.001$ \\
\hline
\hline
$\mathcal{N}$ & 1 & KW \& BAT & \textcolor{red}{0.0} & 0.451 & \textcolor{red}{0.0} & 0.943 \\
$\mathcal{N}$ & 2 & KW \& BAT & \textcolor{red}{1.5} & 0.213 & 19.6 & $<0.001$ \\
$\mathcal{N}$ & 3 & KW \& BAT & 5.0 & 0.037 & 41.2 & $<0.001$ \\
\hline
$\mathcal{SN}$ & 1 & KW \& BAT & 4.3 & 0.053 & 10.4 & 0.005 \\
$\mathcal{SN}$ & 2 & KW \& BAT & 7.8 & 0.009 & 38.0 & $<0.001$ \\
$\mathcal{SN}$ & 3 & KW \& BAT & 13.4 & 0.001 & 67.7 & $<0.001$ \\
\hline
$\mathcal{T}$ & 1 & KW \& BAT & \textcolor{blue}{2.8} & 0.111 & 5.8 & 0.052 \\
$\mathcal{T}$ & 2 & KW \& BAT & 4.3 & 0.053 & 25.4 & $<0.001$ \\
$\mathcal{T}$ & 3 & KW \& BAT & 4.2 & 0.055 & 43.4 & $<0.001$ \\
\hline
$\mathcal{ST}$ & 1 & KW \& BAT & 6.8 & 0.015 & 15.9 & $<0.001$ \\
$\mathcal{ST}$ & 2 & KW \& BAT & 10.3 & 0.003 & 43.5 & $<0.001$ \\
$\mathcal{ST}$ & 3 & KW \& BAT & 15.5 & $<0.001$ & 72.8 & $<0.001$ \\
\hline
\hline
$\mathcal{N}$ & 1 & GBM all & 18.7 & $<0.001$ & 10.8 & 0.003 \\
$\mathcal{N}$ & 2 & GBM all & 6.0 & 0.031 & 16.2 & $<0.001$ \\
$\mathcal{N}$ & 3 & GBM all & 11.7 & 0.002 & 40.0 & $<0.001$ \\
\hline
$\mathcal{SN}$ & 1 & GBM all & 5.4 & 0.041 & 3.5 & 0.101 \\
$\mathcal{SN}$ & 2 & GBM all & 7.8 & 0.012 & 30.1 & $<0.001$ \\
$\mathcal{SN}$ & 3 & GBM all & 20.4 & $<0.001$ & 66.8 & $<0.001$ \\
\hline
$\mathcal{T}$ & 1 & GBM all & 4.9 & 0.053 & \textcolor{red}{0.0} & 0.579 \\
$\mathcal{T}$ & 2 & GBM all & \textcolor{blue}{3.7} & 0.097 & 16.9 & $<0.001$ \\
$\mathcal{T}$ & 3 & GBM all & \textcolor{blue}{3.1} & 0.131 & 34.4 & $<0.001$ \\
\hline
$\mathcal{ST}$ & 1 & GBM all & \textcolor{red}{0.0} & 0.615 & \textcolor{red}{1.2} & 0.318 \\
$\mathcal{ST}$ & 2 & GBM all & 7.0 & 0.019 & 32.3 & $<0.001$ \\
$\mathcal{ST}$ & 3 & GBM all & 26.5 & $<0.001$ & 75.9 & $<0.001$ \\

\hline

\noalign{\smallskip}
\end{tabular}
$$ 
\tablefoot{The acceptable models are shown in red ($\Delta \textrm{IC}< 2$) and blue ($2 < \Delta \textrm{AIC} < 4$) colour.
Cl. provides the number of GRB classes, $\mathcal{N}$ ($\mathcal{SN}$) and $\mathcal{T}$ ($\mathcal{ST}$) denote the symmetric (skewed) Gaussian and Student distributions, correspondingly.}
\end{table}

\begin{table}
\caption[]{Information criteria for the rest-frame hardness-$T_{50}$ duration distributions of the ``KW  trig'' sample.}
\label{tab:T50kw}

$$ 
\begin{tabular}{llrrrrr}
\hline
\noalign{\smallskip}
Statistics & Cl. & Sample & $\Delta_\textrm{AIC}$ & $w_i$ (AIC) & $\Delta_\textrm{BIC}$ & $w_i$ (BIC) \\
\noalign{\smallskip}
\hline
\noalign{\smallskip}

$\mathcal{N}$ & 1 & KW trig & 13.4 & $<0.001$ & 6.9 & 0.029 \\
$\mathcal{N}$ & 2 & KW trig & \textcolor{blue}{3.7} & 0.056 & 16.8 & $<0.001$ \\
$\mathcal{N}$ & 3 & KW trig & 4.2 & 0.044 & 36.8 & $<0.001$ \\
\hline
$\mathcal{SN}$ & 1 & KW trig & \textcolor{red}{0.0} & 0.355 & \textcolor{red}{0.0} & 0.907 \\
$\mathcal{SN}$ & 2 & KW trig & \textcolor{red}{0.9} & 0.227 & 27.0 & $<0.001$ \\
$\mathcal{SN}$ & 3 & KW trig & 8.3 & 0.006 & 60.6 & $<0.001$ \\
\hline
$\mathcal{T}$ & 1 & KW trig & 13.1 & 0.001 & 9.9 & 0.006 \\
$\mathcal{T}$ & 2 & KW trig & \textcolor{blue}{2.1} & 0.124 & 18.5 & $<0.001$ \\
$\mathcal{T}$ & 3 & KW trig & 6.4 & 0.014 & 42.3 & $<0.001$ \\
\hline
$\mathcal{ST}$ & 1 & KW trig & \textcolor{blue}{2.2} & 0.118 & 5.5 & 0.058 \\
$\mathcal{ST}$ & 2 & KW trig & \textcolor{blue}{3.8} & 0.053 & 33.2 & $<0.001$ \\
$\mathcal{ST}$ & 3 & KW trig & 11.0 & 0.001 & 66.5 & $<0.001$ \\

\hline

\noalign{\smallskip}
\end{tabular}
$$ 
\tablefoot{The acceptable models are shown in red ($\Delta \textrm{IC}< 2$) and blue ($2 < \Delta \textrm{AIC} < 4$) colour.
Cl. provides the number of GRB classes, $\mathcal{N}$ ($\mathcal{SN}$) and $\mathcal{T}$ ($\mathcal{ST}$) denote the symmetric (skewed) Gaussian and Student distributions, correspondingly.}
\end{table}

\begin{table}
\caption[]{Information criteria for the rest-frame hardness-$T_{50}$ duration distributions of the ``KW  \& BAT'' sample.}
\label{tab:T50bat}

$$ 
\begin{tabular}{llrrrrr}
\hline
\noalign{\smallskip}
Statistics & Cl. & Sample & $\Delta_\textrm{AIC}$ & $w_i$ (AIC) & $\Delta_\textrm{BIC}$ & $w_i$ (BIC) \\
\noalign{\smallskip}
\hline
\noalign{\smallskip}

$\mathcal{N}$ & 1 & KW \& BAT & \textcolor{red}{0.2} & 0.272 & \textcolor{red}{0.0} & 0.947 \\
$\mathcal{N}$ & 2 & KW \& BAT & \textcolor{red}{1.4} & 0.149 & 19.9 & $<0.001$ \\
$\mathcal{N}$ & 3 & KW \& BAT & \textcolor{red}{0.0} & 0.301 & 37.2 & $<0.001$ \\
\hline
$\mathcal{SN}$ & 1 & KW \& BAT & 4.2 & 0.037 & 10.3 & 0.005 \\
$\mathcal{SN}$ & 2 & KW \& BAT & 5.1 & 0.023 & 36.0 & $<0.001$ \\
$\mathcal{SN}$ & 3 & KW \& BAT & 6.1 & 0.014 & 62.0 & $<0.001$ \\
\hline
$\mathcal{T}$ & 1 & KW \& BAT & \textcolor{blue}{3.1} & 0.064 & 6.0 & 0.047 \\
$\mathcal{T}$ & 2 & KW \& BAT & \textcolor{red}{1.8} & 0.122 & 23.4 & $<0.001$ \\
$\mathcal{T}$ & 3 & KW \& BAT & 13.0 & $<0.001$ & 53.3 & $<0.001$ \\
\hline
$\mathcal{ST}$ & 1 & KW \& BAT & 7.6 & 0.007 & 16.7 & $<0.001$ \\
$\mathcal{ST}$ & 2 & KW \& BAT & 8.0 & 0.006 & 42.1 & $<0.001$ \\
$\mathcal{ST}$ & 3 & KW \& BAT & 8.4 & 0.005 & 67.4 & $<0.001$ \\

\hline

\noalign{\smallskip}
\end{tabular}
$$ 
\tablefoot{The acceptable models are shown in red ($\Delta \textrm{IC}< 2$) and blue ($2 < \Delta \textrm{AIC} < 4$) colour.
Cl. provides the number of GRB classes, $\mathcal{N}$ ($\mathcal{SN}$) and $\mathcal{T}$ ($\mathcal{ST}$) denote the symmetric (skewed) Gaussian and Student distributions, correspondingly.}
\end{table}

\begin{table}
\caption[]{Information criteria for the rest-frame hardness-observer-frame duration distributions.}
\label{tab:obsT90}
$$ 
\begin{tabular}{llrrrrr}
\hline
\noalign{\smallskip}
Statistics & Cl. & Sample & $\Delta_\textrm{AIC}$ & $w_i$ (AIC) & $\Delta_\textrm{BIC}$ & $w_i$ (BIC) \\
\noalign{\smallskip}
\hline
\noalign{\smallskip}

$\mathcal{N}$ & 1 & Total & 66.7 & $<0.001$ & 56.2 & $<0.001$ \\
$\mathcal{N}$ & 2 & Total & \textcolor{red}{0.6} & 0.179 & 14.2 & 0.001 \\
$\mathcal{N}$ & 3 & Total & \textcolor{red}{0.7} & 0.170 & 38.3 & $<0.001$ \\
\hline
$\mathcal{SN}$ & 1 & Total & \textcolor{blue}{2.5} & 0.069 & \textcolor{red}{0.0} & 0.950 \\
$\mathcal{SN}$ & 2 & Total & 5.2 & 0.018 & 34.8 & $<0.001$ \\
$\mathcal{SN}$ & 3 & Total & \textcolor{red}{0.0} & 0.241 & 61.7 & $<0.001$ \\
\hline
$\mathcal{T}$ & 1 & Total & 61.3 & $<0.001$ & 54.8 & $<0.001$ \\
$\mathcal{T}$ & 2 & Total & \textcolor{blue}{4.0} & 0.033 & 21.6 & $<0.001$ \\
$\mathcal{T}$ & 3 & Total & \textcolor{blue}{2.4} & 0.073 & 44.1 & $<0.001$ \\
\hline
$\mathcal{ST}$ & 1 & Total & 4.4 & 0.027 & 5.9 & 0.050 \\
$\mathcal{ST}$ & 2 & Total & 8.9 & 0.003 & 42.5 & $<0.001$ \\
$\mathcal{ST}$ & 3 & Total & \textcolor{red}{0.5} & 0.188 & 66.2 & $<0.001$ \\
\hline

\noalign{\smallskip}
\end{tabular}
$$ 
\tablefoot{The acceptable models are shown in red ($\Delta \textrm{IC} < 2$) and blue ($2 < \Delta \textrm{AIC} < 4$) colour.
Cl. provides the number of GRB classes, $\mathcal{N}$ ($\mathcal{SN}$) and $\mathcal{T}$ ($\mathcal{ST}$) denote the symmetric (skewed) Gaussian and Student distributions, correspondingly.}
\end{table}

\begin{table}
\caption[]{Information criteria for the the samples generated using the Jackknife technique with 10\% and 30\% of the sample omitted and the Bootstrap method. }
\label{tab:sim}
$$ 
\begin{tabular}{lllll}
\hline
\noalign{\smallskip}
Distr. & Cl. & $n$ & IC type & IC value \\
\noalign{\smallskip}
\hline
\noalign{\smallskip}

\multicolumn{5}{c}{Jackknife 10\%} \\
\hline
$\mathcal{N}$ & 1 & 368 & AIC & $1198.2 \pm 13.0$ \\
$\mathcal{N}$ & 2 & 368 & AIC & $\textcolor{ForestGreen}{1162.7 \pm 12.9}$ \\
$\mathcal{N}$ & 3 & 368 & AIC & $\textcolor{ForestGreen}{1164.8 \pm 12.7}$ \\
\hline
$\mathcal{SN}$ & 1 & 368 & AIC & $\textcolor{red}{1156.9 \pm 12.9}$ \\
$\mathcal{SN}$ & 2 & 368 & AIC & $\textcolor{ForestGreen}{1166.9 \pm 12.9}$ \\
$\mathcal{SN}$ & 3 & 368 & AIC & $\textcolor{ForestGreen}{1172.1 \pm 13.1}$ \\
\hline
$\mathcal{N}$ & 1 & 368 & BIC & $1217.8 \pm 12.9$ \\
$\mathcal{N}$ & 2 & 368 & BIC & $\textcolor{ForestGreen}{1205.7 \pm 12.9}$ \\
$\mathcal{N}$ & 3 & 368 & BIC & $1231.2 \pm 12.7$ \\
\hline
$\mathcal{SN}$ & 1 & 368 & BIC & $\textcolor{red}{1184.3 \pm 12.9}$ \\
$\mathcal{SN}$ & 2 & 368 & BIC & $1225.5 \pm 12.9$ \\
$\mathcal{SN}$ & 3 & 368 & BIC & $1262.0 \pm 13.1$ \\
\hline
\multicolumn{5}{c}{Jackknife 30\%}\\
\hline
$\mathcal{N}$ & 1 & 286 & AIC & $\textcolor{ForestGreen}{930.0 \pm 21.6}$ \\
$\mathcal{N}$ & 2 & 286 & AIC & $\textcolor{ForestGreen}{904.3 \pm 21.4}$ \\
$\mathcal{N}$ & 3 & 286 & AIC & $\textcolor{ForestGreen}{907.3 \pm 21.8}$ \\
\hline
$\mathcal{SN}$ & 1 & 286 & AIC & $\textcolor{red}{898.6 \pm 21.6}$ \\
$\mathcal{SN}$ & 2 & 286 & AIC & $\textcolor{ForestGreen}{908.1 \pm 21.0}$ \\
$\mathcal{SN}$ & 3 & 286 & AIC & $\textcolor{ForestGreen}{914.0 \pm 21.1}$ \\
\hline
$\mathcal{N}$ & 1 & 286 & BIC & $\textcolor{ForestGreen}{948.3 \pm 21.6}$ \\
$\mathcal{N}$ & 2 & 286 & BIC & $\textcolor{ForestGreen}{944.5 \pm 21.4}$ \\
$\mathcal{N}$ & 3 & 286 & BIC & $969.5 \pm 21.8$ \\
\hline
$\mathcal{SN}$ & 1 & 286 & BIC & $\textcolor{red}{924.2 \pm 21.6}$ \\
$\mathcal{SN}$ & 2 & 286 & BIC & $\textcolor{ForestGreen}{962.9 \pm 21.0}$ \\
$\mathcal{SN}$ & 3 & 286 & BIC & $998.1 \pm 21.1$ \\
\hline
\multicolumn{5}{c}{Bootstrap}\\
\hline
$\mathcal{N}$ & 1 & 409 & AIC & $\textcolor{ForestGreen}{1329.7 \pm 38.3}$ \\
$\mathcal{N}$ & 2 & 409 & AIC & $\textcolor{ForestGreen}{1282.8 \pm 36.3}$ \\
$\mathcal{N}$ & 3 & 409 & AIC & $\textcolor{red}{1276.5 \pm 36.4}$ \\
\hline
$\mathcal{SN}$ & 1 & 409 & AIC & $\textcolor{ForestGreen}{1281.2 \pm 36.0}$ \\
$\mathcal{SN}$ & 2 & 409 & AIC & $\textcolor{ForestGreen}{1283.4 \pm 35.8}$ \\
$\mathcal{SN}$ & 3 & 409 & AIC & $\textcolor{red}{1277.6 \pm 38.3}$ \\
\hline
$\mathcal{N}$ & 1 & 409 & BIC & $\textcolor{ForestGreen}{1349.8 \pm 38.3}$ \\
$\mathcal{N}$ & 2 & 409 & BIC & $\textcolor{ForestGreen}{1327.0 \pm 36.2}$ \\
$\mathcal{N}$ & 3 & 409 & BIC & $\textcolor{ForestGreen}{1344.7 \pm 36.4}$ \\
\hline
$\mathcal{SN}$ & 1 & 409 & BIC & $\textcolor{red}{1309.3 \pm 36.0}$ \\
$\mathcal{SN}$ & 2 & 409 & BIC & $\textcolor{ForestGreen}{1343.7 \pm 35.8}$ \\
$\mathcal{SN}$ & 3 & 409 & BIC & $\textcolor{ForestGreen}{1369.9 \pm 38.3}$ \\
\hline

\noalign{\smallskip}
\end{tabular}
$$ 
\tablefoot{See Section~\ref{sec:validation} for details. 
The acceptable models are shown in red ($\Delta \textrm{IC} < 2$) and blue ($2 < \Delta \textrm{AIC} < 4$) colour. 
The models that are suitable within the IC uncertainties (standard deviations of the generated samples) are shown
in green.
Cl. provides the number of GRB classes, $n$ shows the number of points in the generated data set, $\mathcal{N}$ ($\mathcal{SN}$) and $\mathcal{T}$ ($\mathcal{ST}$) denote the symmetric (skewed) Gaussian and Student distributions, correspondingly.}
\end{table}

\begin{table}
\caption[]{Information criteria for the the samples generated using the uncertainties of the rest-frame $T_{90}$ and $E_p$.}
\label{tab:sim_err}

$$
\begin{tabular}{lccc}
\hline
\noalign{\smallskip}
Distr. & Cl. & AIC & BIC \\
\noalign{\smallskip}
\hline
\noalign{\smallskip}

$\mathcal{N}$ & 1 & $1356.8 \pm 19.2$ & $\textcolor{ForestGreen}{1376.9 \pm 19.3}$ \\
$\mathcal{N}$ & 2 & $\textcolor{blue}{1316.8 \pm 17.3}$ & $\textcolor{ForestGreen}{1360.9 \pm 17.3}$ \\
$\mathcal{N}$ & 3 & $\textcolor{ForestGreen}{1320.1 \pm 16.9}$ & $1388.3 \pm 16.9$ \\
\hline
$\mathcal{SN}$ & 1 & $\textcolor{red}{1314.1 \pm 18.5}$ & $\textcolor{red}{1342.2 \pm 18.5}$ \\
$\mathcal{SN}$ & 2 & $\textcolor{ForestGreen}{1321.0 \pm 17.0}$ & $1381.2 \pm 17.0$ \\
$\mathcal{SN}$ & 3 & $\textcolor{ForestGreen}{1327.2 \pm 16.6}$ & $1419.5 \pm 16.6$ \\
\hline
$\mathcal{T}$ & 1 & $1355.8 \pm 16.5$ & $1379.9 \pm 16.5$ \\
$\mathcal{T}$ & 2 & $\textcolor{ForestGreen}{1319.1 \pm 17.1}$ & $\textcolor{ForestGreen}{1367.3 \pm 17.1}$ \\
$\mathcal{T}$ & 3 & $\textcolor{ForestGreen}{1322.2 \pm 16.7}$ & $1394.5 \pm 16.7$ \\
\hline
$\mathcal{ST}$ & 1 & $\textcolor{red}{1315.1 \pm 16.6}$ & $\textcolor{ForestGreen}{1347.2 \pm 16.6}$ \\
$\mathcal{ST}$ & 2 & $\textcolor{ForestGreen}{1324.1 \pm 17.1}$ & $1388.3 \pm 17.1$ \\
$\mathcal{ST}$ & 3 & $\textcolor{ForestGreen}{1329.7 \pm 17.0}$ & $1426.0 \pm 17.0$ \\
\hline

\noalign{\smallskip}
\end{tabular}
$$

\tablefoot{See Section~\ref{sec:validation} for details. 
The acceptable models are shown in red ($\Delta \textrm{IC} < 2$) and blue ($2 < \Delta \textrm{AIC} < 4$) colour. 
The models that are suitable within the IC uncertainties (standard deviations of the generated sample), are shown in green.
Cl. provides the number of GRB classes, $\mathcal{N}$ ($\mathcal{SN}$) and $\mathcal{T}$ ($\mathcal{ST}$) denote the symmetric (skewed) Gaussian and Student distributions, correspondingly.}
\end{table}

\begin{table}
\caption[]{Information criteria for the rest-frame hardness-duration distributions of 155 GRBs from the ``GBM all'' sample.}
\label{tab:rest_900keV}
$$ 
\begin{tabular}{llrrrr}
\hline
\noalign{\smallskip}

Statistics & Cl. & $\Delta_\textrm{AIC}$ & $w_i$ (AIC) & $\Delta_\textrm{BIC}$ & $w_i$ (BIC) \\
\noalign{\smallskip}
\hline
\noalign{\smallskip}

\multicolumn{6}{c}{In the $(100-900)/(1+z)$~keV energy window}  \\
\hline
$\mathcal{N}$ & 1 & 7.6 & 0.012 & \textcolor{red}{1.7} & 0.296\\
$\mathcal{N}$ & 2 & 6.4 & 0.021 & 18.7 & $<0.001$\\
$\mathcal{N}$ & 3 & 9.4 & 0.005 & 40.1 & $<0.001$\\
\hline
$\mathcal{SN}$ & 1 & 11.8 & 0.001 & 12.0 & 0.002\\
$\mathcal{SN}$ & 2 & 10.8 & 0.002 & 35.3 & $<0.001$\\
$\mathcal{SN}$ & 3 & 24.0 & $<0.001$ & 72.9 & $<0.001$\\
\hline
$\mathcal{T}$ & 1 & \textcolor{blue}{2.8} & 0.128 & \textcolor{red}{0.0} & 0.693\\
$\mathcal{T}$ & 2 & \textcolor{blue}{2.6} & 0.141 & 18.0 & $<0.001$\\
$\mathcal{T}$ & 3 & 5.8 & 0.029 & 39.4 & $<0.001$\\
\hline
$\mathcal{ST}$ & 1 & 5.5 & 0.033 & 8.7 & 0.009\\
$\mathcal{ST}$ & 2 & \textcolor{red}{0.0} & 0.518 & 27.6 & $<0.001$\\
$\mathcal{ST}$ & 3 & \textcolor{blue}{3.1} & 0.110 & 55.1 & $<0.001$\\
\hline

\multicolumn{6}{c}{In the $(100-900)$~keV energy window}\\
\hline
$\mathcal{N}$ & 1 & 15.0 & $<0.001$ & 3.1 & 0.070\\
$\mathcal{N}$ & 2 & 4.8 & 0.043 & 11.2 & 0.001\\
$\mathcal{N}$ & 3 & \textcolor{red}{0.0} & 0.470 & 24.6 & $<0.001$\\
\hline
$\mathcal{SN}$ & 1 & 5.8 & 0.026 & \textcolor{red}{0.0} & 0.330\\
$\mathcal{SN}$ & 2 & 9.8 & 0.003 & 28.3 & $<0.001$\\
$\mathcal{SN}$ & 3 & 11.1 & 0.002 & 54.0 & $<0.001$\\
\hline
$\mathcal{T}$ & 1 & 8.9 & 0.005 & \textcolor{red}{0.1} & 0.313\\
$\mathcal{T}$ & 2 & \textcolor{red}{0.7} & 0.331 & 10.1 & 0.002\\
$\mathcal{T}$ & 3 & 6.7 & 0.016 & 34.4 & $<0.001$\\
\hline
$\mathcal{ST}$ & 1 & \textcolor{blue}{3.1} & 0.100 & \textcolor{red}{0.3} & 0.284\\
$\mathcal{ST}$ & 2 & 9.8 & 0.003 & 31.4 & $<0.001$\\
$\mathcal{ST}$ & 3 & 13.7 & $<0.001$ & 59.6 & $<0.001$\\
\hline

\noalign{\smallskip}
\end{tabular}
$$ 
\tablefoot{The durations were computed in the observer- and rest-frame energy windows $100-900$~keV, which in the latter case corresponds to the $\frac{100}{1+z}-\frac{900}{1+z}$~keV instrumental bandpass.
In both cases, the durations were corrected for the cosmological time dilation.
The acceptable models are shown in red ($\Delta \textrm{IC} < 2$) and blue ($2 < \Delta \textrm{AIC} < 4$) colour.
Cl. provides the number of GRB classes, $\mathcal{N}$ ($\mathcal{SN}$) and $\mathcal{T}$ ($\mathcal{ST}$) denote the symmetric (skewed) Gaussian and Student distributions, correspondingly.}
\end{table}

\end{appendix}
\end{document}